\documentclass[11pt]{article}

\usepackage[margin=1in]{geometry}
\usepackage{mathtools}
\usepackage[colorinlistoftodos]{todonotes}
\usepackage{xspace}
\usepackage{graphicx}
\usepackage{float}
\usepackage{wrapfig}
\usepackage{comment}
\usepackage{upgreek}
\usepackage{nicefrac}
\usepackage{colordvi}
\usepackage{color}
\usepackage{url}
\usepackage{paralist}
\usepackage{bm}
\usepackage{mathpazo}
\usepackage[normalem]{ulem}
\usepackage{tikz}
\usetikzlibrary{arrows,petri,topaths,automata,backgrounds,petri,positioning,fit}
\usepackage{tkz-berge}
\usepackage[position=top]{subfig}
\usepackage{hyperref}
\usepackage{amsfonts}
\usepackage{amssymb}
\usepackage{amstext}
\usepackage{amsmath}
\usepackage{amsthm}
\usepackage[noabbrev,sort]{cleveref}

\newtheorem{theorem}{Theorem}
\newtheorem{lemma}[theorem]{Lemma}

\newtheorem{observation}{Observation}
\newtheorem{fact}{Fact}

\newtheorem{corollary}[theorem]{Corollary}
\newtheorem{conjecture}[theorem]{Conjecture}
\newtheorem{claim}[theorem]{Claim}
\newtheorem{definition}[theorem]{Definition}
\newtheorem{proposition}[theorem]{Proposition}

\Crefname{claim}{Claim}{Claims}
\Crefname{observation}{Observation}{Observations}
\Crefname{fact}{Fact}{Facts}
\Crefname{proposition}{Proposition}{Propositions}

\newcommand{\strike}[1]{\textcolor{red}{{\ifmmode\text{\sout{\ensuremath{#1}}}\else\sout{#1}\fi}}}

\newcommand{\overbar}[1]{\mkern 1.5mu\overline{\mkern-1.5mu#1\mkern-1.5mu}\mkern 1.5mu}

\newcommand{\tset}{{\mathcal T}}

\newcommand{\aset}{\mathcal{A}}
\newcommand{\sset}{\mathcal{S}}
\newcommand{\cset}{\mathcal{C}}
\newcommand{\dset}{{\mathcal D}}

\newcommand{\Chi}{\mbox{\Large$\chi$}}
\newcommand{\Nu}{{\mathcal V}}
\newcommand{\CL}{{\mathcal CL}}

\newif\ifarxiv
\newif\ifjournal

\arxivtrue

\renewcommand{\l}{\ell}

\title{Multi-transversals for Triangles and the Tuza's Conjecture\thanks{Part of the work was done while PS was visiting Aalto University and KTH, Royal Institute of Technology, and while PS was studying at the University of Maryland. }\ifjournal~\thanks{The full version of our paper is available at \href{https://arxiv.org/abs/2001.00257}{https://arxiv.org/abs/2001.00257}}\fi}
\author{Parinya Chalermsook\thanks{{\tt parinya.chalermsook@aalto.fi}, Aalto University, Espoo, Finland} \and Samir Khuller\thanks{{\tt samir.khuller@northwestern.edu}, Northwestern University, Evanston, USA} \and Pattara Sukprasert\thanks{{\tt pattara.sk127@gmail.com}, Northwestern University, Evanston, USA} \and Sumedha Uniyal\thanks{{\tt sumedha.p.uniyal@gmail.com}, Aalto University, Espoo, Finland}}

\date{}

\begin{document}
\maketitle

\begin{abstract}
In this paper, we study a  primal and dual relationship about triangles:
For any graph $G$, let $\nu(G)$ be the maximum number of edge-disjoint triangles in $G$, and $\tau(G)$ be the minimum subset $F$ of edges such that $G \setminus F$ is triangle-free. It is easy to see that $\nu(G) \leq \tau(G) \leq 3 \nu(G)$, and in fact, this rather obvious inequality holds for a much more general primal-dual relation between $k$-hyper matching and covering in hypergraphs.
Tuza conjectured in $1981$ that $\tau(G) \leq 2 \nu(G)$, and this question has received attention from various groups of researchers in discrete mathematics, settling various special cases such as planar graphs and generalized to bounded maximum average degree graphs, some cases of minor-free graphs, and very dense graphs.
Despite these efforts, the conjecture in general graphs has remained wide open for almost four decades.

In this paper, we provide a proof of a non-trivial consequence of the conjecture;  that is, for every $k \geq 2$, there exist a (multi)-set $F \subseteq E(G): |F| \leq 2k \nu(G)$ such that each triangle in $G$ overlaps at least $k$ elements in $F$.
Our result can be seen as a strengthened statement of Krivelevich's result on the fractional version of Tuza's conjecture (and we give some examples illustrating this.)
The main technical ingredient of our result is a charging argument, that locally identifies edges in $F$
based on a local view of the packing solution.
This idea might be useful in further studying the primal-dual relations in general and the Tuza's conjecture in particular.

\end{abstract}

\newpage

\tableofcontents

\section{Introduction}

The study of the relationship between primal and dual graph problems has been a cornerstone in combinatorial optimization, discrete mathematics, and design of approximation algorithms.
Among the oldest such relations is perhaps the ratio between {\em maximum matching} (denoted by $\nu(G)$) and {\em minimum vertex cover} (denoted by $\tau(G)$), for which  $\nu(G) \leq \tau(G) \leq 2 \nu(G)$ holds for any graph $G$, and this inequality is tight.
In special graph classes, the factor of two can be improved or even removed completely, for instance, in the famous K\"{o}nig's theorem, we learn that $\nu(G) = \tau(G)$ for any bipartite graph $G$ \cite{konig1936theorie}.
This notion of matching and covering can be generalized to any $r$-uniform hypergraphs, where $\nu(G)$ now denotes the cardinality of maximum hyper-matching (a collection of hyperedges that are pairwise disjoint) and $\tau(G)$ is called \textit{transversal} of hypergraphs (a collection of vertices that hit every hyperedge).
It is easy to see that the relation $\nu(G) \leq \tau(G) \leq r \cdot \nu(G)$ still holds, and in general, a factor $r$ is tight: For instance, when $r=3$, a natural set system of Fano plane provides a tight example where $\nu(G) = 1$ but $\tau(G) = 3$.

Knowing K\"{o}nig's theorem, one wonders in what  graph classes, the trivial relation  $\tau(G) \leq r \cdot \nu(G)$ could be improved. An old conjecture, called Ryser's conjecture\footnote{The conjecture was first stated in a thesis of his student, J. Henderson in 1971}, states that $\tau(G) \leq (r-1) \nu(G)$ holds for any $r$-partite hypergraphs (this conjecture can easily be seen as generalizing K\"{o}nig's theorem to any $r$-uniform hypergraphs.)
The conjecture remains wide open for $r \geq 4$ (the $r=3$ case was resolved~\cite{aharoni2001}).
Another direction that has received a lot of attention focuses on hypergraphs that are defined from specific (simple) graph structures: For any simple graph $G=(V,E)$, an $r$-uniform cycle hypergraph is created by having one vertex for each edge in $E$ and a hyperedge for any $r$-cycle $\{e_1,e_2,\ldots, e_r\}$ in $G$.
Now the matching and covering questions turn into studying the ratio between the maximum edge-disjoint $r$-cycle packing and the minimum  edge set that hits all cycles of length $r$.
In particular, when $r=3$, we are interested in the relation between the number of disjoint triangles v.s. the set of edges that hit all triangles in $G$. Denote these numbers by $\nu^c(G)$ and $\tau^c(G)$ respectively.
Tuza conjectured in $1981$ that $\tau^c(G) \leq 2 \nu^c(G)$ and verified the conjecture  for planar graphs~\cite{tuza1981conjecture,T90a}.
Tuza's conjecture has since received attention from various groups of researchers, leading to  confirmations of the conjecture for many special  cases such as $K_{3,3}$-free graphs \cite{K95}, graphs with maximum average degree less than $7$~\cite{Puleo15}, graphs of treewidth at most $6$ \cite{botler2018tuza}, and graphs that contain quadratic-sized edge-disjoint triangles~\cite{yuster2012dense,haxell2001integer}.

In general graphs, Haxell showed that $\tau^c(G) \leq  (3-\epsilon) \nu^c(G)$~\cite{H99}, confirming at least the principles that the trivial ratio of $3$ can really be improved. Haxell's result has remained the best known ratio for general graphs.
Whether Tuza's conjecture holds or not has remained an intriguing open problem.


\subsection{Our contributions}

In this paper, we study a natural question that is a consequence of the Tuza's conjecture and discuss potential applications of our work in making progress towards resolving the conjecture.
From now on, we denote $\tau^c(G)$ and $\nu^c(G)$ simply by $\tau(G)$ and $\nu(G)$ respectively.

Our work is inspired from the mathematical programming perspectives of the Tuza's conjecture.
For each edge $e \in E$, we have a real variable $x_e$.
Define $\tau^*(G)$ to be an optimal value to the following linear program:

\begin{eqnarray*}
 \mbox{(LP)}
  & \min & \sum_e x_e \\
  & \mbox{s.t.} & \sum_{e: e \in E(t)} x_e \geq 1 \mbox{ for all triangles $t$}
\end{eqnarray*}

Clearly, since the above LP is a relaxation of the covering number, we have that $\tau^*(G) \leq \tau(G)$. Observe that the Tuza's conjecture would imply the weaker inequality $\tau^*(G) \leq 2 \nu(G)$, and this can be seen as a fractional variant of the Tuza's conjecture.
Krivelevich~\cite{K95} showed that $\tau^*(G) \leq 2 \nu(G)$ (later strengthened by \cite{chapuy2014packing}).

Another way to phrase Krivelevich's result (without bothering about linear programs) is through the notion of multi-transversals: For some $k \in {\mathbb N}$, there exist a multi-set $F \subseteq E(G): |F| \leq 2k \nu(G) $ such that each triangle in $G$ contains at least $k$ elements of $F$.
We call such set a {\em multi-transversal} set of order $k$ (or $k$-multi-transversal), and define $\tau^*_k(G)$ as the minimum value of $\frac{|F|}{k}$ for any $k$-multi-transversal set $F$.
Observe that the existence of $k$-multi-transversal set $F$ is equivalent to saying that there exists a fractional cover $z: E(G) \rightarrow \{0,1/k, 2/k,\ldots, 1\}$ that is feasible for (LP) such that $\sum_{e \in E(G)}z(e) = \frac{|F|}{k}$. We call such a feasible solution a {\em $(1/k)$-integral solution}.
Notice that, the case when $k=1$ (the $1$-multi-transversal set)  is equivalent to the transversal set in the sense of Tuza's conjecture, and that $(\forall k:) \tau^*(G)  \leq \tau^*_k(G) \leq \tau^*_1(G) = \tau(G)$.

It is then natural to consider the following question:

\begin{conjecture}[Strong Fractional Tuza's Conjecture]
\label{conj: strong frac}
For any $k \in {\mathbb N}$ and $k\geq 2$, we have $\tau^*_k(G) \leq 2 \nu(G)$ for any graph $G$.
\end{conjecture}

Clearly, \Cref{conj: strong frac} is a consequence of the Tuza's conjecture and it implies the standard fractional version of Tuza's conjecture that was resolved by Krivelevich. Krivelevich's proof is based on induction and only implies that $\tau^*_k(G) \leq 2 \nu(G)$ for some (very large) $k \in {\mathbb N}$.
\footnote{We note that the values of $\tau^*_k$ are not monotone in $k$: There is an infinite family of graphs $G$ for which $\tau^*_2(G) \leq (\frac 3 4 +o(1)) \tau^*_3(G)$, and another family of graphs $G$ for which $\tau^*_3(G) \leq (\frac{5}{6} +o(1)) \tau^*_2(G)$. See ~\Cref{subsec:comparison} for more detail.}

In this paper, we resolve the above question in the affirmative.

\begin{theorem}
\label{thm:main}
For all integer $k \geq 2$ and any graph $G$, $\tau^*_k(G) \leq 2 \nu(G)$. Moreover, we can efficiently find the triangle packing solution and the $k$-multi-transversal that together achieve this bound.
\end{theorem}

The proof of this theorem is inspired by the local search technique. Based on an optimal packing solution, we define a collection of edges in the covering solution via the local view of the packing solution. Our proofs are constructive in nature. Starting with any packing solution, we can either find a multi-transversal or improve the current packing solution. Because the packing solution can be improved at most $O(n^3)$ times, hence we get an algorithm to find the triangle packing solution and the $k$-multi-transversal that together achieve this bound.

We believe that the structural insights from this work would be useful in attacking the Tuza's conjecture in the future.

\subsection{Comparisons with Fractional Transversals}
\label{subsec:comparison}
We discuss two showcases which illustrate that our concept of multi-transversal of lower order is stronger than the standard fractional LP solution.

First we show that proving \Cref{conj: strong frac} can be reduced to the cases when $k = 2$ and $k = 3$.

\begin{proposition}
If $\tau^*_2(G) \leq 2 \nu(G)$ and $\tau^*_3(G) \leq 2 \nu(G)$, then $\tau^*_k(G) \leq 2 \nu(G)$ for all $k \geq 2$.
\end{proposition}
\begin{proof}
Consider any graph $G$.
Let $F_2$ and $F_3$ be multi-transversal sets of orders $2$ and $3$ respectively, so any triangle $t$ in $G$ is hit by two elements of $F_2$ and $3$ elements of $F_3$.
We will use (multiple copies of) $F_2$ and $F_3$ to hit every triangle in $G$ at least $k$ times.
For even $k = 2q$, we can simply use $q$ copies of $F_2$, and this would hit each triangle $2q$ times.
For odd $k>3$, we can write $k  = 2 q + 3$, and so we can use $q$ copies of $F_2$ together with $1$ copy of $F_3$.
\end{proof}

Now we show that $\tau^*_2(G)$ and $\tau^*_3(G)$ are not subsumed by each other, so we would need to consider both cases.

\begin{claim}
There is an infinite family of graphs $G$ for which $\tau^*_2(G) \leq (\frac{3}{4}+o(1)) \tau^*_3(G)$
\end{claim}
\begin{proof}
We will illustrate a family of $n$-vertex graphs $G_n$ where $\tau^*_2(G_n) = \frac{n}{2}$ but $\tau^*_3(G_n) = \frac{2}{3}(1-o(1)) n$.
By~\cite{kortsarz2010approximating}, it suffices to construct a triangle-free graph $G_n$ where the gap exists between the fractional and integral optimal solutions of the vertex cover problem.
We consider a triangle-free graph $G_n$ where ${\sf VC}(G_n) = (1-o(1))n$ (this is just a random graph $G(n,p)$ with appropriate parameter and using the alteration steps to remove all triangles).
In any graph, there is a fractional feasible cover of value $\frac n 2$ (by assigning $\frac 1 2$ everywhere), so we have that $\tau^*_2(G) \leq \frac n 2$.

Now we analyze $\tau^*_3(G)$. Consider any $\frac{1}{3}$-integral assignment $z$ on vertices $V(G_n)$. Partition $V(G_n)$ based on the assigned $z$-values into $V_0, V_{1/3}$, $V_{2/3}$, and $V_{1}$.
Notice that $V_0 \cup V_{1/3}$ must be an independent set, for otherwise an edge $e \in G[V_0 \cup V_{1/3}]$ would violate the covering constraint.
Therefore, $|V_0 \cup V_{1/3}| = o(n)$ (since the size of any independent set in $G$ is at most $o(n)$).
This implies that the total assignment must be at least $\frac{2}{3} (1-o(1)) n$.
\end{proof}

\begin{claim}
There is an infinite family of graphs $G$ for which $\tau^*_3(G) \leq \frac{5}{6} \tau^*_2(G)$.
\end{claim}
\begin{proof}
We show that this gap holds in the case of a complete graph on $6$ vertices.
In particular, $\tau^*_3(K_6) \leq 5$ but $\tau^*_2(K_6) \geq 6$.
The first claim is easy, since we can simply assign $\frac 1 3$ on each edge, and there are $15$ edges in the graph.
The second claim relies on \Cref{prop: complete graphs} to say that $\tau^*_2(K_6)\geq 6$.
\end{proof}

Next, we will present evidence that $\tau^*_2(G)$ and $\tau^*_3(G)$ are stronger lower bound of $\tau(G)$ than the standard fractional bound $\tau^*(G)$.


The following proposition shows that  $\tau^*_3(G)$ is strictly stronger than $\tau^*(G)$ in general.

\begin{proposition}
For any graph $G$,
 $\tau(G) \leq 1.5 \tau^*_3(G)$, while there exists a family of graphs $G$ for which $\tau(G) = (2-o(1)) \tau^*(G)$.
\end{proposition}
\begin{proof}
We prove this by turning any LP solution $\{z_e\}_{e\in E}$ that is $\frac{1}{3}$-integral into an integral solution of cost at most $1.5 (\sum_{e \in E} z_e)$. For any subset of edges $S \subseteq E$, let $z(S) := \sum_{e \in S} z_e$.
First, we define $E' = \{e: z_e \geq 2/3\}$.
Notice that $|E'| \leq \frac{3}{2} z(E')$.
We consider the remaining edges $\tilde E \subseteq E - E'$ in the support of $z$, i.e. all edges with non-zero $z$ value. Note that for every $e \in \tilde E$, we have $z_e = 1/3$, which implies $z(\tilde E) = |\tilde E|/3$.
This implies that every triangle $t$ in $G - E'$ must satisfy $|E(t) \cap \tilde E| = 3$.
 Moreover, there exists a collection of edges $E'' \subseteq \tilde E$ whose size is $|E''| \leq |\tilde E|/2$ and whose removal creates a bipartite graph; in particular, $E''$ hits every triangle in $G - E'$.
Notice that $|E''| \leq |\tilde E|/2$ while $z(\tilde E) = |\tilde E|/3$, so we have $|E''| \leq \frac{3}{2} z(\tilde E)$.

The set $E' \cup E''$ is our integral solution.

For the family of graphs $G$, for which $\tau(G) = (2-o(1)) \tau^*(G)$, we refer to \cite{kortsarz2010approximating}.
\end{proof}


Another example illustrates that $\tau^*_2(G)$ can be stronger than $\tau^*(G)$ in complete graphs.

\begin{proposition}
\label{prop: complete graphs}
For any even integer $n$, we have that $\tau(K_n) = \tau^*_2(K_n) = n(n-2)/4 = \frac{n^2}{4}(1-o(1))$, while $\tau^*(K_n) = \frac{1}{3} {\binom{n}{2}} = \frac{n^2}{6}(1-o(1))$.
\end{proposition}

\begin{proof}
Notice that $\tau(K_n) \leq n(n-2)/4$ because we can partition the vertex set into two equal sets of size $n/2$ and remove edges in the same set.
Therefore, we have removed ${\binom{n}{2}} - \frac{n^2}{4} = \frac{n (n-2)}{4}$ edges.

We will now argue that $\tau^*_2(K_n) \geq n(n-2)/4$.
Consider an LP solution $\{z_e\}_{e \in E(K_n)}$ that is half-integral. For any subset of edges $S \subseteq E(K_n)$, let $z(S) := \sum_{e \in S} z_e$. We will prove that the value of $z$ is at least $n(n-2)/4$ by induction. It is easy to see that this is true for the base case of $K_2$.
Now consider $n \geq 4$ and graph $G=K_n$.
If there is no edge in $G$ with zero LP value, we are immediately done, since the total LP value must be at least $\frac{1}{2} {\binom{n}{2}}$.
Otherwise, let $(u,v)$ be such an edge with zero LP value.
Consider $V' = V(G) \setminus \{u,v\}$. The induced subgraph $G[V']$ is $K_{n-2}$, hence by IH, we get that the total LP-value $z(E(G[V']))$ is at least $(n-2)(n-4)/4$. Now we analyze the LP-values on the edges between $V'$ and $\{u,v\}$.
For each $w \in V'$, notice that we have $z_{wu} + z_{wv} \geq 1$ since $z_{uv} = 0$ (and the triangle $\{u,v,w\}$ must be hit).
Therefore, the total LP-values on the edges incident to $u$ or $v$ must be at least
\[\sum_{w \in V'} (z_{wu} + z_{wv}) \geq |V'| = (n-2) \]
which leads to the total of $\frac{(n-2)(n-4)}{4} + (n-2) = \frac{n (n-2)}{4}$.
\end{proof}

\subsection{Open problems}

We propose the notion of multi-transversal of any order and show that the Tuza's conjecture holds in this relaxed setting.
It is interesting to see whether our results and techniques would be useful in making further progress.
Open problems that are most directly related to our work are to characterize the graph classes where $\tau^*_2(G) = \tau(G)$ or $\tau^*_3(G) = \tau(G)$.
This condition would be sufficient for graph $G$ to satisfy the Tuza's conjecture.
The fractional notion may appear weak, but it has in fact been used to obtain non-trivial results of Tuza's conjecture in very dense graphs \cite{yuster2012dense} (through a non-trivial use of Szemeredi regularity lemma).
It is intriguing to see whether our stronger notion of fractional covering would lead  to similar kind of results.

Another open problem is whether our techniques would help in improving the best known duality gap ratio, as shown by Haxell~\cite{H99}.
For instance, can we achieve the bound of $(3-\epsilon)$ using our technique?
Even simpler, can we prove that for every graph $G$, $\tau(G) \leq (1.5 -\epsilon) \max \{ \tau^*_2(G), \tau^*_3(G)\}$? Combining with our result, this would imply the $(3-\epsilon)$ bound.

\subsection{Further related work}
\label{subsec:further-related}
In this paper, we focus on the relation between (a stronger variant of) $\tau^*(G)$ and $\nu(G)$.
Another fractional Tuza problem is the relation between $\nu^*(G)$ (the fractional packing number) and $\tau(G)$.
In that case, Krivelevich also showed that $\tau(G) \leq 2 \nu^*(G)$~\cite{K95}, and this is known to be tight (see,~\cite{kortsarz2010approximating} for a tight example.)

\section{Overview of the proof}
\label{sec:overview}
\subsection{The setup}
\label{subsec:setup}
In this section, we introduce the technical terms that will be used in our proof.
Let $\tset$ be the set of all triangles in $G$ and $\Nu \subseteq \tset$  be an optimal packing solution. Triangles and edges in $\Nu$ are referred to as solution triangles and solution edges respectively.
Notice that each (non-solution) triangle $t \in \tset \setminus \Nu$ overlaps (or conflicts) with at most $3$ triangles in $\Nu$ and that if $t$ conflicts with $\psi \in \Nu$, then $|E(t) \cap E(\psi)| = 1$.

We categorize non-solution triangles in $\tset \setminus \Nu$.

\begin{itemize}
    \item \label{itm:singly}  If $t$ overlaps with exactly one triangle $\psi \in \Nu$, we say that $t$ is \textit{singly-attached}\footnote{The notion of ``attachment'' we use here is motivated by the view that triangles in $\Nu$ are the ``skeleton'' of the graph, and the rest of the graphs are ``attachments'' hanging around the skeleton.} to $\psi$, and the unique edge in $E(t) \cap E(\psi)$ is called a {\em base-edge.} The vertex $anchor(t) := V(t) \setminus V(\psi)$\footnote{ For singleton sets, we abuse notations a bit by interchangeably using $e$ and $\{e\}$.} is called the {\em anchoring vertex} of $t$.

    \item \label{itm:doubly} If $t$ overlaps with exactly two triangles $\psi_1, \psi_2 \in \Nu$, we say that $t$ is {\em doubly-attached} to these triangles.

    \item \label{itm:hollow} Otherwise, if $t$ overlaps with three triangles in $\Nu$, we say that $t$ is a {\em hollow triangle.}
\end{itemize}

\subsubsection*{Types of solution triangles:} For each solution triangle $\psi \in \Nu$, denote the conflict list of $\psi$ by $\CL(\psi)$ which contains triangles in $\tset - \Nu$ whose edges overlap with $\psi$.
We naturally partition $\CL(\psi)$ into $\CL_{sin}(\psi) \cup \CL_{dou}(\psi) \cup \CL_{hol}(\psi)$ which denote the sets of singly attached, doubly-attached, and hollow triangles overlapping with $\psi$.

We denote by $base(\psi)$ the set of all edges in $E(\psi)$ that are base-edges.
The solution triangle $\psi$ is said to be type-$i$ (or $type(\psi) : = i$) if $|base(\psi)| = i$. This naturally partitions the set of solution triangles $\Nu$ into sets of type-$i$ triangles $\Nu_i$ for each $i \in \{0, 1, 3\}$ (next proposition implies $|\Nu_2|=0$).
\paragraph{$\text{Type-}1$ triangles:} For any type-$1$ triangle $\psi^1 \in \Nu_1$, we use $base(\psi^1)$ to refer to the only {\em base-edge} in $\psi^1$ and {\em non-base} edges to refer to edges in $E(\psi^1) \setminus base(\psi^1)$.
If $CL_{sin}(\psi^1)= \{t\}$ (which can only happen for type-$1$ triangles), then we define $anchor(\psi^1) := anchor(t)$.

\subsection*{Figures' drawing convention:} In the figures, we label vertices using roman characters. We will use $\Delta_{abc}$ to refer to a triangle consisting of vertices $\{a, b, c \}$. Moreover, if we want to specify the type of the triangle, we will put it in the superscript, so $\Delta_{abc}^1$ will be a type-$1$ triangle. If the type is not specified, then it can be any type, including singly-attached, doubly-attached or hollow triangle. All the solution triangles will be filled with gray color. All the solution edges are drawn in solid style and the non-solution edges are drawn using dashed style.

\begin{proposition}[Structures of solution triangles]
\label{prop:structures-solution}
The following properties hold:
\begin{enumerate}
    \item \label{itm:type-2} $|base(\psi)| \neq 2$ (there is no type-$2$ solution triangle.)

    \item \label{itm:type-3} If $\psi$ is type-$3$, then $|\CL_{sin}(\psi)|  = 3$ and all triangles in $\CL_{sin}(\psi)$ share a common anchoring vertex.
    In this case, $V(\psi)$ together with such an anchoring vertex induce a $K_4$.

\end{enumerate}
\end{proposition}

The proofs for the proposition follow by ~\Cref{cor:type-2,cor:type-3}.
\begin{claim}
\label{clm:common-anchoring}
For any solution triangle $\psi$, if there are two triangles singly-attached to $\psi$, then these two attached triangles either have the same base, or the same anchoring vertex.
\end{claim}
\begin{proof}
Otherwise if $\psi$ is the solution triangle and $\omega, \omega'$ are the singly-attached triangles of $\psi$ with different base edges and different anchoring vertices, then $\Nu \cup \{\omega, \omega'\} \setminus \{\psi\}$ is a better solution, contradicting the optimality of $\Nu$.
\end{proof}

\begin{corollary}
\label{cor:type-2}
There is no type-$2$ solution triangle.
\end{corollary}
\begin{proof}
By\Cref{clm:common-anchoring}, for any solution type-$2$ triangle $\psi$, the two singly-attached triangles attached to different edges of $\psi$ must share their anchoring vertex and there cannot be any other singly-attached triangle attached to $\psi$. This implies that the vertices of $\psi$ plus the anchoring vertex induce a $K_4$, hence in fact it is a type-$3$ triangle.
\end{proof}

\begin{corollary}
\label{cor:type-3}
For any type-$3$ triangle in $\tset$, there are exactly three attached triangles and one anchoring vertex.
\end{corollary}
\begin{proof}
Follows directly by~\Cref{clm:common-anchoring}.
\end{proof}

\subsubsection*{Common anchoring vertex for type-$3$:} If $\psi \in \Nu$ is a type-$3$ triangle, then we denote the common anchoring vertex for the $CL_{sin}(\psi)$ triangles by $anchor(\psi)$ (see~\Cref{prop:structures-solution}.\ref{itm:type-3}).

\subsubsection*{Types of doubly-attached triangles:} After having defined the types of triangles in $\Nu$, each doubly-attached triangle $t \in \tset - \Nu$ can also be given a type in a natural way, i.e., $type(t)$ is a list of types of triangles in $\Nu$ conflicting with $t$; for instance, if $t$ is a doubly-attached triangle adjacent to solution triangles $\psi_1$ and $\psi_2$, then $type(t) = [type(\psi_1), type(\psi_2)]$ where $type(\psi_1) \leq type(\psi_2)$ (see \Cref{fig:doubly-example}). If there is no restriction on some dimension, then we put a star ($[*]$) there.

\input{tikz_figures/doubly-example}

\begin{proposition}[Structures of doubly-attached triangles]
\label{prop:structures-doubly-attached}
Consider any doubly attached triangle $t$ adjacent to solution triangles $\psi_1,\psi_2 \in \Nu \setminus \Nu_0$.
\begin{enumerate}
    \item $type(t) \neq [3,3]$. So the type is either $[1,3]$ or $[1,1]$.

    \item \label{itm:doubly-1-3}  If $type(t) = [1,3]$ where $type(\psi_1)=1$ and $type(\psi_2) =3$, then (i) $base(\psi_1) \in E(t)$ OR (ii) $|\CL_{sin}(\psi_1)| = 1$ where the only anchoring vertex is in $V(t)$; in this case, $base(\psi_1)$ does not contain the common vertex in $V(\psi_1)\cap V(\psi_2)$.
\end{enumerate}
\end{proposition}
\begin{proof}

Let the doubly-attached triangle $t \in \tset$ be adjacent to the two solution triangles $\psi_1$ and $\psi_2$. Clearly, the three triangles  $t, \psi_1, \psi_2$ share a vertex, say $c$.

First we argue that there is no triangle $t$ of type $[3,3]$.
Assume for contradiction that there is such a triangle. Note that:
\begin{itemize}
    \item The anchoring vertex of $\psi_1$ or $\psi_2$ cannot be in $V(\psi_1) \cup V(\psi_2)$; otherwise, some of the attachments would have been doubly-attached, rather than singly-attached.

    \item Let $e_1$ and $e_2$ be edges in $E(\psi_1)$ and $E(\psi_2)$ not adjacent to $c$. These two edges are vertex-disjoint.
\end{itemize}

There exist two edge-disjoint singly-attached triangles $\omega_1 \ni e_1$ and $\omega_2 \ni e_2$ attached to $\psi_1$ and $\psi_2$ respectively. $\omega_1$ and $\omega_2$ are edge-disjoint because $e_1$ and $e_2$ are vertex-disjoint.
Hence, $\Nu \cup \{\omega_1, \omega_2, t\} \setminus \{\psi_1, \psi_2\}$ is a larger set of edge disjoint triangles, which contradicts the optimality of $\Nu$.

Now consider the case when $type(t) = [1,3]$.
WLOG, let $\psi_1$ be type-$1$. Similar to the previous case, $anchor(\psi_2)$ cannot be in $V(\psi_1) \cup V(\psi_2)$.
If $base(\psi_1) \in E(t)$, then we are done.
For the rest of the proof, we assume that the base $e_1 \notin E(t)$.

First, if $|\CL_{sin}(\psi_1)| > 1$, then there exists a singly-attached triangle (say $\omega_1$), such that $anchor(\omega_1) \notin V(t)$. Notice that $E(\omega_1)$ and $E(t)$ are disjoint.
Since $\psi_2$ is type-$3$, there is a singly-attached triangle $\omega_2$ whose base does not contain $c$.
Notice that $t,\omega_1, \omega_2$ are edge-disjoint triangles (an easy way to see this is that each pair of them has at most one vertex in common).
We could exchange the solution by removing $\{\psi_1, \psi_2\}$ and adding $\{t, \omega_1, \omega_2\}$.
Hence, it must be that $|\CL_{sin}(\psi_1)| = 1$ and that the anchoring vertex is in $V(t)$.

\end{proof}
\input{tikz_figures/nonexisting-doubly-attached}

\input{tikz_figures/2-swap-opt-blue}

\subsection{A Warm-up: Multi-transversal of order six}
We illustrate the essence of our techniques by showing a weaker statement,  that there exists an assignment $f: E \rightarrow {\mathbb R}^+$ such that $f(e) \in \{\frac j {6}: j \in \{0,\ldots, 6\}\}$ where $\sum_e f(e) \leq 2 |\Nu|$ and for all $t \in \tset$, $f(t) := \sum_{e \in E(t)} f(e) \geq 1$.

Our analysis is done locally in the sense that each triangle $t \in \Nu$ will have two credits and distribute these credits to the nearby edges.

\subsubsection*{Charge Distribution:}
For any solution triangle $\psi$, the distribution of credit is done as follows.

\begin{enumerate}
    \item If $type(\psi) = 0$, let $f(e) = \frac 2 3$ for all $e \in E(\psi)$. Total distribution is $2$ credits.

    \item If $type(\psi) =1$, we first define $f(e) =\frac{1}{2}$ for the two non-base edges. Then there are two sub-cases.
    \begin{enumerate}
        \item If $|\CL_{sin}(\psi)| >1$, we give $1$ credit to the base-edge.
        Clearly, total credit distribution for $\psi$ is two.

        \item Otherwise, let $\CL_{sin}(\psi)= \{t\}$. We give $\frac{2}{3}$ credit to the base-edge and $\frac{1}{6}$ each to two other edges of $t$.
        Total credit distribution is two in this case.

    \end{enumerate}

    \item If $type(\psi) = 3$, let $f(e) =\frac{1}{3}$ for every edge $e$ in the $K_4$-subgraph induced by $V(\psi) \cup anchor(\psi)$ (refer to \Cref{prop:structures-solution}.\ref{itm:type-3}).
    Since six edges get $\frac{1}{3}$ each, hence total credits distribution is two.
\end{enumerate}

Refer to \Cref{fig:init-fractional-charging-scheme} for illustration of this scheme.

\input{tikz_figures/initial-fractional-charging-scheme}

\subsubsection*{Analysis:}
It is easy to see that all solution triangles are covered.
We analyze three cases of non-solution triangles.
\begin{itemize}
    \item If $t$ is a hollow triangle, since each solution edge has at least $\frac 1 3$ credit, it is obvious that $t$ is covered.

    \item Now consider the case when $t$ is a singly-attached triangle of $\psi$. If $type(\psi) =3$, we are immediately done since $\frac{1}{3}$ credit would be placed on every edge in $t$.
    Otherwise, if $type(\psi) = 1$, we are also done: Either we have one credit on the base of $\psi$ or we have $\frac{2}{3}+ \frac{1}{6}+ \frac{1}{6}$.

    \item The final case is when $t$ is doubly-attached to $\{\psi_1, \psi_2\}$.
    There is only one case which is not obvious, when $type(\psi_1) = 1$ and $type(\psi_2)=3$, in which it could be possible that two overlapping edges of $t$ only get $\frac{1}{2} + \frac{1}{3}$ credit.
    From \Cref{prop:structures-doubly-attached}, one of the two scenarios must happen:
    \begin{enumerate}
        \item $base(\psi_1) \in E(t)$, in which case we would be done since $base(\psi_1)$ has at least $\frac 2 3$ credits and any solution edge of type-$3$ has $\frac 1 3$ credit.

        \item The anchor of $\psi_1$ is a vertex in $V(t)$, in which case, we have that the edge $E(t) \setminus ( E(\psi_1) \cup E(\psi_2))$ must be an edge of the unique singly-attached triangle to $\psi_1$.
        This edge receives the credit of $\frac{1}{6}$.
    \end{enumerate}

\end{itemize}

\section{Multi-transversals of order three}
\label{sec:order-3}

Recall that in this case we want to find the assignment $f: E \rightarrow {\mathbb R}^+$ such that $f(e) \in \{\frac j {3}: j \in \{0,1,2,3\}\}$ where $\sum_e f(e) \leq 2 |\Nu|$ and for all $t \in \tset$, $f(t) \geq 1$.
This case requires one more idea in comparison to the previous case.
In particular, compared to the proof in the previous section, we need to understand the structures of type-$[1,1]$ doubly-attached triangles.

\begin{proposition}[Structures of doubly-attached type-(1,1) triangles] 
\label{prop:type-[1-1]}
For any doubly-attached $\text{type-}[1,1]$ triangle $t$ adjacent to solution triangles $\psi_1,\psi_2 \in \Nu_1$, one of the following holds (up to renaming $\psi_1$ and $\psi_2$):
    \begin{enumerate}
        \item \label{itm:base-edge} $base(\psi_1) \in E(t)$ (see \Cref{fig:base-edge}).
        \item \label{itm:anchoring-vertex} $|\CL_{sin}(\psi_1)| = 1$ where the only anchoring vertex is in $V(t)$ (see \Cref{fig:lend-structure}).

        \item \label{itm:pair} Both $\psi_1 = \{v,u_1,w_1\}$ and $\psi_2=\{v,u_2,w_2\}$ have exactly one singly-attached triangle each (say, $\omega_1$ and $\omega_2$ respectively) such that $E(\omega_1) \cap E(\omega_2) =\{(v,a)\}$ where $a$ is a common anchoring vertex of $\omega_1$ and $\omega_2$.
(see \Cref{fig:type-11}).
    \end{enumerate}
\end{proposition}
\begin{proof}
Recall that $t$ is a type-$[1,1]$ triangle. Let $\psi_1$ and $\psi_2$ be solution triangles which $t$ is doubly-attached to. Let $e$ be the non-solution edge of $t$ and $c = V(t) \cap V(\psi_1) \cap V(\psi_2)$. 
If $E(t)$ contains $base(\psi_1)$ or $base(\psi_2)$, then $t$ belongs to~\Cref{prop:type-[1-1]}.\ref{itm:base-edge} and we are done.

Now we assume that $\{base(\psi_1), base(\psi_2)\} \cap E(t) = \varnothing$.
If $|CL_{sin}(\psi_1)| \geq |CL_{sin}(\psi_2)| \geq 2$,
then in both cases when $base(\psi_1) \cap base(\psi_2) = \varnothing$ or
$base(\psi_1) \cap base(\psi_2) = \{c\}$,
there exists two singly-attached triangles $\omega_1, \omega_2$ attached to $base(\psi_1), base(\psi_2)$ respectively such that both $anchor(\omega_1) \neq anchor(\omega_2)$ and $\omega_1, \omega_2$ are edge-disjoint from $t$. Hence, we let $\Nu^\prime = (\Nu \setminus \{\psi_1, \psi_2\}) \cup \{t, \omega_1, \omega_2\}$. Since $\Nu^\prime$ has more triangle than $\Nu$, which contradicts the optimality of $\Nu$. 

Now by renaming, let us assume $|CL_{sin}(\psi_1)| = 1$. Let $\omega_1$ be the singly-attached triangle of $\psi_1$. If $anchor(\omega_1) \in V(t)$, then $t$ belongs to~\Cref{prop:type-[1-1]}.\ref{itm:anchoring-vertex} and we are done. 
If not, then if $|CL_{sin}(\psi_2)|>1$ we have an improving swap using the an argument similar to the previous case. If $|CL_{sin}(\psi_2)| = 1$, let $\omega_2$ be the singly-attached triangle. If $anchor(\omega_2) \in V(t)$ then we are done by renaming. Else if $base(\psi_1) \cap base(\psi_2) = \varnothing$ or $anchor(\omega_1) \neq anchor(\omega_1)$, then still $\Nu^\prime = (\Nu \setminus \{\psi_1, \psi_2\}) \cup \{t, \omega_1, \omega_2\}$ is an improving swap because $\omega_1, \omega_2, t$ are edge disjoint since they share at most one vertex with each other. Hence the only remaining possibility is that $anchor(\omega_1) = anchor(\omega_2)$ and $base(\psi_1) \cap base(\psi_2) \neq \varnothing$, which implies $t$ belongs to \Cref{prop:type-[1-1]}.~\ref{itm:pair}. Hence, we conclude the proof.
\end{proof}

\input{tikz_figures/type-1,1}

\subsubsection*{Charge Distribution:}

Starting with $f(e)= 0$ for each edge in $G$, for any solution triangle $\psi \in \Nu_0 \cup \Nu_3$, distribute the credits exactly the same as in the previous section (see~\Cref{fig:init-fractional-charging-scheme}).
The charging for type-$1$ triangles is more subtle in this case (because we used $\frac{1}{2}$ credits in the previous case which we cannot use here.) We will distribute credits to type-$1$ triangles iteratively to make sure that all the type-$[1,1]$ triangles are covered.

\subsubsection*{Charging Scheme for type-$1$ triangles:} Let $\Nu_1^1 \subseteq \Nu_1$ be the subset of type-$1$ triangles having exactly one singly-attached triangle.

For $\psi \in \Nu_1 \setminus \Nu_1^1$, set $f(e) = 1$ if $e$ is the base-edge of $\psi$ and $f(e) = \frac13$ for each $e$ which is a non-base edge of $\psi$. We use exactly $\frac 5 3$ credit for each $\psi \in \Nu_1 \setminus \Nu_1^1$. See~\Cref{fig:init-fractional-charging-scheme-type-1-many-attachments-1-3} for the charging illustration of triangles in $\Nu_1 \setminus \Nu_1^1$.

\input{tikz_figures/initial-fractional-charging-type-1-rich-1-3-int}

Now let $\pi: \Nu_1^1 \to [|\Nu_1^1|]$ be a bijection that denotes an arbitrary permutation of $\Nu_1^1$. We will distribute charge corresponding to triangles in $\Nu_1^1$ in this order.

For $1 \leq i \leq |\Nu_1^1|$, let $\psi_i = \pi^{-1}(i)$ be the $i^{th}$ triangle in the permutation. Let $u_i$ and $v_i$ be two end points of the edge $base(\psi_i)$. Let $e_{u_i}$ and $e_{v_i}$ be the non-base solution edges adjacent to $u_i$ and $v_i$, respectively. Similarly, let $e_{u_i}'$ and $e_{v_i}'$ be the non-solution edges of the unique singly-attached triangle of $\psi_i$ adjacent to $u_i$ and $v_i$, respectively.

Starting from $i=1$, we first assign $f(base(\psi_i)) = \frac23$. We then inspect the charge on $e_{u_i}'$. If $f(e_{u_i}') = 0$, then we set $f(e_{u_i}') = f(e_{u_i}) = \frac13$. Otherwise, we leave $f(e_{u_i}')$ as-is and set $f(e_{u_i}) = \frac23$. Similarly, we distribute the charge to $e_{v_i}$ and $e_{v_i}'$.

We continue assigning charges for $\psi_i$ in order until we have processed all the triangles in $\Nu_1^1$ for every $1 \leq i \leq |\Nu_1^1|$. See~\Cref{fig:fractional-singly-one-attachment-distribution-1-3} for illustration.

The following observation lists the properties of the charging scheme $\{f(e)\}_{e\in E}$ above.

\begin{observation}
\label{obs:charging-properties}
The following properties hold for the charging scheme $\{f(e)\}_{e\in E}$.
\begin{enumerate}
    \item For every solution edge $e \in E(\Nu)$, $\, f(e) \geq \frac13$.
    \item For every $\psi \in \Nu_0$ and for each $e \in E(\psi)$, $f(e) = \frac 23$.
    \item For every $\psi \in \Nu_1$, $f(base(\psi)) \geq \frac 23$. If $|CL_{sin}(\psi)|>1$, then $f(base(\psi)) = 1$.
    \item For every $\psi \in \Nu^1_1$, if $e_1$ and $e_2$ are the non-solution edges of the unique singly-attached triangle of $\psi$, then $f(e_1) = f(e_2) = \frac 13$.
    \item For every $\psi \in \Nu_3$, for each edge $e$ in the $K_4$ structure associated with it, $f(e) = \frac13$.
\end{enumerate}
\end{observation}

As we assign additional $2$ credits per $\psi_i$, it is easy to see that $\sum_e f(e) \leq 2 |\Nu|$.

\begin{figure*}[ht]
\input{tikz_figures/fractional-singly-one-attachment-distribution-1-3}
\end{figure*}

\subsubsection*{Analysis:}

It is easy to see from \Cref{obs:charging-properties} that all solution and singly-attached triangles are covered. We analyze non-solution triangles as follows.

\begin{itemize}
    \item If $t$ is a hollow triangle, since each solution edge has at least $\frac 1 3$ credit, hence it is covered.

    \item If $t$ is a doubly-attached triangle of type-$[0, *]$, by \Cref{obs:charging-properties}, since the edge of $t$ adjacent to the type-$0$ triangle has $\frac23$ credit and the other solution edge has at least $\frac13$ credit, hence it is covered.

    \item If $t$ is a doubly-attached triangle of type-$[1,3]$, let $e_t$ be the non-solution edge of $t$. By \Cref{prop:structures-doubly-attached}, it must be that (i) $E(t)$ contains a base-edge of the type-$1$ triangle or (ii) $e_t$ is an edge of the unique singly-attached triangle of the type-$1$ triangle in $\Nu_1^1$ adjacent to $t$.
    In (i), $t$ is covered by the base-edge which has at least $\frac 2 3$ credit and the other solution edge which has at least $\frac 1 3$ credit. Otherwise, in (ii) $t$ is also covered since any solution edge has at least $\frac 1 3$ credit and $f(e_t) = \frac 1 3$.
    \item If $t$ is a doubly-attached triangle of type-$[1,1]$, then we use \Cref{prop:type-[1-1]} to argue that $t$ is covered. If $t$ contains any base-edge, then since $f(e) \geq \frac23$ for any base-edge of type-$1$ triangles and the other solution edge in $t$ has at least $\frac13$ credit, hence $t$ is covered.
    Otherwise, $t$ does not have any base-edge. Let $\psi_1$ and $\psi_2$ be two solution triangles which $t$ is adjacent to. By renaming we assume that $\pi(\psi_1) < \pi(\psi_2)$.
    Let $e_t$ be the non-solution edge of $t$. If $t$ falls into \Cref{prop:type-[1-1]}.\ref{itm:anchoring-vertex} case, then $f(e_t) = \frac 1 3$ which implies every edge of $t$ has at least $\frac 1 3$ credit, hence $t$ is covered.
    If $t$ falls into the case in \Cref{prop:type-[1-1]}.\ref{itm:pair}, since $anchor(\psi_1) = anchor(\psi_2)$, it must be that $f(E(t) \cap E(\psi_2)) = \frac 2 3$, hence $t$ is covered.
\end{itemize}

\section{Multi-transversals of order two}
\label{sec:order-2-fix}
Again, this is equivalent to finding the assignment $f: E \rightarrow {\mathbb R}^+$ such that $f(e) \in \{\frac j {2}: j \in \{0, 1, 2\}\}$ where $\sum_e f(e) \leq 2 |\Nu|$ and for all $t \in \tset$, $f(t)  \geq 1$.

We will show this result in the following three steps.
\begin{enumerate}
    \item First, we start with an initial charge distribution shown in \Cref{subsec:initial-half-charge}.
    Notice that each $\text{type-}0$ triangle has extra half credit remaining, which will be used later.

    \item Then in \Cref{subsec:type-1-3}, we will discharge credits from type-$1$ solution triangles, using what we call the \emph{alternating chains structure}. After this step, all triangles in $G$ that are not adjacent to any type-$0$ triangle will be covered.

    \item Finally in \Cref{subsec:type-0}, we show an iterative process to use the extra half credits of $\text{type-}0$ triangles along with {\em rotating} the configuration of credit distribution for some type-$1$ and type-$3$ triangles to cover all the triangles in $G$.
\end{enumerate}

\subsection{Initial Charge Distribution}
\label{subsec:initial-half-charge}
For any solution triangle $\psi$, the distribution of credits is done as follows.

\begin{figure*}[ht]

\input{tikz_figures/initial-fractional-charging-scheme-1-2-fix}
\end{figure*}
\begin{figure*}[ht]
\input{tikz_figures/hollow_example.tex}
\end{figure*}

\begin{enumerate}
    \item If $type(\psi) = 0$, let $f(e) = \frac{1}{2}$ for all $e \in E(\psi)$. Total distribution is $1.5$ credits.

    \item If $type(\psi) = 1$, let $f(e) =\frac{1}{2}$ for two non-base edges $e$ and we give $1$ credit to the base edge.
    Total distribution is two credits in this case as well.

    \item If $type(\psi) = 3$, let $H$ be any $C_4$ subgraph of the induced subgraph on $V(\psi)$ and the anchoring vertex of $\CL_{sin}(\psi)$.
    Define $f(e) = \frac{1}{2}$ for every edge in $H$.
    (refer to \Cref{prop:structures-solution} and \Cref{fig:initial-fractional-charging-scheme-1-2-fix}).


\end{enumerate}

We use the following fact several times in the proof of our result.

\begin{fact}[Half-Integral $K_4$ charging]
\label{fact:half-integral-k4}
In $K_4$, assigning $\frac 1 2$ credits to any $C_4$ subgraph covers every triangle.
\end{fact}

We refer to this charge distribution as {\em half-integral $K_4$} charging.


\begin{definition}
    An edge $e$ is called a \emph{null-edge} if $f(e) = 0$, a \emph{half-edge} if $f(e) = \frac{1}{2}$, or otherwise a \emph{full-edge}.
\end{definition}

For convenience, when we say that an edge $e$ is \emph{at least a half-edge}, it means
that $f(e) \geq \frac 1 2$, i.e., $e$ is either a half-edge or a full-edge.
Also, we might drop the dash for brevity, e.g., a null type-$3$ edge $e$ would mean that $e$ is a null-edge and $e$ is a solution edge of a type-$3$ triangle.

\subsubsection*{Types of hollow triangles:} Following a similar naming convention to that of doubly-attached triangles, if $t$ is a hollow triangle adjacent to solution triangles $\psi_1$, $\psi_2$ and $\psi_3$, then $type(t) = [type(\psi_1), type(\psi_2), type(\psi_3)]$ where $type(\psi_1) \leq type(\psi_2)\leq type(\psi_3)$ (see~\Cref{fig:hollow-example}).

The following proposition captures some structural properties of the hollow triangles.

\begin{proposition}[Structures of hollow triangles]
\label{prop:structures-hollow}
Consider any hollow triangle $t$ adjacent to solution triangles $\psi_1$, $\psi_2$ and $\psi_3$.
\begin{enumerate}
    \item $type(t) \neq [3,3,3]$.
    \item If $type(t) = [1,*,*]$\footnote{If there is no restriction on some dimension, then we put a star ($[*]$) there.} where $type(\psi_1)=1$, then $t$ must be adjacent to the base edge of one of the $\psi_1, \psi_2, \psi_3$ which is a type-$1$ triangle or two type-$1$ triangles from $\{ \psi_1, \psi_2, \psi_3\}$ share a sole anchoring vertex.

\end{enumerate}
\end{proposition}
\begin{proof}

First we argue that there is no triangle $t$ of type $[3,3, 3]$.
Assume otherwise that $t$ is attached to type-$3$ triangles $\psi_1,\psi_2,\psi_3$.
Notice that $|\bigcup_{i=1}^{3} V(\psi_i)| = 6$ since these solution triangles cannot share edges.
For any $i$, let $v_i = V(\psi_i) \setminus V(t)$ be an only vertex of $\psi_i$ that is not in $V(t)$. Let $c_{ij} \in V(\psi_i) \cap V(\psi_j)$ be the vertex shared between $\psi_i$ and $\psi_j$.
It is possible that all $a_i = anchor(\psi_i)$ are the same vertex for all $i$. Consider the set $S$ of four disjoint triangles $\{\Delta_{v_1 c_{12} a_1} , \Delta_{v_2 c_{23} a_2}, \Delta_{v_3 c_{13} a_3}, t \}$.
Since $ \Nu^\prime =  (\Nu \setminus \bigcup_{i=1}^{3} \psi_i) \cup S$ is a set of disjoint triangle of size $|\Nu| + 1$, then it contradicts the fact that $\Nu$ is optimal.

Now we consider the case where $type(\psi_1) =1$. There are three sub-cases.

The first sub-case is when $type(t) = [1, 3, 3]$. Assume that $base(\psi_1)$ is not in $t$. In this case, $anchor(\psi_2) \neq anchor(\psi_3)$ or the base edge of $\psi_1$ will be in $t$.
WLOG, assume that the $base(\psi_1)$ is $\overbar{v_1 c_{12}}$.
We then can select the set $S$ of four disjoint triangles $\{\Delta_{v_2 c_{23} a_2}, \Delta_{v_3 c_{23} a_3}, \omega_1 \in \CL_{sin}(\psi_1), t\}$.
Since $\Nu^\prime =  (\Nu \setminus \bigcup_{i=1}^{3} \psi_i) \cup S$ is a set of disjoint triangle of size $|\Nu| + 1$, then it again is a contradiction.

The second sub-case is when $type(t) = [1, 1, 3]$.
Assume that $\{base(\psi_1), base(\psi_2)\} \cap E(t) = \varnothing$.
Let $\omega_1 \in \CL_{sin}(\psi_1)$ and $\omega_2 \in \CL_{sin}(\psi_2)$ be two disjoint singly-attached triangles. There exists such triangles or $\psi_1$ and $\psi_2$ will share a single anchoring vertex. Let $\omega_3$ be a singly-attached triangle of $\psi_3$, which is disjoint from $\omega_1$ and $\omega_2$. There exists such $\omega_3$ or $\omega_1$ and $\omega_2$ would share their anchoring vertex.
Let $S= \{\omega_1, \omega_2, \omega_3, t\}$ be a set of four disjoint triangles.
Since $ \Nu^\prime =  (\Nu \setminus \bigcup_{i=1}^{3} \psi_i) \cup S$ is a set of disjoint triangle of size $|\Nu| + 1$, then it again is a contradiction.

Now we consider the third sub-case when $type(t) = [1, 1, 1]$.
Assume that $\{base(\psi_1), base(\psi_2), base(\psi_3)\} \cap E(t) = \varnothing$. If no two out of these three base edges share a vertex, then it is easy to see that there exists three disjoint singly-attached triangles $\omega_1, \omega_2, \text{ and } \omega_3$, attaching to $\psi_1, \psi_2, \text{ and } \psi_3$ in respective order.
Let $S= \{\omega_1, \omega_2, \omega_3, t\}$ be a set of four disjoint triangles.
Since $ \Nu^\prime =  (\Nu \setminus \bigcup_{i=1}^{3} \psi_i) \cup S$ is a set of disjoint triangle of size $|\Nu| + 1$, then it again is a contradiction.

Otherwise, WLOG, assume that $base(\psi_1)$ and $base(\psi_2)$ share the vertex $c_{12}$. If these base edges share an anchoring vertex, we are done;  otherwise, we can select two disjoint singly-attached triangle $\omega_1 \in \CL_{sin}(\psi_1)$ and $\omega_2 \in \CL_{sin}(\psi_2)$. Let $S= \{\omega_1, \omega_2, \omega_3 \in \CL_{sin}(\psi_3), t\}$ be a set of four disjoint triangles.
Since $ \Nu^\prime =  (\Nu \setminus \bigcup_{i=1}^{3} \psi_i) \cup S$ is a set of disjoint triangles of size $|\Nu| + 1$, then it again is a contradiction.

This concludes the proof.
\end{proof}

\subsection{Type-$1$ and type-$3$ instances}
\label{subsec:type-1-3}
In this section we assume that $\Nu_0 = \varnothing$ and show a valid charging scheme to cover all triangles $G$. First we do the initial charge distribution as shown in \Cref{subsec:initial-half-charge}.

\subsubsection*{Covered triangles after initial charge distribution:} We first argue about all triangles that are covered by the initial charge distribution.
Using \Cref{prop:structures-doubly-attached}, we can infer the following:

\begin{lemma}
Using the initial charge distribution, if triangle $t$ is not covered, then $t$ is doubly-attached triangle of type-$[1,3]$, and $t$ must be adjacent to null type-$3$ edges.
\end{lemma}
\begin{proof}
The proof is by case analysis.
Type-$1$ solution triangles and their corresponding singly-attached triangles are covered since the base edge is a full-edge.
Type-$3$ solution triangle and all their corresponding  singly-attached triangles are covered by \Cref{fact:half-integral-k4}.
Type-$[1, 1]$ doubly-attached triangles are covered since each solution edge of type-$1$ is at least a half-edge.
Type-$[1, 3]$ doubly-attached triangles containing base edge of type-$1$ triangle are covered since any base-edge of type-$1$ triangle is a full-edge.
Type-$[1, 1, *]$ hollow triangles are covered since each solution edge of type-$1$ is at least a half-edge.
Type-$[1, 3, 3]$ hollow triangles are covered since it contains the base-edge of type-$1$ triangle which is a full-edge. \Cref{prop:structures-hollow} implies no other hollow triangles exist.
\end{proof}

\subsubsection*{Twin-doubly-attached triangles and lending structures}




\begin{definition}[Lend relation and Gain function]
    \label{def:lend-and-gain}
Let $t_1$ and $t_2$ be doubly-attached triangles that are attached to the same pair of type-$1$ triangle $\psi^1$ and a triangle $\psi \in \Nu_1 \cup \Nu_3$ such that:
\begin{itemize}
\item $V(\psi^1) \cup anchor(\psi^1) = V(t_1) \cup V(t_2)$, and $V(t_1) \cup V(t_2)$ induces a $K_4$.

\item $|\CL_{sin}(\psi^1)| = 1$.
\end{itemize}

Then, we say that $t_1$ and $t_2$ are {\bf twin-doubly-attached triangles} and that $(\psi^1, \psi)$ has a {\bf lend relation}.
The overlapping edge $E(t_1) \cap E(t_2)$ is called a {\bf gain edge}.
(see~\Cref{fig:lending-structure,subfig:before-lend-discharge-type-13}).
    \end{definition}

That is, the lend relation is a relation $lend \subseteq \Nu_1 \times (\Nu_1 \cup \Nu_3)$.
The gain function can be define as $gain((\psi^1, \psi)) = E(t_1) \cap E(t_2)$.

From now on we develop a discharging strategy to cover twin-doubly-attached type-$[1, 3]$ triangles while maintaining the coverage of the rest of the triangles.
In fact, we will guarantee some additional stronger properties which would help in the subsequent section where we deal with the general case.

For brevity, we will drop a pair of parenthesis and use the notation $gain(\psi^1, \psi)$ instead of $gain((\psi^1, \psi))$.

\input{tikz_figures/lending-structure.tex}

The following structural lemma is crucially used to define a valid charging.

\begin{lemma}
\label{lem:lend-outdegree-one} In the $lend$ relation, any $\psi^1 \in \Nu_1$ is related to at most one $\psi \in \Nu$.
\end{lemma}
\begin{proof}
  Let $v$ be the vertex not adjacent to the base-edge of $\psi^1$ and
  let $a = anchor(\psi^1)$ be the anchoring vertex.
  For any $\psi$ such that $(\psi^1, \psi) \in lend$, it must be that
  $\overbar{va} \in E(\psi)$. Since any solution edge must be a part of exactly
  one triangle, this lemma is true.
\end{proof}

\begin{fact}
\label{fact:lend-structure}
If $(\psi^1, \psi) \in lend$ and $t_1, t_2$ are the corresponding twin-doubly-attached triangles and $\omega_1$ is the singly-attached triangle of $\psi^1$, then each of $t_1, t_2$ shares a different non-solution edge with $\omega_1$ and a different non-base edge of $\psi^1$. 
\end{fact}

\subsubsection{Alternating Chains of Triangles}
First we start with the high-level overview of our charging strategy and its proof. Our structures may be viewed as a natural extension of the alternating paths structure for matching and the property that for any optimal matching there cannot exist any {\em augmenting path} in the graph. We will contrast and compare them at different points in the paper. We refer to our alternating structure of triangles as {\em alternating chain}. We will use the alternating chain structure and the optimality of the solution $\Nu$ in two ways.
Informally speaking, the alternating chains have (in certain way) a ``direction'' and based on that we will refer the first triangle of the chain as {\em head} and the last triangle as {\em tail}.
Our strategy is as follows.

First we use the alternating chain of solution and non-solution triangles to {\em discharge} credits from type-$1$ triangles participating in this chain from tail to head which will be a type-$3$ triangle. Secondly, we argue that the tail of any alternating chain cannot be adjacent to some special kind of doubly-attached or hollow triangles, otherwise it would lead to an {\em augmenting} structure for the solution $\Nu$ violating its optimality. \footnote{These special doubly-attached or hollow triangles can be thought of as an unmatched edge between the two odd leveled vertices from two different alternating trees in the alternating forest in the blossom's algorithm.}
The exclusion of such structures allows us to discharge credits from tail triangle iteratively to the type-$3$ triangle to cover the type-$[1, 3]$ triangles that were not covered while maintaining the coverage for all the other triangles in the graph.

\subsubsection*{Discharging-via-Chains construction:}
We start with any pair of twin-doubly-attached type-$[1,3]$ triangles such that the type-$3$ triangle has one null-solution edge. Let the type-$3$ triangle be $\psi_0$ and the type-$1$ triangle be $\psi_1$.
Note that $(\psi_1, \psi_0) \in lend$ (see~\Cref{fig:lend-charging-type-1-3}).
We cover them by {\em discharging} $\frac{1}{2}$ credit from the type-$1$ triangle to the type-$3$ triangle by making any solution edge and any two non-solution edges in the graph induced by $V(\psi_0) \cup \{anchor(\psi_0)\}$ half-edges. This ensures that $f(gain(\psi_1, \psi_0)) = \frac{1}{2}$.
Then we complete the assignment of credits in the graph induced by $V(\psi_1) \cup \{anchor(\psi_1)\}$ such that all edges except a matching of $G[V(\psi_1) \cup \{anchor(\psi_1)\}]$ are half-edges (as shown in~\Cref{fig:lend-charging-type-1-3}). Note that $f(gain(\psi_1, \psi_0)) = \frac{1}{2}$ ensures that $f(base(\psi_1))=\frac{1}{2}$.
Also in total we still use exactly $4$ credits for $\psi_0$ and $\psi_1$. We call the base-edge of discharged type-$1$ triangle as half-base edge and the solution edge without any credit as null non-base edge. After this discharging, it is clear that the solution and the singly-attached triangles remain covered, triangles in $G[V(\psi_1) \cup \{anchor(\psi_1)\}]$ are covered by~\Cref{fact:half-integral-k4}. But there could be some doubly-attached or hollow triangles adjacent to $\psi_1$'s null non-base edge which may not be covered. We will later show that all such triangles, except some type-$[1, 1]$ triangles, are either already covered or cannot exist by using optimality of our packing solution $\Nu$.
We continue growing the current chain, if there exists any $\psi_2 \in \Nu_1$ corresponding to twin-doubly-attached type-$[1, 1]$ triangles having a lend relationship to $\psi_1$ (See~\Cref{fig:lend-charging-type-1-1}), such that $gain(\psi_2, \psi_1) \neq base(\psi_1)$.
Hence, $(\psi_2, \psi_1) \in lend$ and $gain(\psi_2, \psi_1)$ is a non-base edge of $\psi_1$ (which may not be the one which is the null-edge).
We again discharge half credit from $\psi_2$ to the null-non-base edge of $\psi_1$ by making it a half-edge and then completing the assignment of credits in the graph induced by $V(\psi_2) \cup \{anchor(\psi_2)\}$ such that all edges except a matching of $G[V(\psi_2) \cup \{anchor(\psi_2)\}]$ are half-edges (as shown in~\Cref{fig:lend-charging-type-1-1}).
For any $i \geq 1$, we continue discharging credits by finding any arbitrary type-$1$ triangle $\psi_{i+1}$ corresponding to $(\psi_{i+1}, \psi_i) \in lend$ and $gain(\psi_{i+1}, \psi_i) \neq base(\psi_i)$, until we reach some type-$1$ triangle (say $\psi_k$) which does not have any twin doubly-attached triangles adjacent to any of its non-base edges.


\input{tikz_figures/lend-charging-type-1,3.tex}

\input{tikz_figures/lend-charging-type-1,1.tex}

 Let $\cset_1 = \{\psi_k, \psi_{k-1}, \dots, \psi_0\}$ be our first finished chain. For any such chain, we define $head(\cset_1) := \psi_0$ and $tail(\cset_1) := \psi_k$. We will also refer to $\psi_0$ as the {\em head} of $\cset_1$ and $\psi_k$ as the {\em tail} of $\cset_1$. We define the {\em size} of such a chain to be $k$.

 \begin{observation}
 Any size-$k$ chain contains one type-$3$ triangle as its head and $k \geq 1$ type-$1$ triangles.
 \end{observation}

 We repeat this procedure starting with another pair of twin-doubly-attached type-$[1,3]$ triangles which are not covered to form $\cset_2$. We keep on repeating this procedure until we cannot find any uncovered type-$[1, 3]$ triangle. Let $\Chi := \{ \cset_1, \cset_2, \dots \cset_\ell\}$ be the set of all finished chains which we get.
Since at any iteration $i \geq 1$, we continue discharging credits every time when there exists some $\psi \in \Nu_1$ such that $(\psi, \psi_i) \in lend$, we get the following property.

\begin{fact}
\label{fact:no-lending-to-tail-property} For any chain $\cset = \{\psi_k, \psi_{k-1}, \dots, \psi_0\} \in \Chi$, there is no $\psi \in \Nu_1$ such that $(\psi, \psi_k) \in lend$ and $gain(\psi, \psi_k) \neq base(\psi_k)$. This implies that any arbitrary $K_4$ half-integral charging for $V(\psi_k) \cup \{anchor(\psi_k)\}$ such that $f(gain(\psi_k, \psi_{k-1})) =\frac{1}{2}$ for the tail triangle covers all the twin doubly-attached triangles adjacent to it.
\end{fact}

 The following claim shows that the Discharging-via-Chains construction is well-defined. Moreover, for any two chains $\cset, \cset' \in \Chi$,  $\cset \cap \cset' = \emptyset$.

\begin{claim}
\label{clm:type-1-outdeg-one} Any type-$1$ triangle will be required to discharge credits at most once.
\end{claim}
\begin{proof}
The proof directly follows by~\Cref{lem:lend-outdegree-one} and the way we construct the chains.
\end{proof}

The following observation lists the properties of the charging $\{f(e)\}_{e \in E}$ at end of the Discharging-via-Chains procedure.
\begin{observation}
\label{obs:discharging-chains-property}
The charging $\{f(e)\}_{e \in E}$ at end of the Discharging-via-Chains procedure has the following properties:
\begin{enumerate}
    \item For any type-$3$ triangle $\psi \in \Nu_3$ such that $head(\cset)=\psi$ for some $\cset \in \Chi$ and , $\forall e \in E(\psi)$, $f(e) = \frac{1}{2}$.
    \item For any type-$1$ triangle $\psi \in \Nu_1$ such that $\psi \in \cset$ and $tail(\cset) \neq\psi$ for some $\cset \in \Chi$, $\forall e \in E(\psi)$, $f(e) = \frac{1}{2}$.
    \item For any type-$1$ triangle $\psi \in \Nu_1$  such that $tail(\cset)=\psi$ for some $\cset \in \Chi$, the base-edge and one non-base edge have $\frac{1}{2}$ credits and there is one null-non-base edge with no credit.
    \item Except the null type-$3$ edges and null-non-base $\text{type-}1$ edge of any type-$1$ triangle which is a tail, for any solution edge $e$, $e$ is at least a half-edge, i.e., $f(e) \geq \frac{1}{2}$.
\end{enumerate}

\end{observation}

The following observation pin points the edges for which there is some {\em credit loss}.

\begin{observation}[Credit Loss]
\label{obs:credit-loss}
There are two types of edges for which the credits reduce after the Discharging-via-Chains construction.
Non-base edge of every tail type-$1$ triangle is reduced from a half-edge to a null-edge.
Base-edge of every type-$1$ triangle participating in any chain is reduced from a full-edge to a half-edge.
\end{observation}

\subsubsection*{Covered triangles after the Discharging-via-Chains construction:} After the Discharging-via-Chains construction, the covered triangles are as follows.
Any type-$1$ triangle and its singly-attached triangles are covered either by the half-integral $K_4$ charging (\Cref{fact:half-integral-k4}), used during the Discharging-via-Chains construction or the initial charge distribution.
Type-$3$ triangles and their singly-attached triangles are covered as they did not lose any credit (\Cref{obs:credit-loss}).
Type-$[1, 1]$ twin-doubly-attached triangles are covered. The coverage of intermediate twin-doubly-attached triangles of any chain follows from the fact that both the solution edges are half-edges. Twin-doubly-attached triangles not involved in any chain remain covered. By~\Cref{fact:no-lending-to-tail-property}, twin-doubly-attached triangles adjacent to any tail triangle are covered.
Type-$[1, 3]$ twin-doubly-attached triangles are covered since the procedure terminated are covered.
Type-$[1, 1, 1]$ hollow triangles with at most one null-edge or two null-edges and one full-base edge are covered.
Type-$[1, 1, 3]$ hollow triangles with no null-non-base edges for type-$1$ or one null-edge and one full-base edge of type-$1$ are covered.
Type-$[1, 3, 3]$ hollow triangle containing the full-base edge of type-$1$ are covered.

\subsubsection*{Potentially-uncovered triangles after the Discharging-via-Chains construction:} Now we enumerate the triangles which may not remain covered because of the reduction in credits for these edges.
To do so we use \Cref{obs:charging-properties} and \Cref{prop:structures-doubly-attached,prop:type-[1-1],prop:structures-hollow}.
Type-$[1, 1]$ doubly-attached triangles containing at least one null-non-base edge of some type-$1$ triangle which is a tail of some chain. The second solution edge can either be a half-base edge or some non-base edge.
Type-$[1, 3]$ doubly-attached triangles containing half-base edge of some type-$1$.
Type-$[1, 1, 1]$ hollow triangles with at least two null-non-base edges of two type-$1$ triangles which are tails of two chains.
Type-$[1, 1, 3]$ hollow triangle containing at least one null-non-base edge of some type-$1$ which is a tail of some chain. Another type-$1$ solution edge can either be a half-base edge or a non-base edge.
Type-$[1, 3, 3]$ hollow triangle containing half-base edge of type-$1$ triangle and null-edges of the other two type-$3$ triangles.

We will use the following lemma which implies that at the end of Discharging-via-Chains procedure we get a valid assignment
\ifjournal
(see Appendix A.1.1 in the full version for the full proof).
\fi
\ifarxiv
(see~\Cref{subsubsec:proof-lem:main-type-1-3} for the full proof).
\fi

\begin{lemma}
\label{lem:main-type-1-3}
 For any optimal solution $\Nu$ of our graph $G$, such that $\Nu_0 = \varnothing$, starting with the initial charging distribution and then applying the Discharging-via-Chains construction, any potentially-uncovered triangle cannot exist in $G$. In other words, the construction covers every triangle in $G$.
\end{lemma}




\subsubsection*{Proof-of-Concept:} We look at the solution edges adjacent to some potentially-uncovered triangle $t$. The main idea is to show the existence of an improving swap involving $t$ to derive a better packing solution, thus contradicting the fact that $\Nu$ is optimal. Notice that one of these solution edges is either a null-non-base edge $g_{k+1}$ of the type-$1$ tail triangle of some $k$-sized chain $\cset$ ($k >1$) or a half-base edge of any type-$1$ triangle participating in some chain.
We use the structure of chains that allows for several alternate solutions which {\em frees-up} the $g_{k+1}$ edge and/or the half-base edge without making the triangle packing solution $\Nu$ worse.
We observe that there exists simple alternate solutions which frees-up the other solution edges of $t$, while maintaining an optimal solution. If we flip the solution triangles with these alternate solutions, then we can free-up all the edges of $t$ and then include it to the new solution thereby getting a solution of higher cardinality.
This is not hard to see if all the non-solution edges which we deal with are different. Things become complex when these structures start sharing vertices leading to shared non-solution edges. To deal with such situations, we crucially exploit the rich structure of our alternating chains which allows for various choices of alternate solutions. In some sub-cases, using the charging scheme for chains, we can deduce that $t$ is covered, leading to a different contradiction.

Moreover we get the following observations for this charging which will be crucially used in the next section to prove the general case. They follow from the fact that we arbitrary fix any $K_4$ half-integral charging for every tail triangle and for every type-$3$ triangle which is not part of any chain.

\begin{observation}
\label{obs:type-1-rotation}
\Cref{lem:main-type-1-3} remains true for any arbitrary but fixed $K_4$ half-integral charging for any tail triangle $\psi_k$ of any $k$-sized chain $\cset \in \Chi$ such that $f(base(\psi_k)) = f(gain(\psi_k, \psi_{k-1}))=\frac{1}{2}$.
\end{observation}

\begin{observation}
\label{obs:type-3-rotation}
\Cref{lem:main-type-1-3} remains true for any arbitrary but fixed $K_4$ half-integral charging for any type-$3$ triangle which is not part of any chain.
\end{observation}

\subsection{Covering everything using the remaining half credits and rotations}
\label{subsec:type-0}
For any fixed optimal solution $\Nu$ for our graph $G$, we first tentatively do the initial charge distribution for all triangles as shown in \Cref{subsec:initial-half-charge} and then we run the Discharging-via-Chains construction described in \Cref{subsec:type-1-3}.

\subsubsection*{Covered triangles after the Discharging-via-Chains construction:} \Cref{lem:main-type-1-3} implies that all triangles not adjacent to any type-$0$ triangles are covered. Using the fact that all type-$0$ solution edges are half-edges implies that type-$0$ solution triangles, all the doubly-attached triangles of type-$[0,0]$ and hollow triangles of type-$[0,0,*]$ are also covered. By \Cref{prop:structures-doubly-attached,prop:structures-hollow}, the rest of the possibly uncovered triangles in $G$ are type-$[0,3]$, type-$[0, 1]$, type-$[0, 1, 1]$, type-$[0, 1, 3]$ and type-$[0, 3, 3]$. Out of these triangles, the triangles adjacent to any solution edge of non-tail type-$1$ triangles, any solution edge of head type-$3$ triangles or any base edge of type-$1$ triangles are surely covered because they get at least half credit from these type-$1$ or type-$3$ solution edges and half credit from the type-$0$ edge.

In this section, we show a strategy to cover the \emph{rest} of these triangles while maintaining the coverage of the previously argued triangles. Let $\aset$ be the set of \emph{free} solution triangles consisting of (1) all type-$0$ triangles, (2) tail type-$1$ triangles, (3) type-$3$ triangles which are not part of any chain. Let $\aset_i = \Nu_i \cap \aset$ be the set of type-$i$ triangles in $\aset$. Also, let $\aset_{13}:=\aset_1 \cup \aset_3$.

\subsubsection*{Demanding triangles:}
Let $\dset$ be the set of all the non-solution triangles such that each triangle contains exactly one type-$0$ solution edge, does not contain any solution edge outside $E(\aset)$, and may only contain non-base edges of type-$1$ triangles in $\aset_1$. Note that, these are precisely the triangles of type-$[0,3]$, type-$[0, 1]$, type-$[0, 1, 1]$, type-$[0, 1, 3]$ and type-$[0, 3, 3]$ minus the ones which we argue to be covered. We refer to any triangle in $\dset$ as a {\em demanding triangle}.

We will show that apart from a few very special cases, any type-$0$ triangle can share at most one edge $e$ with any triangle in $\dset$. 

For any solution triangle $\psi$, let $\dset(\psi) \subseteq \dset$ be the set of all the triangles in $\dset$, adjacent to $\psi$.

We sketch the proof idea of the following technical \Cref{lem:demanding-pairwise};
\ifarxiv
the full proof can be found in~\Cref{subsubsec:proof-demanding-pairwise}.
\fi
\ifjournal
the full proof can be found in~Appendix A.2.1 in the fullversion.
\fi
The proof technique used is very similar to the proof of \Cref{lem:main-type-1-3}.
We prove it by contradiction by showing that for any type-$0$ triangle $\psi \in \Nu_0$, if any two triangles $t_1, t_2 \in \dset(\psi)$ are edge disjoint, then we can find an improving swap by finding alternate solutions using the triangle(s) adjacent to $t_1$ (but not to $\psi$) and the triangle(s) adjacent to $t_2$  (but not to $\psi$) and finally swapping in both $t_1, t_2$ instead of $\psi$.

\begin{lemma}
\label{lem:demanding-pairwise} For any type-$0$ triangle $\psi \in \Nu_0$, for any ${t, t' \in \dset(\psi)}$, $|E(t') \cap E(t)| = 1$.
\end{lemma}

Now we prove a generic claim which will be handy to prove the next lemma.

\begin{claim}
  \label{clm:pairwise-intersecting-triangles}
  Given a set $\sset$ of triangles with $|\sset| > 1 $, if it holds that for any $t, t' \in \sset$, $|E(t') \cap E(t)| = 1$ then one of the two cases is true. Either $|\cap_{t \in \sset} E(t)| = 1$ or $|\sset| \leq 4$, such that $\cup_{t \in \sset}V(t)$ induce a $K_4$.
\end{claim}
\begin{proof}

Let us assume that $|\cap_{t \in \sset} E(t)| \neq 1$, i.e., $|\cap_{t \in \sset} E(t)| = 0$.
It implies that there exists some triangles $t_1, t_2, t_3 \in \sset$, such that $E(t_1) \cap E(t_2) \neq E(t_1) \cap E(t_3)$.
Let $E(t_1) \cap E(t_2) = e_{12}$, $E(t_1) \cap E(t_3) = e_{13}$ and $E(t_2) \cap E(t_3) = e_{23}$ Since $e_{12}, e_{23}$ belongs to $t_2$ and $e_{13}, e_{23}$ belongs to $t_3$, it implies that $e_{23}$ is incident to the common vertex between $e_{12}, e_{13}$, say $u$.
Also, let the other end points of $e_{23}, e_{13}, e_{12}$ be $v_1, v_2, v_3$ respectively.
Clearly $V(t_1) = \{ u, v_2, v_3 \}$, hence $V(t_1) \cup \{v_1\}$ induces a $K_4$. This implies that there is another triangle $t_4 = \Delta_{v_1v_2v_3}$ which shares edge $\overbar{v_2 v_3}, \overbar{v_1 v_3}, \overbar{v_1 v_2}$ with $t_1, t_2, t_3$ respectively.
Hence, any three triangles pair-wise sharing different edges, share a common vertex and induces a $K_4$. Moreover, $t_4$ is a valid triangle to be in $\sset$.
If there exists triangle $t \in \sset \setminus \{t_1, t_2, t_3, t_4\}$, then by the pairwise-intersection property, it has to share one edge each with all the three triangles $t_1, t_2, t_3$.
Since there is no common edge between all three triangles, it should contain at least two edges (say $f_1, f_2$) of the $K_4$ graph induced by $V(t_1) \cup \{v_1\}$. Now $f_1, f_2$ cannot be adjacent otherwise $t$ will be one of $t_1, t_2, t_3, t_4$ otherwise they cannot be part of same triangle $t$.
Hence, no such $t$ exists in this case, which implies our claim.
\end{proof}
Using \Cref{lem:demanding-pairwise} and the claim above we prove the following structural lemma which is the key for defining our final discharging process.

\begin{lemma}
\label{lem:demanding} For any type-$0$ triangle $\psi \in \Nu_0$ and the corresponding set $\dset(\psi)$ only one of the following three cases are possible:
    \begin{enumerate}
        \item \label{itm:demanding-common-edge} Either there exists a common edge $e$, such that $\{e\} = \bigcap_{t \in \dset(\psi)} E(t) \subseteq E(\psi)$ (see~\Cref{subfig:demand-structure-common}),
      \item  \label{itm:demanding-two-doubly-one-hollow} or there are precisely two doubly-attached triangles $t_1, t_2 \in \dset(\psi)$ and one hollow triangle $t_3  \in \dset(\psi)$ adjacent to each solution edge of $\psi$ respectively (see~\Cref{subfig:demand-structure-doubly}),
      \item \label{itm:demanding-three-hollow} or there are precisely three hollow triangles $t_1, t_2, t_3 \in \dset(\psi)$  adjacent to each solution edge of $\psi$ respectively (see~\Cref{subfig:demand-structure-hollow}).
  \end{enumerate}
\end{lemma}
\begin{proof}
\Cref{lem:demanding-pairwise} implies that for any $t, t' \in \dset(\psi) \cup \{\psi\}$, $|E(t) \cap E(t')| = 1$.
Let us assume that \Cref{itm:demanding-common-edge} is not true. This implies that there exists at least two triangles $\{t_1, t_2\} \in \dset(\psi)$ such that $E(t_1) \cap E(\psi) \neq E(t_2) \cap E(\psi)$.
Using \Cref{clm:pairwise-intersecting-triangles}, we deduce that the graph induced by $V(\psi) \cup V(\dset(\psi))$ is a $K_4$. We need to prove that the fourth triangle in the $K_4$ structure also belongs to $\dset(\psi)$ and that the type of triangles which we get here falls in \Cref{itm:demanding-two-doubly-one-hollow}~or~\ref{itm:demanding-three-hollow}.
Let $t_3$ be the fourth triangle in the induced $K_4$ graph. Let the vertices of $\psi$ be $\{v_1, v_2, v_3\}$ such that it shares $\overbar{v_2v_3}, \overbar{v_1v_3}$ with triangles $t_1$ and $t_2$ respectively.
Let the fourth vertex in $V(\psi) \cup V(\dset(\psi))$ be $u$. This implies $V(t_1) = \{u,v_2,v_3\}$, $V(t_2) = \{u,v_1,v_3\}$ and $V(t_3) = \{u,v_1,v_2\}$. Clearly, $\psi$ share edge $\overbar{v_1v_2}$ with $t_3$.
Now since $t_1, t_2 \in \dset(\psi)$, this implies that the edges $\{\overbar{uv_1}, \overbar{uv_2}, \overbar{uv_3}\}$ are either non-solution edges or solution edges adjacent to some triangles in $\aset_{13}$. Hence $t_3 \in \dset(\psi)$.

Note that at least two edges out of $\{\overbar{uv_1}, \overbar{uv_2}, \overbar{uv_3}\}$ should be solution edges as $\psi$ is a type-$0$ triangle. This implies that at least one of $t_1, t_2, t_3$ is a hollow triangle. If the third edge is a non-solution edge, then the other two triangles will be doubly-attached triangles else all three will be hollow triangles which proves the two cases.
\end{proof}

\input{tikz_figures/demanding-structure}

\subsubsection*{Discharging overview:} Using \Cref{lem:demanding}, we describe our discharging scheme by giving some high-level intuition before the formal description.

This discharging scheme aims at covering the set of demanding triangles $\dset$, while maintaining the coverage of the triangles outside $\dset$. Note that even if some of these triangles may be initially covered, we still put them in $\dset$ and {\em declare} them to be {\em covered} only during our process to make sure that we do not uncover any of them along the way. We also make sure that the process does not uncover any of the triangles outside of $\dset$.

Remember that after the initial charge distribution, each type-$0$ triangle has an unused half credit. The first natural attempt is to allocate these half credits in a clever way to cover all triangles in $\dset$. Clearly it does not affect the coverage of any triangle outside of $\dset$, but unfortunately only using this allocation does not work. If a type-$0$ triangle is in the second or the third sub-case of \Cref{lem:demanding}, then it is actually impossible to put this half credit anywhere to cover all $t_1, t_2, t_3$ at the same time (see~\Cref{fig:discharge-impossible}). Then arises the need to exploit credits from type-$3$ and type-$1$ triangles in $\aset_{13}$ to cover these demanding triangles.
\input{tikz_figures/discharge-impossible.tex}
For any type-$1$ triangle $\psi^1 \in \aset_1$ we use the property in \Cref{obs:type-1-rotation} that allows us to freely {\em rotate} the allocated credits by picking any one of the non-base edge to be the null-edge. Hence, if for $\psi^1$ there is a non-base solution edge $e^1$ that is not adjacent to any demanding triangle in $\dset$, then we can choose the charging that makes $e^1$ a null-edge without affecting the coverage of triangles outside of $\dset$.

Similarly, for any type-$3$ triangle $\psi^3 \in \aset_3$ (by \Cref{obs:type-3-rotation}), we are free to \emph{rotate} the allocated credits by specifying the one null-solution edge. Hence, if there is a solution edge $e^3$ of $\psi^3$ that is not adjacent to any demanding triangle in $\dset$, then we can make $e^3$ a null-edge without affecting the coverage of triangles outside of $\dset$.

But we cannot use these operation for the $\psi \in \aset_1$ or $\psi' \in \aset_3$ triangles which have demanding triangles sharing both the non-base or all three solution edges respectively. Again, in such a situation we can try to use the extra half credit from a type-$0$ to cover some of these triangles.

Note that the two rotation operations have a very similar structure. Also in the $\dset$ these type-$1$ and type-$3$ triangles in $\aset_1$ and $\aset_3$ are structurally very similar. The only difference is that, for type-$3$ we have three possibilities for the null-edge whereas for the tail type-$1$ there are only two (non-base edges) possibility for the null-edge. But also the triangles in $\dset$ are not adjacent to the base-edges of triangles in $\aset_1$, hence we can perform the rotation operations analogously everywhere.

At this point, the main difficulty lies in taking advantage of the unused half credit from type-$0$ triangles in $\aset_0$ and the rotation operations for triangles in $\aset_{13}$ in a coherent and synchronized way. Fortunately, we can show that a very simple and natural greedy procedure can cover all the demanding triangles without affecting any triangle outside of $\dset$. We call it {\em Discharge-and-Pin} algorithm.


Our procedure works in iterations as follows. We maintain a dynamic set (initially $D = \dset$) which contains our potentially uncovered demanding triangles and remove from $D$ the triangles which we {\em declare} as {\em covered}. We also keep track of triangles in $\aset$ for which we have not discharged or rotated the credits yet, so that we perform these operations at most once for any triangle in $\aset$. We repeat until $D = \varnothing$ and declare that all triangles in $G$ are covered.

First, for any type-$0$ triangle $\psi^0 \in \aset_0$, if any triangle in $D$ is adjacent to at most one solution edge $e \in \psi^0$ (similar to  \Cref{lem:demanding}~\Cref{itm:demanding-common-edge}), then we assign the extra half credit of $\psi^0$ to $e$ making it a full-edge. Then we remove from $D$ all triangles adjacent to $e$ and go to the next iteration.


Otherwise, it would be true that the demanding triangles adjacent to the type-$0$ triangle $\psi^0$ must lie in \Cref{lem:demanding} \Cref{itm:demanding-two-doubly-one-hollow}~or~\ref{itm:demanding-three-hollow}.

Then we start an iteration by fixing a rotation of any arbitrary triangle $\psi \in \aset_{13}$, which covers all the demanding triangles except the ones adjacent to the only null-edge $e \in E(\psi)$. We first try to find an edge of $\psi$ (excluding the base-edge in case when $\psi \in \aset_1$) that is not adjacent to any triangle in $D$. If we can find such an edge, then we fix $e$ to be this edge, remove covered triangles from $D$ and move on to the next iteration.

Else, we fix an edge $e$ arbitrarily and rotate the credits of $\psi$ so that $e$ becomes a null-edge. We look at any demanding triangle $t$ adjacent to the null-edge $e$ of $\psi$ and {\em discharge} half credit from the type-$0$ triangle (say $\psi^0$), adjacent to $t$, to $e$ by making it a half-edge.
Remove all the triangles adjacent to $\psi$ from $D$, as they are covered. Now for the type-$0$ triangle $\psi^0$, once it spends the extra half credit, two out of three demanding triangles adjacent to it are covered. For the third demanding triangle $t'$, we pick any triangle $\psi' \in \aset_{13}$ which is adjacent to $t'$ via edge $e'$.
To make sure that $t'$ gets covered, we fix any one of the rotations for $\psi'$ in which $e'$ will be a half-edge. Then we continue with the iteration as we did for $\psi$, via the null-edge $e''$ of $\psi'$, covering and removing triangles from $D$ using discharge and rotate operations, until we arrive at a case where we reach a null-edge which is not adjacent to any triangle in $D$. Then we start with a new iteration. We show that we can repeat this natural depth-first-search like process until every triangle is covered without ever getting stuck. We already know that any of these operations cannot affect the coverage of any triangle outside of $\dset$. We can show that any step of the procedure does not affect the coverage of any triangle in $\dset \setminus D$ and we discharge or rotate any triangle in $\aset$ at most once. Also, note that we only move around initially assigned credits and unused credits of type-$0$ triangles, hence we do not overcharge. All these properties together imply that the procedure is well-defined and gives a valid charging.

Now we give the formal description of our algorithm.

\subsubsection*{Discharge-and-Pin Operations:} We define three operations, one for type-$0$ triangles, one for type-$1$ triangles in $\aset_1$ and one for type-$3$ triangles in $\aset_3$.
\input{tikz_figures/discharge-operation.tex}
\begin{enumerate}
    \item For any type-$0$ triangle $\psi^0$ with extra half credit, we want to utilize this half credit to cover some demanding triangles closed to $\psi^0$ in $\dset$, the set of demanding triangle adjacent to $\psi^0$. Let $discharge(\psi^0, e)$ be the function that, when called, discharge the extra half credit of $\psi^0$ to the edge $e$ (not necessarily in $E(\psi^0)$, see~\Cref{fig:discharge-operation}).
    \item For type-$1$ triangle $\psi^1 \in \aset_1$, we want to rotate the credit of $\psi^1$ to cover some triangles in $\dset(\psi^1)$. For a non-base edge $e \in E(\psi^1)\setminus base(\psi^1)$, let $pin(\psi^1, e)$ be the function that, when called, rotate the credit of $\psi^1$ in a way that $e$ becomes a null-edge, (see~\Cref{subfig:pin-operation-type-1}).
    To be more specific, let $v = V(\psi^1) \setminus V(e)$ and let $a = anchor(\psi^1)$, we put half credit on four edges in of $K_4$ induced by $V(\psi^1) \cup \{a\}$ except $e$ and $\overbar{v a}$.
    \item For type-$3$ triangle $\psi^3 \in \aset_3$, we want to rotate the credit of $\psi^3$ to cover some triangles in $\dset(\psi^3)$. For an edge $e \in E(\psi^3)$, let $pin(\psi^3, e)$ be the function that, when called, rotate the credit of $\psi^3$ in a way that $e$ becomes a null-edge, (see~\Cref{subfig:pin-operation-type-3}).
    To be more specific, let $v = V(\psi^3) \setminus V(e)$ and let $a = anchor(\psi^3)$, we put half credit on four edges in of $K_4$ induced by $V(\psi^3) \cup \{a\}$ except $e$ and $\overbar{v a}$.
\end{enumerate}
\input{tikz_figures/pin-operation.tex}

By \Cref{obs:type-1-rotation,obs:type-3-rotation} the pin operation does not uncover anything. Combining this with the fact that the discharge operation for type-$0$ does not reduce credits on any edge, we get the following lemma.

\begin{lemma}
  \label{lem:covered_triangles_are_still_covered}
  Let $\sigma$ be any sequence of Discharge-and-Pin operations called on each triangle in $\aset$ at most once. Let $t$ be a triangle not in $\dset$, then $t$ is covered after the initial charge distribution and is still covered after applying $\sigma$.
\end{lemma}

Now we prove the following lemma by showing an iterative algorithm to find the sequence $\sigma$ and cover the triangles in $\dset$ along the way, such that $\sigma$ performs discharge or pin operation on any triangle in $\aset$ at most once.

\begin{lemma}
  \label{lem:iterative-discharging}
  There exists a sequence of Discharge-and-Pin operations $\sigma$ which performs these operations on each triangle in $\aset$ at most once, such that applying $\sigma$ covers all triangles in $\dset$.
\end{lemma}

\subsubsection*{Discharge-and-Pin Algorithm:} This algorithm will find the sequence $\sigma$ which satisfies \Cref{lem:iterative-discharging}.

Let $D = \dset$ be the current demanding triangles that we need to cover.
For brevity, we define $D(e) = \{t \in D : e \in E(t)\}$ to be the set of triangles in $D$ containing the edge $e$.
Also, for any triangle $t$, let $D(t) = \bigcup_{e \in E(t)} D(e)$.

Recall the set of triangles $\aset$ consisting of (1) all type-$0$ triangles, (2) tail type-$1$ triangles, (3) type-$3$ triangles which are not part of any chain. Let $A=\aset$ be the set of triangles for which we have not performed the pin or discharge operation yet.
 At every step, we will find a triangle $\psi \in A$ to perform discharge or pin operation (based on the type of triangles) to cover some triangles in $D$. Then we remove $\psi$ from $A$ and the covered triangles from $D$. We can show that until $D$ is empty, we can always find some $\psi \in A$ to perform discharge or pin operation to cover some triangles in $D$. This together with the fact that we just pin or discharge any triangle in $\aset$ at most once, implies that we cover all the triangles in $D$. In the description of the algorithm we assume that we can always find some triangle in $A$ at every step, which we prove later. Each while loop corresponds to an iteration which performs a sequence of discharge and pin operations in a greedy depth-first way starting from some triangle in $A$. At any point in the algorithm, we use $A_i$ to refer to $A \cap \aset_i$ and $A_{13}$ to refer to $A \cap \aset_{13}$.

while $D \neq \varnothing$, we will do the following.
\begin{enumerate}
    \item  If for any type-$0$ triangle $\psi^0 \in A_0$, there exists exactly one edge $e \in E(\psi^0)$ such that $D(e) \neq \varnothing$, then we can apply $discharge(\psi^0, e)$~(\Cref{subfig:discharge-to-demanding-edge}). Once the operation is applied on $\psi^0$, we remove $\psi^0$ from $A$ and remove $D(e)$ from $D$ and repeat.
    \item Else we pick an arbitrary triangle $\psi_1 \in A_{13}$ to be the starting point of the current iteration. Then we greedily alternate between triangles in $A_0$ and $A_{13}$ while fixing their distribution. $\psi_i$ would be the $i^{th}$ inspected triangle in $A_{13}$. Similarly, $\psi^0_i$ would be the $i^{th}$ inspected triangle in $A_0$. We inspect $\psi^0_i$ right after $\psi_i$.
    To make sure that we cover all the triangles, we keep track of some {\em critical} edge $p_i \in \psi_{i+1}$ which needs to be a half-edge, based on the charging of previously fixed triangles $\psi^0_i$ and $\psi_i$.

    To begin with the current iteration, let $p_0 = \varnothing$. For any $i>0$, based on the information in $p_{i-1}$ we will decide to pin one edge of $\psi_i$.
    \begin{enumerate}
        \item If $\psi_i \in A_3$ and there exists an edge $e_i \neq p_{i-1}$, such that $D(e_i) = \varnothing$, then we can pin this edge and end this iteration.
        Similarly, If $\psi_i \in A_1$ and there exists a non-base edge $e_i \neq p_{i-1}$, such that $D(e_i) = \varnothing$, then we can pin this edge and end this iteration. That is, we apply $pin(\psi_i, e_i)$~(\Cref{subfig:pin-operation-type-3}), set $D \leftarrow D \setminus D(\psi_i)$ and remove $\psi_i$ from $A$ and set $A \leftarrow A - \psi_i$ and start a new iteration.

        Observe that any demanding triangle in $D(\psi_i)$ is now covered since they are adjacent to a half-solution edge of $\psi_i$ ($D(e_i) = \varnothing$) and a half-solution edge of a type-$0$ triangle.

        \item Otherwise in case $\psi_i \in A_3$, for each $e \in E(\psi_i)$, $D(e) \neq \varnothing$. We pick $e^3_i$ to be any one of at least two edges in $E(\psi_i) \setminus p_{i-1}$. Similarly, in case $\psi_i \in A_1$, for each $e \in E(\psi_i) \setminus base(\psi_i)$, $D(e) \neq \varnothing$.
        We pick $e_i$ to be the non-base edge in $E(\psi_i) \setminus \{p_{i-1}, base(\psi_i)\}$. We apply $pin(\psi_i, e_i)$ (~\Cref{subfig:pin-operation-type-1}), set
        $D \leftarrow D \setminus (D(\psi_i) \setminus D(e_i))$ and remove $\psi_i$ from $A$ and set $A \leftarrow A - \psi_i$.

         In this case, since $D(e_i) \neq \varnothing$ and $e_i$ is a null-edge, hence we continue with the current iteration, as there will be some uncovered demanding triangles adjacent to $e_i$.  Pick $t_i$ arbitrarily from $D(e_i)$. Let $e^0_i$ be the solution edge of a type-$0$ triangle in $E(t_i)$. Let $\psi^0_i$ be the type-$0$ triangle containing $e^0_i$. Here we assume that $\psi^0_i \in A$.
         Let ${e'}^0_i$ be the edge of $\psi^0_i$ not adjacent to $V(\psi_i) \cap V(\psi^0_i)$.
         Once we have $\psi^0_i, e_i, e^0_i, {e'}^0_i$, we will discharge half credit of $\psi^0_i$.
         We call $discharge(\psi^0_i, e_i)$~(\Cref{subfig:discharge-to-other-edge}). Now we cover all triangles in $D(e_i)$ since they will be adjacent to one half-solution edge of a type-$0$ triangle and $e_i$ which is now a half-edge. In particular, we cover the two demanding triangles adjacent to $E(\psi^0_i) \setminus {e'}^0_i$.
         Hence, we can remove $D(e_i)$ from $D$ and $\psi^0_i$ from $A$.
         \begin{enumerate}
             \item If $D({e'}^0_i) = \varnothing$, then we are done with this iteration. This could happen when the triangle $D({e'}^0_i)$ got covered when one of the type-$3$ triangle(s) adjacent to this non-solution triangle got pinned previously.
             \item Else, we let ${t'}_i = D({e'}^0_i)$. Note that there is exactly one such triangle, since for any type-$0$ belonging to \Cref{lem:demanding}.\ref{itm:demanding-common-edge}, the algorithm would finish discharging them in the initial iterations.
             We let $p_i \in E({t'}_i)$ be a solution edge of some $\psi_{i+1} \in A_{13}$ (there can be one or two such triangles, breaking ties arbitrarily). Also here we assume that there is such $\psi_{i+1} \in A$~(\Cref{subfig:discharge-to-other-edge}). Continue with the current iteration using $p_i$ and $\psi_{i+1}$. Note that, fixing this edge as $p_i$ ensures that we cover the last triangle ${t'}_i \in D$ adjacent to $\psi^0_i$.
         \end{enumerate}
    \end{enumerate}
\end{enumerate}

\subsubsection{Proof of \Cref{lem:iterative-discharging}}
\label{subsubsec:proof-iterative-discharging}

By the way we define the algorithm, it is clear that we discharge or pin any triangle in $\aset$ at most once.
Now we show a series of claims to prove this lemma.

First we show that at every step when we remove any triangle in $D$, they are covered. For triangles $D(e)$ falling in  \Cref{lem:demanding}.\ref{itm:demanding-common-edge} for some $\psi \in \aset_0$ and a common demanding edge $e \in \psi$, we call $discharge(\psi, e)$ which makes $e$ a full edge, hence all triangles in $D(e)$ are covered.

For the other steps we prove the following claims.

\begin{claim}
\label{clm:covered-demanding-triangles-type-13}
In any iteration, for any $i>0$, $D(\psi_i) = \varnothing$ after calling $discharge(\psi^0_i, e_i)$ and removing $D(e_i)$ from $D$.
\end{claim}
\begin{proof}
  We did call $pin(\psi_i, e_i)$. Hence, any triangle in $D(\psi_i) \setminus D(e_i))$ is covered and removed from $D$.
  When we call $discharge(\psi^0_i, e_i)$, we cover $D(e_i)$ and remove them.
\end{proof}

\begin{claim}
\label{clm:covered-demanding-triangles-type-0}
In any iteration, for any $i>1$, $D(\psi^0_{i-1}) = \varnothing$ after calling $pin(\psi_i, e_i)$ and setting $D \leftarrow D \setminus (D(\psi_i) \setminus D(e_i))$.
\end{claim}
\begin{proof}
  Since we have already covered and removed triangles from $D$ falling in \Cref{lem:demanding}.\ref{itm:demanding-common-edge}, it implies that, $|D(e)| = 1$ for all $e \in E(\psi^0_{i-1})$.
  When we call $discharge(\psi^0_{i-1}, e_{i-1})$, two triangles in $D(\psi^0_{i-1}) \setminus D({e'}^0_i)$ are covered and removed from $D$. Calling $pin(\psi_i, e_i)$ cover the triangle in $D({e'}^0_i)$.
\end{proof}

Now we prove the two assumptions we made in the algorithm.

\begin{claim}
\label{clm:existance-in-A-type-13}
 In any iteration, for any $i>0$, we can find some $\psi_{i+1} \in A_{13}$, such that $p_i \in \psi_{i+1}$.
\end{claim}
\begin{proof}
  This is true as $D(\psi_{i+1}) \neq \varnothing$ when we choose $\psi_{i+1}$ but $D(\psi) = \varnothing$ for any $\psi \in \aset_{13} \setminus A_{13}$, because we remove them from $D$ when we remove $\psi$ from $A$.
\end{proof}

\begin{claim}
\label{clm:existance-in-A-type-0}
  In any iteration, for any $i>0$, we can find $\psi^0_i \in A_0$ which contains $e^0_i$.
\end{claim}
\begin{proof}
  This is true as $D(\psi^0_i) \neq \varnothing$ when we choose $\psi^0_i$ but $D(\psi) = \varnothing$  for any $\psi \in \aset_{0} \setminus A_{0}$, because we remove them from $D$ when we remove $\psi$ from $A$.
\end{proof}

Now we show that triangles in $\dset \setminus D$ remain covered which implies that when $D = \varnothing$, everything is covered.

\begin{claim}
  \label{clm:demanding-remain-covered} After any discharge or pin operation, every triangle in $\dset \setminus D$ is covered.
\end{claim}
\begin{proof}
By the above reasoning, it is clear that whenever we remove any triangle from $D$, it is covered at that point in time. Also, discharge operation cannot uncover any triangle as it does not reduce credits on any edge. What remains to argue is that every triangle $\dset \setminus D$ {\em stay covered} after any pin operation. Recall that we start with $D = \dset$ to be all the triangles which may not be covered, which implies that they are oblivious to the initial tentative charging of any triangle in $\aset_{13}$. Now by the way we proceed, we only rely on the half credits of type-$0$ triangles edges, the half credits of the edges of the triangles which are pinned or the credits of the edges to which a type-$0$ triangle discharge credits to cover and remove triangles from $D$. Since we only discharge or pin at most once, the algorithm never removes credits from these edges hence triangles in $\dset \setminus D$ stay covered during the execution of the algorithm.
\end{proof}

The following claim finishes the proof by implying that the Discharge-and-Pin algorithm does not terminate until $D = \varnothing$.

\begin{claim}
\label{clm:DP-algorithm-termination}
As long as $D \neq \varnothing$, we can always find a triangle $\psi \in A$ as the starting point of the next iteration.
\end{claim}
\begin{proof}
  First note that for any triangle in $t \in D$ which is adjacent to any type-$0$ triangle $\psi$,
  such that it lies in \Cref{lem:demanding}.\ref{itm:demanding-common-edge}, the algorithm will always find $\psi \in A$, discharge it, cover $t$ and remove it from $D$.

Now any remaining triangle $t \in D$, is adjacent to some triangle $ \psi \in A_{13}$, which implies $D(\psi) \neq \varnothing$. But this implies that $\psi \in A$ because for any triangle $\psi' \notin A$, $D(\psi') = \varnothing$ by \Cref{clm:covered-demanding-triangles-type-13,clm:covered-demanding-triangles-type-0}.
\end{proof}

~\Cref{lem:covered_triangles_are_still_covered,lem:iterative-discharging} together imply that we do not overcharge any triangle and we cover every triangle in $G$ which finishes the proof for the main result of this section.


\subsubsection*{Acknowledgement:} Part of this work was done while PC and SK were visiting the Simons Institute for the Theory of Computing. It was partially supported by the DIMACS/Simons Collaboration on Bridging Continuous and Discrete Optimization through NSF grant \#CCF-1740425. PC is currently supported by European Research Council (ERC) under the European Union's Horizon 2020 research and innovation programme (grant agreement No 759557) and by Academy of Finland Research Fellows, under grant number 310415. SK is currently supported on research grants by Adobe and Amazon.
We also thank the anonymous reviewers for their detailed comments and suggestions.

\bibliographystyle{plain}
\bibliography{references}

\ifarxiv
\appendix

\section{Omitted proofs for \Cref{sec:order-2-fix}}

\subsection{Omitted proofs for \Cref{subsec:type-1-3}}
\label{subsec:proofs-type-1-3}

We start by defining some notations for any chain structure $\cset \in \Chi$.
\paragraph{More Notations for Chains}
For any chain $\cset = \{\psi_k, \psi_{k-1}, \dots, \psi_0\}$, we define for each $i>0$, $c_i$ to be the common vertex between $\psi_i, \psi_{i-1} \in \cset$.
Note that, $c_i$ will always be the non-base vertex of $\psi_i$ by the way the $lend$ structure is defined.
Let $\omega_i$ be the singly-attached triangle of $\psi_i$ and $b_i$ be the corresponding base edge and $a_i:=anchor(\psi_i)$.
For $\psi_0$ let $\omega^1_0, \omega^2_0, \omega^3_0$ be the three singly-attached triangles of $\psi_0$, such that $\omega^3_0$ be the one not containing $c_1$. Let $\omega_0 := \omega_0^3$.
Finally let $t^1_i, t^2_i$ be the two twin-doubly-attached triangles adjacent to $\psi_{i-1}, \psi_i$. Let $g_i := gain(\psi_{i}, \psi_{i-1})$ be the gain edge corresponding to $(\psi_{i}, \psi_{i-1})$.
Additionally, we define $g_{k+1}$ to be the null non-base edge of $\psi_k = tail(\cset)$. Think of $g_{k+1}$ as the {\em potential} gain edge where $\cset$ would have been extended. Using this, we define $c_0 := V(\psi_0) \setminus V(g_1)$. Let $h_i$ be the edge in $E(\psi_i) \setminus \{b_i, g_{i+1}\}$. Notice that both $t^1_i, t^2_i$ contain vertex $c_i$ and the $g_i$ edge.
Also, one of $t^1_i, t^2_i$ contains $g_{i+1}$. By renaming we assume that $g_{i+1} \in E(t^2_i)$.  Let $e^1_i, e^2_i$ be the non-solution edges of $\omega_i$ contained in $t^1_i, t^2_i$ respectively. Clearly $g_{i+1}$ and $e^2_i$ are adjacent.

\input{tikz_figures/chain-structure.tex}

The following facts will be used many times later.

\begin{fact}[Solution edges]
\label{fact:chain-solution-edges}
In any chain $\cset = \{\psi_k, \psi_{k-1}, \dots, \psi_0\}$, let $e$ be any solution edge in $E(\psi_i)$ for some $i \in [k]$. Among the set of twin-doubly-attached triangles $\{t^1_i, t^2_i \}_{i \in [k]}$ and singly-attached triangles $\{\omega_i\}_{i \in [k]}$, if $e = b_i$, then it is contained in only $\omega_i$. In the case when $e = h_i$, then it is contained in only $t^1_i$, otherwise $e = g_{i+1}$ will be contained in $t^2_i, t^1_{i+1}, t^2_{i+1}$.
\end{fact}
\begin{proof}
  It is straight forward to see that the mentioned triangle(s) will contain $e$.
  If $e = b_i$, i.e., $e$ is a base-edge, then clearly $e \notin E(\omega_j)$ for all $j \neq i$, by disjointness of $\psi_i, \psi_j$.
  Also, by  \Cref{def:lend-and-gain} and by the restriction that $gain(\psi_{i+1}, \psi_i) \neq base(\psi_i)$ in the chain construction, $e$ is not in any twin doubly-attached triangle.

  By similar argument, we can say that if $e$ is not base-edge, then $e$ is not in
  any $\omega_j$ for all $j$.
  Now suppose $e$ is not a base-edge. If $e$ is in a doubly-attached triangle $t \in \{t^1_i, t^2_i \}_{i \in [k]}$
  not captured by the fact, then it implies $t$ is adjacent to $\psi_i$ and $\psi_j$
  where $|i-j| > 1$. But this is not possible since it would imply that
  $(\psi_i, \psi_j) \in lend$ or $(\psi_j, \psi_i) \in lend$ which contradicts
  \Cref{lem:lend-outdegree-one}.
\end{proof}

\begin{fact}[Non-solution edges]
\label{fact:chain-non-solution-edges}
For any chain $\cset$, and any $i>1$, both the non-solution edges of $\omega_i$ and $\omega_{i-1}$ are different.
\end{fact}
\begin{proof}
   If not, then that would imply $(\psi_{i-1}, \psi_i) \in lend$, but $\psi_{i-1}$ must discharge credits to another triangle $\psi_{i-2} \neq \psi_{i}$ which requires $(\psi_{i-1}, \psi_{i-2}) \in lend$, which contradicts \Cref{lem:lend-outdegree-one}.
\end{proof}

The following fact is true by our naming convention.
\begin{fact}[Adjacency edges]
\label{fact:chain-adjacent-edges}
For any chain $\cset$, and any $i>0$, the non-solution $e^1_i$ is incident to the common vertex of edges $b_i, h_i$ and the non-solution $e^2_i$ is incident to the vertex $c_i$ which is also the common vertex of edges $b_i, g_{i+1}$.
\end{fact}


\paragraph{Alternating Solutions using Chains}
A possible alternate solution to free-up and make the null non-base edge $g_{k+1}$ a non-solution edge is to choose $\omega_k$ instead of $\psi_k$. Similarly, to free-up a half-base edge of some triangle $\psi_i$, a straight forward way is to swap out $\psi_i, \psi_{i-1}$ with $t^1_i, \omega_{i-1}$. The following claim ensures that the solution will still be valid.

\begin{claim}
\label{clm:chain-consecutive-disjoint-singly-twins}
For any chain $\cset = \{\psi_k, \psi_{k-1}, \dots, \psi_0\}$, and any $i>0$,
\begin{enumerate}
    \item If $i=1$, then for any $t \in \{t^1_1, t^1_2, \omega_1\}$, there is an $\omega \in \{\omega_0^1, \omega_0^2, \omega_0^3\}$ such that $t$ and $\omega$ are edge-disjoint.
    \item For $i>1$, $\omega_i$ is edge-disjoint from any triangle in $\{ \omega_{i-1}, t^1_{i-1}, t^2_{i-1} \}$.
    \item $t^1_{i-1}$ is edge-disjoint from any triangle in $\{\omega_i, t^1_i, t^2_i \}$
    \item Any $t \in \{t^1_i, t^2_i \}$ is edge-disjoint from any triangle in $\{ t^1_{i-1}, \omega_{i-1} \}$.
\end{enumerate}
\end{claim}
\begin{proof}
Follows from \Cref{fact:chain-solution-edges,fact:chain-non-solution-edges}.
\end{proof}

The above claim implies the following corollary.

\begin{corollary}
\label{cor:small-alternate-solution}
For any chain $\cset = \{\psi_k, \psi_{k-1}, \dots, \psi_0\}$, there exists the following valid swaps leading to an alternate optimal packing solution such that:
\begin{enumerate}
    \item \label{itm:free-g_k+1} Swapping $\{ \psi_k, \psi_{k-1} \}$ with $\{ t_k^1, \omega_{k-1} \}$ makes the null non-base edge $g_{k+1}$ a non-solution edge.
    \item \label{itm:free-semi-rich-base} For any $i>0$, swapping $\{ \psi_i, \psi_{i-1} \}$ with $\{ t_i^1, \omega_{i-1} \}$ makes the half base edge $b_i$ of $\psi_i$ a non-solution edge.
\end{enumerate}
\end{corollary}

The above swap operations are equivalent to flipping the edges which are in and out of a matching $M$ along an even length alternating path. Note that such a flip does not affect the size of $\Nu$ and $M$ respectively. The advantage of these alternating chains of triangles over the augmenting path for matching is that we have three choices for triangles which can be swapped in (namely $\omega_i, t^1_i, t^2_i$).
But these swap operations are not enough to get the full argument.
As compared to augmenting paths in matching, these augmenting chains are difficult to deal with as they may not be simple and can {\em entangle} with itself. They may also overlap with other chains/singly-attached triangles belonging to the structure from other solution edge side of the potentially-uncovered triangle $t$. These overlaps happen at the non-solution edges making it difficult to get an overall independent alternate solution on the two sides of $t$. We can overcome the complication coming from overlaps by exploiting the rich structure of our alternating chains and showing a strategy to choose the non-solution triangles such that whatever way the chains share the non-solution edges, we can always make the flip while keeping all the triangles edge-disjoint. In particular, we take advantage of the three choices for triangles which can be swapped in (namely $\omega_i, t^1_i, t^2_i$) at the $i$th point of the chain structure.

\subsubsection{Proof of \Cref{lem:main-type-1-3}}
\label{subsubsec:proof-lem:main-type-1-3}
In this section we prove \Cref{clm:excluding-type-11}-\ref{clm:excluding-type-133} that exclude all the possibly uncovered triangles, which finishes the proof of \Cref{lem:main-type-1-3}. In each claim we do an exhaustive case analysis to show that either such a triangle is already covered or there is an improving swap contradicting its existence. This is the most technical part of the paper where we essentially exploit the various choices of alternate triangles present in the chain structure to find a swap in each possible configuration.

Recall that in \Cref{subsec:type-1-3}, we deal with the graph $G$ and an optimal packing solution $\Nu$ without any type-$0$ triangle.
We inspect each type of potentially-uncovered triangle and prove that they cannot exist in the graph. For a full intuitive description of the proof strategy we refer the readers to the ''proof-of-concept`` paragraph after \Cref{lem:main-type-1-3}.

\paragraph{Simplification:} Using the {\em vanilla} version of the Discharging-via-Chains construction
(described in \Cref{sec:order-2-fix}),
led to very tedious case-analysis based proofs for
the technical \Cref{clm:excluding-type-11}-\ref{clm:excluding-type-133}.

To get around this, we develop a {\em fine-tuned} Discharging-via-Chains construction,
which does not affect the coverage of triangle
and the properties in \Cref{obs:type-1-rotation,obs:type-3-rotation}. This version of the proof is easier to verify as it involves more structure and less case analysis.

\subsubsection*{Satisfy-and-Truncate Discharging-via-Chains construction:}
\label{subsubsec:satisfy-truncate-chains}

First note that the following observation is true.
\begin{observation}
\label{obs:chain-intermediate-twin-doubly-coverage}
For any $k$-sized chain $\cset$, $g_i$ is a half-edge for any $i \leq k$. This implies that for $0<i<k$, the twin-doubly attached triangle $t^1_i$ is covered by half-edges $h_i, g_i$ and $t^2_i$ is covered by half-edges $g_i, g_{i+1}$.
\end{observation}
The above observation implies the following observation which we will crucially use in the development of the new process.
\begin{observation}
\label{obs:chain-intermediate-non-solution-charge-flexibility}
For any $k$-sized chain $\cset$ and $0 < i < k$, the half credit on the non-solution edge of $\omega_i$ can be {\em freely} assigned to any of the edge of $\omega_i$, while maintaining the coverage of the solution and the twin-doubly attached triangles in $\cset$.
\end{observation}

Another, two easy to see observations are stated below which we use several times in our proof.
These observations are true by the symmetry of our structure.

\begin{observation}
\label{obs:head-type-3-flexibility}
For $\psi_0 \in \cset$, any arbitrary non-solution edge
in the $K_4$ induced by $V(\psi_0) \cup anchor(\psi_0)$ can be the null-edge.
\end{observation}


\paragraph{Overview:} Using the above observations, we modify the Discharging-via-Chains construction to get some strong structural properties. 
To maintain the flexibility in \Cref{obs:type-1-rotation,obs:type-3-rotation} for the chains construction, we do not fix the credits for the $\psi_i$ but only until $\psi_{i-1}$ while constructing any chain $\cset$. 
We do this until the very end, when all the chains have been constructed and there in no type-$[1, 3]$ uncovered triangle in our graph $G$.

The idea is (during the $i$th step of the construction of a chain $\cset$) to {\em fix} the assignment of the flexible credits for the singly-attached triangle $\omega_{i-1}$ of $\psi_{i-1}$ (see~\Cref{obs:chain-intermediate-non-solution-charge-flexibility}) by assigning it to a specific edge (in particular $e^1_{i-1}$ for $i>1$).
For $i=1$, we would fix the null-non-solution edge for $\psi_0$ (see~\Cref{obs:head-type-3-flexibility}) to be a specific edge (in particular $e^3_0$) and the other two non-solution edges to be half-edges (For instance see~\Cref{fig:new-chain-charging}).
This imposes a strong structure which we crucially exploit in several surprising ways.

We start by defining some additional notations which we need for our new construction.

\begin{definition}[Satisfied and unsatisfied triangles]
    Any solution triangle $\psi \in \Nu$ is a \emph{satisfied} triangle iff every edge $e$ of $\psi$ is at least a half-edge.
    Otherwise, we say that $\psi$ is an \emph{unsatisfied} triangle.
\end{definition}

\begin{definition}
    A non-solution edge $e$ is a non-solution edge of a chain $\cset = \{\psi_0, \psi_1, \ldots, \psi_k\}$ iff
    $e \in \{e_i^1, e_i^2 : 0 \leq i \leq k\} \cup \{e_0^3\}$.
\end{definition}

Now we define {\em half-non-solution} edges of a chain $\cset$, which exclusively gets exactly half-credit from a solution triangle in $\cset$ during its construction. 

\begin{definition}
\label{def:half-edge-of}
    A non-solution edge $e$ is a half-non-solution edge of a chain $\cset = \{\psi_0, \psi_1, \ldots, \psi_k\}$ iff
    $e \in \{e_i^1 : 0 < i < k\} \cup \{e_0^1, e_0^2\}$.
\end{definition}

For any two chains $\cset, \cset' \in \Chi$,
when we say that $\cset$ {\em contains} a half-edge $e$ of $\cset'$, it means that $e$ is a half-non-solution edge of $\cset'$ and is also a non-solution edge of $\cset$.

When we say that $\cset$ and $\cset'$ {\em share} a half-edge $e$, it means
that $e$ is a half-non-solution edge of both $\cset$ and $\cset'$.
In this case, it is true that $f(e) \geq 1$ (since it takes half credits from both $\cset$ and $\cset'$).
Notice that if $\cset$ and $\cset'$ share an edge $e$, then it is true that $\cset$ contains $e$ of $\cset'$, however, if $\cset$ contains $e$ of $\cset'$, it is not necessarily true that $\cset$ and $\cset'$ share $e$.

\input{tikz_figures/satisfy-itself.tex}
\paragraph{Truncate-and-Satisfy:} Imagine a situation where during the $i^{th}$ step of growing some chain $\cset$, 
$\omega_i$ contains some half-non-solution edge for which the half credit was fixed
earlier due to some previously grown chain $\cset'$. In this case, we can {\em satisfy}
$\psi_i$ by making each solution-edge of $\psi_i$ a half-edge (see~\Cref{fig:chain-satisfying-other-chain}).
This way we ensure the coverage of $\omega_i$ while getting rid of the requirement for some
new triangle $\psi_{i+1}$ to $lend$ credit to $\psi_i$. After this there is no need to grow
$\cset$ anymore since there is no null-edge in $\psi_i$. We can do the same, when it shares some
half-non-solution edge contained in one of its own earlier singly-attached triangles (see~\Cref{fig:satisfy-itself}).
Once $\psi_i$ is satisfied, the chain $\cset \ni \psi_i$ becomes a {\em satisfied} chain.
In contrast, any tail triangle (of any existing chain) containing a null-edge will be an {\em unsatisfied} tail triangle. 

Using this, we segregate chains in $\Chi$ into {\em satisfied or unsatisfied} chains using the following definition.

\begin{definition}[Satisfied and unsatisfied chains]
    A chain $\cset \in \Chi$ is a \emph{satisfied} chain iff $tail(\cset)$ is satisfied.
    Otherwise, we say that $\cset$ is an \emph{unsatisfied} chain.
\end{definition}

\input{tikz_figures/satisfy-non-head-type-3.tex}
\input{tikz_figures/chain-satisfying-other-chain.tex}

\paragraph{Satisfying the unsatisfied:} Imagine the situation when one of the half-non-solution edge of $\omega_{i-1}$ of the currently growing chain $\cset$ is the same as some edge of $\omega_{k'}$ for some $\psi_{k'} = tail(\cset')$, such that $\psi_{k'}$ is unsatisfied. Now, we can {\em satisfy} $\psi_{k'}$ (and in turn $\cset'$) as well by making each solution-edge of $\psi_{k'}$ a half-edge (see~\Cref{fig:chain-satisfying-other-chain}). Note that, we can do a similar operation of satisfying any type-$3$ triangle (which is not a head so far) if any of its singly-attached triangles overlaps with the half-non-solution edge of the growing chain (see~\Cref{fig:satisfy-non-head-type-3}). Such a type-$3$ triangle can be thought of as as {\em zero-sized satisfied} chain. We satisfy all such triangles in this step, fix credits for their non-solution edges (if any) and add them as {\em zero-sized chains without a tail} in $\Chi$.

If the current chain $\cset$'s tail triangle is satisfied, then we terminate the current iteration for growing $\cset$ and start growing a new chain as before.
Else, we continue growing the current chain $\cset$ as before. If a chain terminates naturally (as before), since we do not find a further lend relation $(\psi_{i+1}, \psi_i)$, then for the $\psi_i$ tail triangle, we only fix the half credit on the gain and the base edges $g_i, b_i$ (recall $g_i \notin E(\psi_i$)). The extra one credit of $\psi_i$ we assign only when it gets satisfied by a newly growing chain or at the very end when the chains growing phase stops.

We partition the set of chain $\Chi =\{\cset_1, \cset_2, \dots, \cset_\l\}$
into $\Chi_{sat}$ (including zero-sized chains) and $\Chi_{unsat}$ corresponding to
chains containing satisfied and unsatisfied tail triangles respectively.
For any subset of chains $\Chi' \subseteq \Chi$, we define $head(\Chi')$ and $tail(\Chi')$ as the set of head and tail triangles of the chains in $\Chi'$ respectively.
We also define $\Chi^3_{sat} \subseteq \Chi_{sat}$ as
the set of zero-sized chains, each containing exactly one satisfied type-$3$ triangle.
As the elements of $\Chi^3_{sat}$ are singleton sets $\{\psi\}$,
so sometimes we will abuse the notations by using
$\Chi^3_{sat} \cup \{\psi\}$ instead of $\Chi^3_{sat} \cup \{\{\psi\}\}$ and
$\psi \in \Chi^3_{sat}$ instead of $\{\psi\} \in \Chi^3_{sat}$.

Building our chains by incorporating these ideas give us the following structural lemma.

\begin{lemma}
\label{lem:chains-half-edge-interactions}
For the set of chains in $\Chi$ constructed using the Satisfy-and-Truncate Discharging-via-Chains construction, before fixing the extra one credit for triangles in $tail(\Chi_{unsat})$, we get the following properties:
\begin{enumerate}
    \item \label{itm:two-chains-no-common-half-edge} For any two chain $\cset', \cset \in \Chi$ no two half-non-solution edges are the same. Applies also for $\cset' = \cset$ for any two half-non-solution edges of singly-attached triangles of different triangles in the chain.
    \item \label{itm:unsat-singly-no-share-before} Any singly-attached triangle $\omega_i$ for $i<k$ of some triangle $\psi_i \in \cset \in \Chi_{unsat}$ cannot contain any half-non-solution edge of any chain in $\Chi$ constructed before (including $\cset$), except its own half-edges $e^1_i$ for $i >0$ and $e^1_0, e^2_0$ for $i = 0$.
    \item \label{itm:unsat-tail-no-share} The singly-attached triangle $\omega_k$ of some tail triangle $\psi_k \in \cset \in \Chi_{unsat}$ cannot contain any half-non-solution edge of any chain in $\Chi$.
    
    \item \label{itm:sat-tail-share} The singly-attached triangle of a tail triangle $\psi_k \in \cset \in \Chi_{sat}$ shares 
    at least one half-non-solution edge of some chain $\cset' \in \Chi$ constructed before $\cset$
    (possibly  $\cset' = \cset$).
    \item \label{itm:unsat-type-3} Any singly-attached triangle of some type-$3$ triangle $\psi \in \Nu_3 \setminus head(\Chi)$ cannot contain any half-non-solution edge of any chain in $\Chi$.

    Note that chains might share null-non-solution edges.
\end{enumerate}
\end{lemma}
\begin{proof}
    This lemma holds directly by our construction.

    For item~\ref{itm:two-chains-no-common-half-edge}, suppose for $\cset$ and $\cset'$, two half-non-solution edges are in common. 
    Let this edge be $e$. 
    WLOG, let $\cset$ be the chain that gets constructed first. 
    Suppose $e$ gets half-credit from $\psi_i \in \cset$ and $\psi_j' \in \cset'$.
    Then it must be that $\cset'$ gets truncated at $\psi_j'$, i.e., $tail(\cset') = \psi_j'$.
    In this case, $e$ cannot get half-credit from $\psi_j'$ by construction.
    
    Using the a similar reasoning as above, if \Cref{itm:unsat-singly-no-share-before} is not true, then $\cset$ would become a satisfied chain with $\psi_i$ as its satisfied tail triangle, which contradicts the $i<k$ condition.

    For \Cref{itm:unsat-tail-no-share}, suppose an $e \in E(\omega_k)$ is a half-non-solution edge of some chain $\cset' \neq \cset$.
    Then $\psi_k$ must be satisfied and $\cset$ should have been in $\Chi_{sat}$.

    If \Cref{itm:sat-tail-share} is not true, then $\psi_k$ will not be satisfied by the construction.

    For \Cref{itm:unsat-type-3}, if $\psi$ share a half-non-solution edge of a chain $\cset \in \Chi$,
    then $\psi$ must have been a zero-sized chain.

\end{proof}

\input{tikz_figures/new-chain-charging.tex}

\paragraph{Using half-edges as a certificate of disjoint alternate solutions:} We get the above lemma by fixing the half-non-solution edges in any arbitrary way.
We will exploit the flexibility given by \Cref{obs:chain-intermediate-non-solution-charge-flexibility}, by fixing them in a special way, we get more structural properties, used later in the proof of our technical claims. 

The following proposition (conditional generalization of~\Cref{cor:small-alternate-solution}) led us to the way we will fix the half credits.

\begin{proposition}[Alternate Solutions for Chains]
\label{prop:chains-alternate-solution}
For any chain $\cset = \{\psi_k, \psi_{k-1}, \dots, \psi_0\}$,
if for every $0 < i \leq k$, the edges $\{e^1_i, e^2_i\}$ are not in $\{e_j^1 : 0 \leq j < i\} \cup \{e_0^2\}$,
then there exists the following
valid swaps leading to N alternate optimal packing solution such that:
\begin{enumerate}
    \item \label{itm:free-semi-rich-base-ij-1} For any $k \geq i> 0$,
    swapping $\{ \psi_i, \psi_{i-1}, \dots, \psi_0 \}$
    with $\{t_i^1, t_{i-1}^1, \dots, t_1^1, \omega_0 \}$
    makes the half base-edge $b_i$ and the gain edge $g_{i+1}$ of $\psi_i$ a non-solution edge.
    \item \label{itm:free-semi-rich-base-ij-2} For any $k \geq i>0$,
    swapping $\{ \psi_i, \psi_{i-1}, \dots, \psi_0 \}$
    with $\{t_i^2, t_{i-1}^1, t_{i-2}^1, \dots, t_1^1, \omega_0 \}$
    makes the half base-edge $b_i$ and the edge $h_i$ of $\psi_i$ a non-solution edge.

\end{enumerate}
\end{proposition}
\begin{proof}
  It is clear that solution edges used in the alternate solution set are disjoint
  (or there exist $i \neq j$ such that $\psi_i = \psi_j$ which contradicts our construction).

  For the non-solution edges, it is clear from the assumption that all the $e^1_i$ edges and  $e^2_i, e^2_0$ edges are distinct, hence both the swaps are valid.
  
  By~\Cref{fact:chain-solution-edges}, since $g_{i+1}, b_i$ are both not contained in $t^1_i$, hence the consequence for the first swap follows.
  
  Also, since $h_i, b_i$ are both not contained in $t^2_i$, hence the consequence for the second swap follows.

\end{proof}

Note that in \Cref{lem:chains-half-edge-interactions}, we ensure that most of the half-non-solution edges are distinct. 

This together with the above proposition makes it a natural choice to fix the half credits on the $e^1_i$ edges and on the edges of $\omega_0$.
Hence, at the $i^{th}$ step of a chain $\cset$ construction,
if $i=1$, we fix the non-solution edges of $\omega_0$ as half-edges. 
Else for $i>1$, we fix $e^1_{i-1}$ non-solution edge of $\omega_{i-1}$ as the half-edge. 
This way of charging gives us a {\em certificate} to locate alternate solution for $\Nu$ using multiple chains at the same time. These alternate solutions will be the central in proving the final technical claims.

\paragraph{Fully-satisfied type-$1$ triangles:} Finally, we describe the last simple but important observation we make to get more structure. Suppose at any $i^{th}$ iteration while growing a chain $\cset$, we get a potential triangle $\psi_i$ have a lend relationship to $\psi_{i-1}$, such that both the non-solution edges of its only singly-attached triangle $\omega_i$ are already half-edges due to some previously fixed chains structure. In this case, $\psi_i$ no longer needs to {\em lend} credits to $gain(\psi_i, \psi_{i-1})$ in order to cover the twin-doubly attached triangles adjacent to $\omega_i$, as they are already covered. Hence, we declare such a $\psi_i$ to be a {\em fully-satisfied} triangle and do not add it to $\cset$. 
In the actual construction, we will try to find a potential $\psi_i$ that is not fully-satisfied first (so that we can extend the chain). If all potential $\psi_i$'s are fully-satisfied, then we terminate the iteration for growing the chain $\cset$ and set $\psi_{i-1}$ to be the tail of $\cset$.
We formally define such triangles below.

\begin{definition}[Fully-satisfied triangles]
\label{def:fully-satisfied} A type-$1$ triangle $\psi \in \Nu_1$ which has exactly one unique singly-attached triangle $\omega$ is fully-satisfied if and only if:
\begin{enumerate}
    \item $\psi$ is satisfied.
    \item The base-edge $base(\psi)$ is a full-edge.
    \item Both the non-solution edge of its unique singly-attached triangle $\omega$ are half-edges.
\end{enumerate}
\end{definition}

\paragraph{The chains construction algorithm:} Here we formally describe the full algorithm for the Satisfy-and-Truncate Discharging-via-Chains construction which we motivate and develop above. We repeat some notations definitions and steps to make this algorithm self-sufficient.\\

\noindent \textbf{\em Initialization:} We start by the initial charging distribution as described in \Cref{subsec:initial-half-charge}.

Note that all the type-$1$ triangles are satisfied and all the type-$3$ triangles are unsatisfied after the initial charge distribution. Let the set of satisfied and unsatisfied chains are $\Chi_{sat}$ and $\Chi_{unsat}$ respectively. We also maintain $\Chi^3_{sat} \subseteq \Chi_{sat}$ the set of zero-sized tailless chains containing only one satisfied type-$3$ triangle each. Since they do not have a tail, they are vacuously satisfied.

In the beginning let $\Chi_{sat}, \Chi_{unsat}, \Chi^3_{sat} = \varnothing$. 

\noindent \textbf{\em Chain growing phase:} We iteratively build chains one at a time and add them to the relevant sets. In the process of building any chain,
we will be moving chains which gets satisfied from $\Chi_{unsat}$ to $\Chi_{sat}$ and
add any type-$3$ getting satisfied to $\Chi^3_{sat}$ and $\Chi_{sat}$. We never remove any fixed credit, hence any structure which is covered and/or satisfied remains so till the end of the process. At any moment of the algorithm, we use $\Chi$ to refer to the current set of chains $\Chi_{sat} \cup \Chi_{unsat}$ and we maintain the relation $\Chi^3_{sat} \subseteq \Chi_{sat}$. 

We start growing a new chain with $\cset_j= \varnothing$ and $i=1$. Each {\em iteration} corresponds to growth of such a chain. Each {\em step} $i= 1, 2,\dots $ corresponds to growth of the current chain $\cset$ resulting in fixing some credit edges and adding a triangle $\psi_i$ to $\cset$ while satisfying other triangles (including itself) whenever possible. 

We keep on growing new chains, until we cannot find any valid type-$[1,3]$ triangle to begin with.
\begin{itemize}
    \item \noindent \textbf{\em Beginning of a chain ($i=1$):} We start by identifying a type-$[1,3]$ triangle adjacent to solution triangles $\psi_1 \in \Nu_1$ and $\psi_0 \in \Nu_3$ such that $(\psi_1, \psi_0) \in lend$,  $\psi_3$ is not satisfied and \textbf{$f(e) = 0$ for at least one non-solution edge $e$ of $\omega_1$, the singly-attached triangle of $\psi_1$}, in other words $\psi_i$ is not fully-satisfied. Initiate the chain $\cset = \{\psi_1, \psi_0\}$ as the currently growing chain. Let $head(\cset) := \psi_0$.

    \noindent \textbf{\em Fixing credits:} Let $g_1: = gain(\psi_1, \psi_0)$,
    $c_1 := V(\psi_1 \cap \psi_0)$, $a_0 := anchor(\psi_0)$,
    $a_1 := anchor(\psi_1)$ and $b_1 := base(\psi_1)$.
    Let $e^1_0, e^2_0$ be the non-solution edges in $K_4$ induced by $V(\psi_0) \cup \{a_0\}$ not incident to $c_1$.
    Let $\omega_1$ be the singly-attached triangle of $\psi_1$ and $\omega_0^1,\omega_0^2,\omega_0^3=\omega_0$ be the singly-attached triangles of $\psi_0$, such that $e_0^1, e_0^2 \in \omega_0$ (see~\Cref{fig:new-chain-charging}).
    Fix the credits such that $f(b_1) = f(e^1_0) = f(e^2_0) = \frac{1}{2}$ and all the edges of $\psi_0$ are half-edges, making $\psi_0$ a satisfied triangle. We keep the extra one credit with $\psi_1$ which is fixed in the next step. Hence $\psi_1$ is unsatisfied at this moment.

    \item \noindent\textbf{\em Grow-or-Terminate ($i>1$):}
    At the beginning of this step the currently growing chain is
    $\cset = \{\psi_{i-1}, \psi_{i-2}, \dots, \psi_0\}$.
    Find some $\psi_i \in \Nu_1$, such that $(\psi_i, \psi_{i-1}) \in lend$,
    $gain(\psi_i, \psi_{i-1}) \neq b_{i-1}$,
    and \textbf{$f(e) = 0$ for at least one non-solution edge $e$ of $\omega_i$, the singly-attached triangle of $\psi_i$}, in other words $\psi_i$ is not fully-satisfied.
    Set $\cset \leftarrow \{\psi_i\} \cup \cset$.
    If no such $\psi_i$ exists, then we terminate this iteration by
    adding the chain $\cset = \{\psi_{i-1}, \psi_{i-2}, \dots, \psi_0\}$
    to $\Chi_{unsat}$ and set $tail(\cset):=\psi_{i-1}$.
    Note that the extra one credit of $tail(\cset)$ is not fixed yet.
    Now start growing a new chain.

    \noindent\textbf{\em Fixing credits:}
    Let $g_i: = gain(\psi_i, \psi_{i-1})$, $c_i := V(\psi_i \cap \psi_{i-1})$,
    $a_i := anchor(\psi_i)$ and $b_i := base(\psi_i)$.
    Let $h_{i-1} := E(\psi_{i-1}) \setminus \{b_{i-1}, g_i\}$.
    Let $e^1_{i-1}, e^2_{i-1}$ be the non-solution edges of $\omega_{i-1}$
    such that $e^1_{i-1}$ is adjacent to $h_{i-1}$ and $e^2_{i-1}$ is adjacent to $g_i$.
    Let $\omega_i$ be the singly-attached triangle of $\psi_i$.
    Fix the credits such that $f(b_i) = f(h_{i-1}) = f(g_i) = f(e^1_{i-1}) = \frac{1}{2}$ (see~\Cref{fig:new-chain-charging}).
    Note that in this step we fix the one credit of $\psi_{i-1}$ and keep the extra one credit with $\psi_i$ which is fixed in the next step.
    Hence both non-base edges of $\psi_i$ are null-edges and $\psi_i$ is unsatisfied at this moment.

    \item \noindent \textbf{\em Satisfying chains in $\Chi_{unsat}$:} If any of the half-edges fixed in the current step belongs to the singly-attached triangle $\omega'_k$ of the tail triangle $\psi'_k = tail(\cset')$ for any chain $\cset' \in \Chi_{unsat}$, then satisfy $\psi'_k$ by assigning its one unused credit to the two edges in $E(\psi'_k) \setminus base(\psi'_k)$ and move $\cset'$ from $\Chi_{unsat}$ to $\Chi_{sat}$ (see~\Cref{fig:chain-satisfying-other-chain}). Note that $\omega'_k$ will be covered by this charging.

    \item \noindent \textbf{\em Satisfying unsatisfied triangles in $\Nu_3 \setminus head(\Chi)$:} If any of the half-edges fixed (say $e$) in the current step belongs any singly-attached triangle of any type-$3$ triangle $\psi^3 \in \Nu_3$ which is unsatisfied, then we satisfy it by assigning its $1.5$ credits to all three solution triangles and $1/2$ credit to any non-solution edge other than $e$ (see~\Cref{fig:satisfy-non-head-type-3}).
    Add $\{\psi^3\}$ to $\Chi^3_{sat}$ and $\Chi_{sat}$, set $head(\{\psi^3\}) = \psi_3$ and $tail(\{\psi^3\}) = \varnothing$.
    Note that all triangles in $G[V(\psi^3) \cup anchor(\psi^3)]$ are covered by this charging.

    \item \noindent \textbf{\em Satisfy-and-Truncate:} If $\omega_i$ (singly-attached triangle of $\psi_i$) contains a half-non-solution edge, then {\em satisfy} $\psi_i$ by using its one unused credit and fixing $f(e) = \frac{1}{2}$ for both edges in $E(\psi_i) \setminus \{b_i\}$ (see~\Cref{fig:chain-satisfying-other-chain}). In this case, we terminate this iteration by adding the chain $\cset = \{\psi_i, \psi_{i-1}, \dots, \psi_0\}$ to $\Chi_{sat}$ and set $tail(\cset):=\psi_i$. Now start growing a new chain. Note that $\omega_i$ will be covered by this charging.

Else, continue growing the chain $\cset = \{\psi_i, \psi_{i-1}, \dots, \psi_0\}$ by going to the $i+1$ step.
\end{itemize}

At the end of this phase of the algorithm, we get the following fact (analogous to~\Cref{fact:no-lending-to-tail-property}).

\begin{fact}
\label{fact:only-fully-satisfied-lending-to-tail-property} For any $k$-sized unsatisfied chain $\cset = \{\psi_k, \psi_{k-1}, \dots, \psi_0\} \in \Chi_{unsat}$, 
if there exist $\psi \in \Nu_1$ such that $(\psi, \psi_k) \in lend$ and $gain(\psi, \psi_k) \neq base(\psi_k)$, 
then $\psi$ is fully-satisfied.
This implies that any arbitrary $K_4$ half-integral charging for $V(\psi_k) \cup \{a_k\}$ such that $f(gain(\psi_k, \psi_{k-1})) =\frac{1}{2}$ for the tail triangle covers all the twin doubly-attached triangles adjacent to it. 

Similarly, any $\psi \in \Nu_1$ such that $\psi^3 \in \Nu_3 \setminus head(\Chi)$ and  $(\psi, \psi^3) \in lend$, is fully-satisfied.
\end{fact}

\noindent \textbf{\em Assigning the unused credits of $tail(\Chi_{unsat})$ triangles:}
Using the above fact, in this final phase of the algorithm, we assign the unused one credit of $tail(\Chi_{unsat})$ triangles as we do in the vanilla construction.
For any $\psi_k \in tail(\Chi_{unsat})$ which is a tail unsatisfied triangle of a $k$-sized chain, we assign half credit to a non-base edge of $\psi$ and call it $h_k$
and another half credit to the non-solution edge of $\omega_k$ which is not adjacent to $h_k$ and call it $e^2_k$.
Then we name the edge in $E(\psi_k) \setminus \{b_k, h_k\}$ as $g_{k+1}$
and the other non-solution edge of $\omega_k$ as $e^1_k$ (see~\Cref{fig:chain-structure}).

\paragraph{Chains charge distribution:} As~\Cref{lem:chains-half-edge-interactions} applies to the above algorithm, the following fact holds for the charging fixed by the algorithm (see \Cref{subfig:new-chain-charging,fig:chain-satisfying-other-chain,fig:satisfy-itself,fig:satisfy-non-head-type-3} for illustration). \Cref{def:half-edge-of} does not say anything about credit on the half-edge $e$ of a chain $\cset$.
But the following observation captures the exact distribution of the credit on these edges.

\begin{observation}
\label{obs:half-edge-of}
Let $e$ be a half-edge of a chain $\cset$ as in \Cref{def:half-edge-of}. By our construction,  $f(e) = \frac 1 2$ and $e$ is not a half-edge of any other chain $\cset' \neq \cset$.
\end{observation}

We also get some more properties for containment (in addition to \Cref{lem:chains-half-edge-interactions}) of the half-solution edges of any chain $\cset \in \Chi$ in some other chain $\cset' \in \Chi$ based on whether $\cset$ was build before or after $\cset'$.
\begin{lemma}
\label{lem:ordered-chains-half-edge-interactions}  
In addition to properties stated in \Cref{lem:chains-half-edge-interactions},
the following properties holds for chains in $\Chi$.
\begin{enumerate}
    \item \label{itm:unsat-half-edge-after} For any $\cset \in \Chi_{unsat}$, then it can only contain half-edge for any other chain $\cset' \in \Chi$ if $\cset'$ is built after $\cset$. Moreover, the tail triangle's singly-attached triangles contains none such edge.
    \item  \label{itm:sat-half-edge-after} For any $\cset \in \Chi_{sat}$, then apart from exactly one half-edge contained in the tail triangle $\omega_k$ mentioned in \Cref{lem:chains-half-edge-interactions}.\ref{itm:sat-tail-share}, it can only contain half-edge for any other chain $\cset' \in \Chi$ if $\cset'$ is built after $\cset$.
\end{enumerate}
\end{lemma}
\begin{proof}
    If \Cref{itm:unsat-half-edge-after} is not true, then $\cset$ would contain a half-edge of some already built chain (including  $\cset$), which mean that $\cset$ should have been satisfied. This contradicts the assumption that $\cset$ is unsatisfied. The moreover part also follows from the same contradiction.
    
    If \Cref{itm:sat-half-edge-after} is not true,
    let $\cset \in \Chi_{sat}$ be a $k$-sized chain that does not go along with \Cref{itm:sat-half-edge-after}.
    One possibility is that there exists a singly attached triangle $\omega_{i<k}$ which contains a half-edge of some already built chain (including $\cset$). This means we would have satisfied $\psi_i$ and truncated the chain $\cset$ there, contradiction the fact that $\cset$ is a $k$-sized chain.
    Another possibility is that $\omega_k$ contains two half non-solution edges for some already built chains (including $\cset$). This means $\psi_k$ would have been fully-satisfied and shouldn't be added into $\cset$ during the construction. This is contradicting the condition for choosing $\psi_k$ to add to the chain $\cset$.
    
\end{proof}

All in all, the following fact encapsulates the precise charge distribution for the chains in $\Chi$.

\begin{fact}
\label{fact:chains-edges-credits} We get the following charge distribution for any chain $k$-sized chain $\cset \in \Chi$:
\begin{enumerate}
    \item \label{itm:half-edges-head} $e^1_0, e^2_0$ are half-edges of exactly one chain $\cset$.
    \item \label{itm:half-edges-intermediate} Every non-solution edge $e^1_i$ for $0 <i < k$ is a half-edge of exactly one chain $\cset$.
    \item \label{itm:atleast-half-edges-tail-unsat} If $\cset \in \Chi_{unsat}$ then $e^2_k$ is at least a half-edge.
    \item \label{itm:atleast-half-edges-tail-sat} If $\cset \in \Chi_{sat}$ then one of the $e^1_k, e^2_k$ edges is a half-edge of some chain $\cset' \in \Chi$ (possibly $\cset = \cset'$).
    \item \label{itm:solution-half-edges-unsat} If $\cset \in \Chi_{unsat}$ then every solution edges except $g_{k+1}$ is a half-edge.
    \item \label{itm:solution-half-edges-sat} If $\cset \in \Chi_{sat}$ then all the solution edges are half-edges.
\end{enumerate}
\end{fact}

The following fact follows from the above fact. 
\begin{fact}
\label{fact:coverage-intermediate-twins-chains}
 For any $k$-sized chain $\cset \in \Chi$ and for $0 < i < k$ all the twin-doubly attached triangles $t^1_i, t^2_i$ are covered by half-edges $g_{i+1}, g_i, h_i$.
\end{fact}
In addition, we get the following observation which follows from \Cref{fact:only-fully-satisfied-lending-to-tail-property} for the chains in $\Chi_{unsat}$.
\begin{observation}
\label{obs:unsat-tail-type-1-twins}
For any $k$-sized chain $\cset \in \Chi_{unsat}$ we can rename $h_k, g_{k+1}$ arbitrarily,
fixing the names $e^1_k, e^2_k$, such that $h_k, e^2_k, b_k, g_k$
are the half-edges and $g_{k+1}, e^1_k$ are the null-edges in $G[V(\psi_k) \cup \{a_k\}]$,
while maintaining the coverage of $\psi_k$, $\omega_k$ and
the twin-doubly attached triangles $t^1_k, t^2_k$.
\end{observation}
For chains in $\Chi_{sat}$ we get the following.
\begin{observation}
\label{obs:sat-tail-type-1-twins} For any $k$-sized chain $\cset \in \Chi_{sat}$ the the twin-doubly attached triangles $t^1_k, t^2_k$ are covered by half-edges $g_{k+1}, g_k, h_k$.
\end{observation}

Combining \Cref{fact:chains-edges-credits} with \Cref{prop:chains-alternate-solution}, we can state the following two corollaries for our new construction.

\begin{corollary}
\label{cor:unsat-chains-alternate-solution}
For any chain $\cset = \{\psi_k, \psi_{k-1}, \ldots, \psi_0\} \in \Chi_{unsat}$, there exists the following valid swaps leading to an alternate optimal packing solution $\Nu'$ such that:
\begin{enumerate}
    \item \label{itm:unsat-swap-t1} For any $k \geq i> 0$,
    swapping $\{ \psi_i, \psi_{i-1}, \dots, \psi_0 \}$
    with $\{t_i^1, t_{i-1}^1, \dots, t^1_1, \omega_0 \}$
    makes the half base-edge $b_i$ and the gain edge $g_{i+1}$ of $\psi_i$ a non-solution edge. 
    Moreover, the non-solution edge $e^2_i$ of $\omega_i$ adjacent to $g_{i+1}$ remains a non-solution edge in $\Nu'$.
    
    \item \label{itm:unsat-swap-t2} For any $k \geq i>0$,
    swapping $\{ \psi_i, \psi_{i-1}, \dots, \psi_0 \}$
    with $\{t_i^2, t_{i-1}^1, t_{i-2}^1, \dots, t^1_1, \omega_0 \}$
    makes the half base-edge $b_i$ and the edge $h_i$ of $\psi_i$ a non-solution edge. 
    Moreover, the non-solution edge $e^1_i$ of $\omega_i$ adjacent to $h_i$ remains a non-solution edge in $\Nu'$.
\end{enumerate}
\end{corollary}

\begin{proof}
  To prove this, we just need to ensure that our construction gives us the assumption
  needed by \Cref{prop:chains-alternate-solution}.
  In the other word, we have to show that for any $i$,
  $\{e_i^1, e_i^2\}$ are not in $\{e_j^1 : j < i\} \cup \{e_0^2\}$.
  Suppose not, let $e_i$ be an edge in $\{e_i^1, e_i^2\} \cap (\{e_j^1 : j < i\} \cup \{e_0^2\})$.
  In the $i^{th}$ step, $\psi_i$ must be satisfied and truncated, so $\cset$ should not be in $\Chi_{unsat}$ in the first place.
 
    The moreover parts are also true from our construction. We show the proof of this part for \Cref{itm:unsat-swap-t1} and
    the moreover part of \Cref{itm:unsat-swap-t2} follows analogously.
    If $e_i^2$ becomes a solution edge in $\Nu'$, then it must be that $e_i^2$ is a half-non-solution edge of $\cset$, i.e., either $e_i^2 \in E(t_j^1)$ for $0 < j < i$ or $e_i^2 \in E(\omega_0)$, which cannot happen by \Cref{lem:chains-half-edge-interactions}.\ref{itm:unsat-singly-no-share-before}.
\end{proof}

\begin{corollary}
\label{cor:sat-chains-alternate-solution}
For any chain $\cset = \{\psi_k, \psi_{k-1}, \ldots, \psi_0\} \in \Chi_{sat}$, there exists the following valid swaps leading to an alternate optimal packing solution $\Nu'$ such that:
\begin{enumerate}
    \item \label{itm:sat-swap-t1} For any $k > i> 0$,
    swapping $\{ \psi_i, \psi_{i-1}, \dots, \psi_0 \}$
    with $\{t_i^1, t_{i-1}^1, \dots, t^1_1, \omega_0 \}$
    makes the half base-edge $b_i$ and the gain edge $g_{i+1}$ of $\psi_i$ a non-solution edge. 
    Moreover, the non-solution edge $e_i^2$ of $\omega_i$ adjacent to $g_{i+1}$ remains a non-solution edge in $\nu'$.
    \item \label{itm:sat-swap-t2} For any $k > i>0$,
    swapping $\{ \psi_i, \psi_{i-1}, \dots, \psi_0 \}$
    with $\{t_i^2, t_{i-1}^1, t_{i-2}^1, \dots, t^1_1, \omega_0 \}$
    makes the half base-edge $b_i$ and the edge $h_i$ of $\psi_i$ a non-solution edge. 
    Moreover, the non-solution edge $e_i^1$ of $\omega_i$ adjacent to $h_i$ remains a non-solution edge in $\nu'$.
    \item \label{itm:sat-swap-tail}
    swapping $\{ \psi_k, \psi_{k-1}, \dots, \psi_0 \}$
    with at least one of the sets
    $\{t_k^1, t_{k-1}^1, \dots, t^1_1, \omega_0 \}$ or \newline
    $\{t_k^2, t_{k-1}^1, t_{k-2}^1, \dots, t^1_1, \omega_0 \}$ 
    (using the one out of $t_k^1, t_k^2$ which does not contains a half-edge of chains built before $\cset$ (including $\cset$ itself))
    
    is valid and
    makes the half base-edge $b_k$ of $\psi_k$ a non-solution edge.
\end{enumerate}
\end{corollary}
\begin{proof}
    The proof for the first two items follow directly from the same proof of \Cref{cor:sat-chains-alternate-solution}. 

    For the third item, we can only ensure that at most one of the swaps is valid, as we cannot determine which non-solution edge of $\omega_k$ has credit, and where does it come from. But by our construction, at least one of the swaps must work, since at the time when we add $\psi_k$ to $\cset$, exactly one of the non-solution edges of $\omega_k$ is a half-edge, which in the worst case may belong to the set $\{e^1_i\}_{i=0}^{k-1} \cup \{e^2_0\}$. Hence, at least one of the non-solution edges of either $t^1_k$ or $t^2_k$ should be disjoint from the set $\{e^1_i\}_{i=0}^{k-1} \cup \{e^2_0\}$ of half-non-solution edges of $C$.  
\end{proof}

\paragraph{Triangle types with respect to packing solution} So far,
we have been using a fixed packing solution $\Nu$. In the subsequent parts of this paper, we will need to argue about alternate packing solutions to
show some structural properties.
To get an alternate packing solution $\Nu'$, we will identify a set of disjoint solution triangles $S$ and another set of disjoint triangles $S'$ of the same cardinality.
$\Nu'$ is obtained by replacing $S$ with $S'$ in $\Nu$.
Note that the type of some triangle $\psi$ might gets changed as a result.
When it is not clear from the context, we use $type_{\Nu'}(\psi)$ to refer to the type of $\psi$ in the packing solution $\Nu'$.
Similarly, we say that $\psi$ is $\text{type}_{\Nu'}\text{-}i$ if $type_{\Nu'}(\psi) = i$. We will also use this notation for doubly-attached triangles and hollow triangles.
The subscript will be omitted (as before) when the packing solution is the initial packing solution.

\begin{observation}
\label{obs:make-tail-type-3}
    Let $\Chi' = \{\cset_1, \cset_2, \ldots, \cset_\l\} \subseteq \Chi_{unsat}$ be a subset of unsatisfied chains.
    Applying \Cref{cor:unsat-chains-alternate-solution}.\ref{itm:unsat-swap-t1} with $i = k-1$ to each chain $\cset \in \Chi'$ gives us an alternate packing $\Nu'$ such that:
    \begin{enumerate}
        \item For any $\cset \in \Chi'$, $type_{\Nu'}(tail(\cset)) = 3$.
        \item For any $\psi^3 \in \Nu_3 \setminus head(\Chi)$, $type_{\Nu'}(\psi^3) = 3$.
    \end{enumerate}

\end{observation}

\begin{proof}
    
  We first show that these swaps can be applied simultaneously. Since each solution triangle belongs to exactly one chain, the set of solution edges used for different chains will be disjoint. If any non-solution edge $e$ is contained in two triangles which appear in swaps corresponding to two different chains $\cset$ and $\cset'$, then $e$ must be a half-edge of both $\cset$ and $\cset'$ which contradicts \Cref{obs:half-edge-of}.

  We show that $type_{\Nu'}(tail(\cset)) = 3$ for each $\cset \in \Chi'$.
  This is true as the gain-edge $g_k$ of each chain becomes a non-solution edge by the swap and by ~\Cref{lem:chains-half-edge-interactions}.\ref{itm:unsat-tail-no-share}
  no non-solution edges of $\omega_k$ of a tail triangle is used in the swaps.

  Also each unsatisfied type-$3$ triangle $\psi^3 \in \Nu_3 \setminus head(\Chi)$ remained type-$3$ in $\Nu'$. This follows from  ~\Cref{lem:chains-half-edge-interactions}.\ref{itm:unsat-type-3} which implies that no non-solution edges of any singly-attached triangle of $\psi^3$ is used in any of the swaps.
  \end{proof}

\paragraph{Proving technical claims while maintaining the flexibility for credits:} Now we are ready to prove our technical claims.
We maintain the {\em flexibility} (similar to \Cref{obs:type-1-rotation,obs:type-3-rotation}) for credits of the type -$1$ tail triangles corresponding to chains in
$\Chi_{unsat}$ and the non-head type-$3$ triangles in $\Nu_3 \setminus head(\Chi)$. More precisely, we will get the following modified versions.

\begin{observation}
\label{obs:unsat-type-1-rotation}
\Cref{lem:main-type-1-3} remains true for any arbitrary but fixed $K_4$ half-integral charging for $G[V(\psi_k) \cup anchor(\psi_k)]$, where $\psi_k$ is the tail triangle of any $k$-sized chain $\cset \in \Chi_{unsat}$ such that $f(base(\psi_k)) = f(gain(\psi_k, \psi_{k-1}))=\frac{1}{2}$.
\end{observation}

\begin{observation}
\label{obs:unsat-type-3-rotation}
\Cref{lem:main-type-1-3} remains true for any arbitrary but fixed $K_4$ half-integral charging for $G[V(\psi^3) \cup anchor(\psi^3)]$, where $\psi^3$ is any type-$3$ triangle which is not part of any chain, i.e. in $\Nu_3 \setminus head(\Chi)$.
\end{observation}

Any other triangle does not need this flexibility, since they are satisfied. These properties will be crucially used to cover all the triangles in the presence of type-$0$ triangles.
For the non-head triangles in $\Nu_3$, we tentatively fix the $K_4$ half-integral charging which covers itself and all its singly-attached triangles. Apart from that we do not use these credits to cover any other triangle.
Similarly, for any tail triangle $\psi_k \in \cset \in \Chi_{unsat}$, we tentatively fix the $K_4$ half-integral charging such that its base edge $b_k$ and the gain edge $g_k = gain (\psi_k, \psi_{k-1})$ are half-edges and covers itself, singly-attached triangle $\omega_k$ and twin-doubly-attached triangles $t^1_k, t^2_k$. Apart from that, we allow to use only the credits for $b_k, g_k$ (which will remain fixed) to cover any other triangle in $G$.

This implies that no matter how we fix the flexible credits (as in \Cref{obs:unsat-type-1-rotation,obs:unsat-type-3-rotation}) for these,
the charging still covers all the triangles in any instance without type-$0$ triangles.
This flexibility allows us to {\em freely} assign the two credits of non-head type-$3$ triangles respecting $K_4$ half-integral charging in the three possible ways.
For unsatisfied tail triangles, it allows to {\em freely} rotate the non-base solution edge and non-solution edge credits in two possible ways.
Later on in the general instance, we crucially use this flexibility to cover any general instance.

To capture this, we define the flexible and fixed credits.
\begin{definition}[Flexible and fixed credits]
\label{def:flexible-fixed-credits}
With respect to the current charge function \{$f(e)\}_{e \in E}$ and chains $\Chi$, we define a {\em fixed} charge function $f_{fix}: E \rightarrow \mathbb{R}^+$ based on whether the credit for any edge $e \in E$ is {\em flexible} or {\em fixed}. For any edge $e \in E$, we say that its credit is {\em flexible} if:
\begin{enumerate}
    \item either $e$ belongs to $G[V(\psi_k) \cup anchor(\psi_k)] \setminus \{b_k, g_k\}$, where $\psi_k$ is the tail triangle of any $k$-sized chain $\cset \in \Chi_{unsat}$ structure.
    \item or $e \in G[V(\psi^3) \cup anchor(\psi^3)]$, where $\psi^3$ is any type-$3$ triangle in $\Nu_3 \setminus head(\Chi)$.
\end{enumerate}
 If $e$ falls in any of the above cases, we set $f_{fix}(e) = 0$.
Otherwise, we say that credit of $e$ is {\em fixed} and set $f_{fix}(e) = f(e)$.
\end{definition}

We get the following fact for the above defined charge function $f_{fix}$.
\begin{fact}
\label{fact:f_fix}
For the fixed charge function $f_{fix}$ defined for edges of $G$, the following properties hold:
\begin{enumerate}
    \item \label{itm:f_fix-half-edges-of} For any edge $e$ which is a half edge any chain in $\Chi$, $f_{fix}(e) = \frac 1 2$.
    \item \label{itm:type-1-not-in-chain} For any edge $e \in E(\psi)$, where $\psi \in \Nu_1 \setminus tail(\Chi_{unsat})$, $f_{fix}(e) = f(e)$.
\end{enumerate}
\end{fact}
The next observation follows from the fact that  $f_{fix}(h_k)=f_{fix}(g_{k+1})=f_{fix}(e^1_k)=f_{fix}(e^2_k)=0$ for any $k$-sized chain $\cset \in \Chi_{unsat}$.

\begin{observation}
\label{obs:tail-renaming}
For any $k$-sized chain $\cset \in \Chi_{unsat}$, we can rename $h_k$ and $g_{k+1}$ (and accordingly $e^1_k$ and $e^2_k$) without  changing the chains structure in $\Chi$ and the charge function $f_{fix}$.
\end{observation}

Like before, all the solution triangles and singly attached triangles are covered after the chains construction. Some type-$[1,*]$ or type-$[1,*,*]$ triangles adjacent to null edges of tail triangles or unsatisfied type-$3$ triangles may not be covered. This together with \Cref{obs:unsat-tail-type-1-twins,obs:sat-tail-type-1-twins} and assuming \Cref{clm:excluding-type-11,clm:excluding-type-13,clm:excluding-type-111,clm:excluding-type-113,clm:excluding-type-133}, we get \Cref{lem:main-type-1-3}. Moreover, \Cref{obs:unsat-type-1-rotation,obs:unsat-type-3-rotation} holds by the way we define $f_{fix}$.

Finally, we will prove our technical claims by contradiction by every time starting with some triangle $t$ which is not covered by the charge function $f_{fix}$.

\begin{claim}
\label{clm:excluding-type-11}
All the type-$[1, 1]$ triangles in $G$, except for the twin-doubly attached triangles covered in \Cref{obs:unsat-tail-type-1-twins}, are covered by charge function $f_{fix}$.
\end{claim}
\begin{proof}
Let $t$ be a type-$[1, 1]$ triangle in $G$ which is considered in the claim and is not covered by $f_{fix}$. This implies that $t$ should contain at least one null non-base edge (by \Cref{obs:tail-renaming} assume this edge is $g_{k+1}$) with respect to the charge function $f_{fix}$ of some chain $\cset = \{\psi_k, \psi_{k-1}, \dots, \psi_0\} \in \Chi_{unsat}$'s tail type-$1$ triangle $\psi_k$. We can make this assumption as because all the other solution edges of type-$1$ triangles are at least a half-edge w.r.t to the charge function $f_{fix}$. The second solution edge could be either some other non-base edge or a half base edge of a type-$1$ triangle which takes part in some chain (possibly in $\cset$). Since $t$ contains a null-edge of $\psi_k$, it cannot be the case that $t$ belongs to \Cref{obs:sat-tail-type-1-twins}.

Let $\psi$ be the other type-$1$ triangle adjacent to $t$ and $c = V(\psi) \cap V(\psi_k)$ be the common vertex. Let $e$ be the non-solution edge in $t$.

First we deal with the case when $t$ is adjacent to non-base edge of $\psi$.
If $t$ belongs to \Cref{prop:type-[1-1]}.\ref{itm:anchoring-vertex},  it implies either $(\psi, \psi_k) \in lend$ or $(\psi_k, \psi) \in lend$.
But then $t$ is covered, in former case by \Cref{fact:only-fully-satisfied-lending-to-tail-property}.
In the latter case when $(\psi_k, \psi=\psi_{k-1}) \in lend$, $t$
belongs to \Cref{obs:unsat-tail-type-1-twins},
hence $t$ is excluded from this claim.
Note that since $\cset \in \Chi_{unsat}$, $t$ cannot be in \Cref{obs:sat-tail-type-1-twins} in this case.

\input{tikz_figures/uncovered-type-11-paired}

Otherwise $t$ belongs to \Cref{prop:type-[1-1]}.\ref{itm:pair}, where singly-attached triangles $\omega, \omega_k$ of $\psi, \psi_k$ respectively share a non-solution edge and the only anchoring vertices $a:= anchor(\psi)$ and $a_k$ are the same. Note that, $t$ is edge-disjoint from both $\omega$ and $\omega_k$ (see \Cref{fig:uncovered-type-11-paired}).

Clearly $f_{fix}(t)$ can be $0$ or $\frac 1 2$. We will first rule out the case where $f_{fix}(t) = 0$.
Suppose $f_{fix}(t) = 0$. Then $\psi = \psi'_{k'} =  tail(\cset')$ for some $\cset' = \{\psi'_{k'}, \psi'_{k'-1}, \ldots, \psi'_{0}\} \in \Chi_{unsat}$.
In this case, we apply
\Cref{cor:unsat-chains-alternate-solution}.\ref{itm:unsat-swap-t1} to both $\cset, \cset'$ using $i=k,k'$ respectively to get alternate packing solution $\Nu'$. Note that by \Cref{obs:half-edge-of}, $\Nu'$ is a valid packing solution, such that $|\Nu'|=|\Nu|$. Also, any triangle in $\Nu'$ does not contain $e$ as $e$ has $0$ credit, hence $e$ remains a non-solution edge with respect to $\Nu'$. Altogether, all edges of $t$ become non-solution in $\Nu'$, so we can add $t$ to our packing solution $\Nu'$, which contradicts the optimality of $\Nu$.

In the other case when $f_{fix}(t) = \frac 1 2$, there are two sub-cases, one when $f_{fix}(e) = \frac 1 2$ and $f_{fix}(e') = 0$; and another when $f_{fix}(e) = 0$ and $f_{fix}(e') = 0$, where $e' = E(\psi \cap t)$.

In the first sub-case when $f_{fix}(e) = \frac 1 2$, $\psi$ is an unsatisfied tail triangle such that $f_{fix}(e') = 0$. Again let, $\psi = \psi'_{k'} =  tail(\cset')$ for $\cset' = \{\psi'_{k'}, \psi'_{k'-1}, \ldots, \psi'_{0}\} \in \Chi_{unsat}$.
By the argument used for $f_{fix}(t) = 0$ case, $e$ must be a half-edge of either $\cset$ or $\cset'$ (not both by \Cref{obs:half-edge-of}).
By renaming, suppose $e$ is a half-edge of $\cset'$.
Then we apply 
\Cref{cor:unsat-chains-alternate-solution}.\ref{itm:unsat-swap-t1} with $i=k$ to $\psi_k$ to get an alternate packing solution $\Nu'$, such that $g_{k+1}$ becomes a non-solution edge and the common non-solution edge of $\omega, \omega_k$ remains a non-solution edge.
By \Cref{lem:chains-half-edge-interactions}.\ref{itm:unsat-tail-no-share}, $\omega$ cannot contain any other non-solution edge (which are precisely set of half-edges of $\cset$) used by \Cref{cor:unsat-chains-alternate-solution}.\ref{itm:unsat-swap-t1}.
Hence we can replace $\psi$ with $\omega$ and $t$ to get a larger packing solution, which contradicts the optimality of $\Nu$.

In the second sub-case we have $f_{fix}(e)=0$ and $f_{fix}(e') = \frac 1 2$.
If the edge $\hat e = \overbar{va}$ (where $v = V(\psi) \setminus V(t)$) is not in $t_i^1$ for some $i>0$ and not in $\omega_0$ of chain $\cset$, i.e. not one of the half-solution edges of $\cset$, then by the same reasoning as above we arrive at a contradiction.

Hence $\hat e $ is one of the half-non-solution edges of $\cset$.

In the special case when $\psi = \psi_i$ for some $i>0$ in the same chain $\cset$, once we 
apply
\Cref{cor:unsat-chains-alternate-solution}.\ref{itm:unsat-swap-t1} to get $\Nu'$, we claim that $t$ would become available to be added to $\Nu'$, contradicting optimality of $\Nu$. The claim is true since $\hat e = e^1_i$, hence $t^1_i$ will not contain $e'$, which implies $e'$ will become a non-solution edge. Moreover, the other two edges of $t$ will be non-solution edges by the same argument as in the last case.

Now we deal with the case when $\hat e$ is a half-edge of $\cset$ and $\psi$ is not in $\cset$.

Consider $\psi_{k-1}$ and $\omega_{k-1}$. Since $c_k, a \in V(\psi_{k-1})$ (as $g_k = \overbar{c_k a}$) hence they dominate vertices in $V(\psi) \cup V(\psi_k) \cup \{a\}$, which implies the third vertex $v_{k-1}$ should be a new vertex (see \Cref{fig:uncovered-type-11-paired}). As $\hat e$ is adjacent to $g_k$ it cannot be $e^1_{k-1}$ which together with  \Cref{lem:chains-half-edge-interactions}.\ref{itm:unsat-singly-no-share-before}, implies that $\hat e \notin \omega_{k-1}$.
If $\omega_{k-1}$ does not contain $e$, then we can replace $\{\psi_{k-1}, \psi_k, \psi\}$
with $\{\omega_{k-1}, t^1_k, t, \omega\}$ to get a larger packing solution, contradicting optimality of $\Nu$.

Hence $e \in E(\omega_{k-1})$. The only way to have this containment is when $d = V(e) \cap V(e')$ is the same as the anchor vertex $a_{k-1}$. Also, $\overbar{v_{k-1} d}$ is the other non-solution edge of $\omega_{k-1}$. For $k-1=0$,  $\omega_0$ cannot contain $e$ (since $f_{fix}(e)=0$) hence it cannot be the case.

Hence, there exists $\psi_{k-2}, \omega_{k-2}$, such that $\overbar{ad}$ is the gain edge $g_{k-1}$ contained in $\psi_{k-2}$.
This implies that the third vertex $v_{k-2}$ of $\psi_{k-2}$ cannot be in $V(\psi) \cup V(\psi_k) \cup  V(\psi_{k-1})$, hence it is a new vertex.

If $\omega_{k-2}$ contains either $\hat e$ or $\overbar{ca}$, then either $v$ or $c$ will be same as the anchor vertex $a_{k-2}$, which implies $(\psi_{k-2}, \psi) \in lend$, which is a contradiction since we are in the case when $\psi \notin \cset$.

If $\omega_{k-2}$ does not contain $e^1_k = \overbar{v_ka}$ (which is always true by \Cref{lem:chains-half-edge-interactions}.\ref{itm:unsat-tail-no-share} in case $k-2=0$), then we can swap $\{\psi_{k-2}, \psi_{k-1}, \psi_k, \psi\}$ with $\{\omega_{k-2}, t^1_{k-1}, t^1_k, t , \omega \}$ to arrive at a contradiction.
Notice that $\omega_{k-2}$ cannot contain $e$ or $\overbar{v_{k-1}d}$ because of \Cref{fact:chain-non-solution-edges}.

\input{tikz_figures/uncovered-type-11-paired-detail-k-2}
Hence it has to be that $k-2>0$ and $\omega_{k-2}$ contains $e^1_k$. Notice that in this case, $v_k = a_{k-2}$ (see \Cref{fig:uncovered-type-11-paired-detail-k-2}).
Now consider $\psi_{k-3}$. The gain edge $g_{k-2}$ is $\overbar{v_k d}$. Because $v_k$ and $d$ dominate $V(\psi_k) \cup V(\psi) \cup V(\psi_{k-1}) \cup V(\psi_{k-2})$, $v_{k-3}$ is a new different vertex.
If $\omega_{k-3}$ is edge-disjoint from $\omega_k$ and $e$,
then we can replace $\{\psi_{k-3}, \psi_{k}\}$ with $\{\omega_k, \Delta_{dv_kc_k},\omega_{k-3}\}$ to arrive at a contradiction. Now $\overbar{ca}$ cannot be in $\omega_{k-3}$ since none of its vertices are in $\psi_{k-3}$. By \Cref{fact:chain-non-solution-edges} $e^1_k$ cannot be in $\omega_{k-3}$. Finally, $\omega_{k-3}$ cannot contain $e$ otherwise $a_{k-3} = c_k$, which implies $(\psi_{k-3}, \psi_k) \in lend$ which contradicts our chain construction. Hence indeed the swap can be performed in this case which concludes the proof for this claim.

\input{tikz_figures/uncovered-type-11-half-base}

Now we deal with the case when $t$ is adjacent to the half base edge of $\psi$ (see \Cref{fig:uncovered-type-11-half-base}), which implies $\psi$ is some solution triangle $\psi'_i$ also contains in some chain $\cset' = \{\psi'_{k'}, \psi'_{k'-1}, \dots \psi'_0\} \in \Chi$ (possibly $\cset' = \cset$) and also $f_{fix}(e) = 0$.

We apply \Cref{cor:unsat-chains-alternate-solution} or \Cref{cor:sat-chains-alternate-solution} (based of whether $\cset'$ is satisfied or unsatisfied), starting from one of the twin-doubly-attached triangles ${t'}_i^1, {t'}_i^2$ containing $\overbar{da}$ to get an alternate packing solution $\Nu'$. This solution cannot contain $e$ as $f_{fix}(e) = 0$ and also none of the edges of $\omega_k$ by \Cref{lem:chains-half-edge-interactions}.\ref{itm:unsat-tail-no-share}. Hence, we can replace $\psi_k$ with $\omega_k$ and add $t$ to $\Nu'$ to obtain a larger packing solution, contradiction the optimality of $\Nu$.

With this last case we conclude the proof for this claim.

Since all possible cases lead to a contradiction,
we conclude that having an uncovered type-$[1,1]$ triangle is not possible.

\end{proof}

\begin{claim}
\label{clm:excluding-type-13}
All the type-$[1, 3]$ triangle in $G$, except for the twin-doubly attached triangles covered in \Cref{obs:unsat-tail-type-1-twins}, are covered by charge function $f_{fix}$.
\end{claim}
\begin{proof}


Suppose $t$ is a type-$[1,3]$ triangle in $G$ which is considered in the claim and is not covered by $f_{fix}$.

First we show that the type-$1$ triangle cannot have a lend relationship
with the type-$3$ triangle. If not, then either both of $t$'s solution edges are half-edge in $f_{fix}$, hence $t$ has to be covered. Otherwise, $t$ falls in \Cref{fact:only-fully-satisfied-lending-to-tail-property} (fully-satisfied triangle), hence again covered by $f_{fix}$.
All the other type-$[1,3]$ triangles in $G$ fall in  \Cref{obs:unsat-tail-type-1-twins},
hence are excluded from this claim.

The only other possibility (by \Cref{prop:structures-doubly-attached}.\ref{itm:doubly-1-3}) is that $t$
contains the half base edge $b_i$ of the type-$1$ triangle
$\psi_i$ which is part of some chain $\cset$, otherwise $t$ will be covered by $f_{fix}$.
Note that in this case, the anchoring vertices of $\psi_i$ and $\psi^3$ must be different, otherwise there exists a non-solution triangle without any solution edge, which contradicts the optimality of $\Nu$.
Also, the type-$3$ triangle $\psi^3$ must not be satisfied or $t$ is covered, hence $\psi^3 \in \Nu_3 \setminus head(\Chi)$.

\input{tikz_figures/uncovered-type-13}

Let $e$ be the non-solution edge of $t$. Note that $f_{fix}(e) = 0$, otherwise $t$ will be covered. Let $v^3$ be the vertex $V(\psi^3) \setminus V(t)$.
Let $d$ be the third vertex of $\psi^3$ and let $a^3 = anchor(\psi^3)$.

If $\psi_i$ is in an unsatisfied chain,
then we can use \Cref{cor:unsat-chains-alternate-solution}.\ref{itm:unsat-swap-t1}
on $\psi_i$ to free up $b_i$, the base-edge of $\psi_i$ in the alternate solution $\Nu'$.
Otherwise, $\psi_i$ is in a satisfied chain, then we can apply
 \Cref{cor:sat-chains-alternate-solution} (based on the structure of $\psi_i$) to free up $b_i$ in the alternate solution $\Nu'$.
By \Cref{lem:chains-half-edge-interactions} with the fact that $f_{fix}(e)=0$ and; the fact that $e \notin E(\omega_i)$, triangles in $\Nu'$ cannot contain $e$.
Also, $\Nu'$ does not contain any non-solution edge of any singly-attached triangle of $\psi^3$ either (by \Cref{lem:chains-half-edge-interactions}.\ref{itm:unsat-type-3} and the fact that $e$ cannot belong to any singly-attached triangle of $\psi^3$).
Hence we can replace $\psi^3$ with $\Delta_{a^3 c v^3}$ and $t$ in $\Nu'$ to obtain a larger packing solution, so this is not possible.
\end{proof}

\begin{claim}
\label{clm:excluding-type-111}
All the type-$[1, 1, 1]$ triangle in $G$ are covered by charge function $f_{fix}$.
\end{claim}
\begin{proof}
    Let $t$ be a type-$[1, 1, 1]$ triangle in $G$ which is considered in the claim and is not covered by $f_{fix}$. It must be that at least two edges in $E(t)$
    are null-edges and it does not contain any full-base edge.

    Let $\psi_k$ and $\psi$ be two triangles such that $e_k = E(\psi_k) \cap E(t)$ and $e = E(\psi) \cap E(t)$ are null w.r.t. $f_{fix}$, i.e. $f_{fix}(e)=f_{fix}(e_k) = 0$. Let $\psi^1$ be the third type-$1$ triangle. The edge $e^1 = E(\psi^1) \cap E(t)$ can be a null non-base edge, half non-base edge, or half base-edge.
    In all these cases, $|CL_{sin}(\psi_k)|=|CL_{sin}(\psi)| =1$.
    Also, $|CL_{sin}(\psi^1)| = 1$ unless $e^1$ is a half non-base edge.
    Let $a_k, a, a^1$ be the anchoring vertices of $\psi_k, \psi, \psi^1$, respectively
    (in the case that $|CL_{sin}(\psi^1)| > 1$ choose any arbitrary anchoring vertex to be $a^1$). Now we name vertices of $t$.
    Let $c = V(\psi_k) \cap V(\psi), c^1 = V(\psi) \cap V(\psi^1), c^2 = V(\psi_k) \cap V(\psi^1)$.
    Let the base-edges of $\psi_k, \psi, \psi^1$ be $b_k, b, b^1$ respectively.
    We also name the third vertex of each type-$1$ triangle. Let $v_k = V(\psi_k) \setminus V(t), v = V(\psi) \setminus V(t), v^1 = V(\psi^1) \setminus V(t)$.

    Let us first consider the case when $e^1$ is a non-base edge of $\psi^1$.
    In this case, $f_{fix}(t)$ can be $0$ or $\frac 1 2$.
    We will first rule out the case where $f_{fix}(t) = 0$ (see \Cref{fig:uncovered-type-111-f0}).

    \input{tikz_figures/uncovered-type-111-f0}

    Suppose $f_{fix}(t) = 0$, it means that $\psi_k, \psi, \psi^1$ are tails of three different chains in $\Chi_{unsat}$.

    We use \Cref{cor:unsat-chains-alternate-solution}.\ref{itm:unsat-swap-t1}
    on all three chains to get an alternate packing solution $\Nu'$. These swaps will free up $t$, which we can add to our  solution $\Nu'$ to get a larger packing solution, which  contradicts the optimality of $\Nu$.

    \input{tikz_figures/uncovered-type-111-e1-nonbase-3anchoring}

    Now we can assume that $f_{fix}(t) = \frac 1 2$. By renaming, let $b^1 = \overbar{c^2 v^1}$.
    Notice that if $|\{a_k, a, a^1\}| = 3$, then we can replace $\{\psi_k, \psi, \psi^1\}$ with $\{\omega_k, \omega, \omega^1, t\}$ to get a larger packing solution, so this is not possible.

    \input{tikz_figures/uncovered-type-111-e1-nonbase-1anchoring}

    The case where $a_k = a = a^1$ is also not possible since that would imply at least one of three triangles is type-$3$ (see \Cref{fig:uncovered-type-111-e1-nonbase-1anchoring}).

    Now we can assume $|\{a_k, a, a^1\}| = 2$.
    It can be deduced that $a \neq a^1$, otherwise $\overbar{cc^2}$ becomes a base edge of $\psi_k$ contradicting the fact that $e_k$ is a non-base null edge of $\psi_k$.
    Let $\cset$ be the chain such that $tail(\cset) = \psi_k$. By \Cref{obs:tail-renaming} assume $g_{k+1} = e_k$.

    \input{tikz_figures/uncovered-type-111-e1-nonbase-ak=a1}

    Suppose $a_k = a^1$. This implies that, out of $\overbar{v_k c}, \overbar{v_kc^2}$, it should be the case that  $b_k =  \overbar{v_kc^2}$. If not, then $\psi_k$ will become a type-$3$ triangle (see \Cref{fig:uncovered-type-111-e1-nonbase-ak=a1}).
    Applying
    \Cref{cor:unsat-chains-alternate-solution}.\ref{itm:unsat-swap-t1} with $i=k$ on
    $\cset$ we get $\Nu'$, such that solution triangles in $\Nu'$ cannot contain the edge $e^2_k = \overbar{c^2a_1}$. Suppose triangles in $\Nu'$ do not contain the edge $\overbar{v^1 a^1}$ as well. Then we can 
    replace $\{\psi, \psi^1\}$
    with $\{\omega, \omega^1, t\}$ 
    in $\Nu'$
    to get a larger packing solution,
    so this is not possible.
    Note that, by \Cref{lem:chains-half-edge-interactions}.\ref{itm:unsat-tail-no-share}, 
    $\omega$ cannot contain any edge of any solution triangle in $\Nu'$ excepts $\psi$'s base edge $b$.
    Now we can assume that the edge $\overbar{v^1 a^1}$ is a half-edge of chain $\cset$. By \Cref{fact:f_fix}, $f_{fix}(\overbar{v^1 a^1}) =\frac 1 2$.

    Consider $\psi_{k-1}$, $\omega_{k-1}$ for chain $\cset$. Note that, $g_k = \overbar{c a^1}$ is the gain edge contained in $\psi_{k-1}$.
    As $\overbar{v^1 a^1}$ is adjacent to $g_k$ it cannot be $e^1_{k-1}$ which together with  \Cref{lem:chains-half-edge-interactions}.\ref{itm:unsat-singly-no-share-before}, implies that $\overbar{v^1 a^1} \notin \omega_{k-1}$.
    Suppose $\omega_{k-1}$ is disjoint from $\omega$.
    This together with the fact that $\omega_{k-1}$ is disjoint from $\omega_k$ by \Cref{fact:chain-non-solution-edges}, implies that we can replace
    $\{\psi_k, \psi_{k-1}, \psi, \psi^1\}$ with
    $\{t, t_k^1, \omega_{k-1}, \omega, \omega^1\}$ to get a larger packing solution, contradicting the optimality of $\Nu$.
    Hence, $\omega_{k-1}$ must share some edge with $\omega$. Since the vertices $c, a^1 \in V(\psi_{k-1})$ dominate all the vertices in $V(\psi)\cap V(\psi_k)\cap V(\psi^1) \cap \{a, a^1\}$, hence the third vertex (say $v_{k-1} \in V(\psi_{k-1})$) has to be a new vertex. This implies that the only possible way that $\omega_{k-1}$ and $\omega$ share an edge is that $\overbar{ca} \in E(\omega_{k-1})$ and the anchor vertex $a_{k-1} = a$. Hence $\overbar{ca} = e^2_{k-1}$.

    \input{tikz_figures/uncovered-type-111-e1-nonbase-ak=a1-detail-k-1}

    In this case, we apply 
    \Cref{cor:unsat-chains-alternate-solution}.\ref{itm:unsat-swap-t1} with $i=k'$ on the $k'$-sized chain $\cset'$
    (such that $tail(\cset')=\psi$ and assuming $g'_{k'+1} = e$ by \Cref{fact:f_fix}) to get an alternate solution $\Nu'$.
    Note that the edge $e^2_{k'} = \overbar{ca}$ cannot be contained by triangles in $\Nu'$ (see \Cref{fig:uncovered-type-111-e1-nonbase-ak=a1-detail-k-1}).
    Also, they cannot contain the other (half-)edge $e^1_{k-1}$ of $\cset$ contained in $\omega_{k-1}$ by \Cref{lem:chains-half-edge-interactions}.\ref{itm:two-chains-no-common-half-edge}.
    In addition to that, any triangle in $\Nu'$ cannot contain any of the non-solution edges, with respect to $\Nu$, of $\omega_k$ (by \Cref{lem:chains-half-edge-interactions}.\ref{itm:unsat-tail-no-share}).
    Now to show that any triangle in $\Nu'$ excepts $\psi^1$ does not contain any edge of $\omega^1$, we just need to argue that $\overbar{a^1v^1}$ is not a solution edge of any triangle in $\Nu'$.
    Recall that $\overbar{a^1v^1}$ is a half-edge of $\cset$, hence by \Cref{lem:chains-half-edge-interactions}.\ref{itm:two-chains-no-common-half-edge}, $\overbar{a^1v^1}$ will not become a solution edge of any triangle in $\Nu'$.
    All in all, we can replace $\{ \psi_k, \psi_{k-1}, \psi^1\}$ with $\{t_k^1, \omega_{k-1}, \omega^1, t\}$ in the new solution $\Nu'$ to get a packing solution that is strictly larger then $\Nu$, which is a contradiction. So it is not possible that $a_k = a^1$.

    \input{tikz_figures/uncovered-type-111-e1-nonbase-ak=a}

    Suppose $a_k = a$.
    Note that this is only possible when $V(b) \cap V(b_k) = c$, which implies $b = \overbar{v c }$ and $b_k = \overbar{c v_k}$.
    In the case when both $b, b_k$ do not contain $c$, $e^1$ will support a singly-attached triangle.
    In the other cases, when exactly one of $b, b_k$ contains $c$, then either $e$ or $e_k$ will support a singly-attached triangle.
    All in all, it contradicts the fact that we are in the case when all three edge are non-base edges (see \Cref{fig:uncovered-type-111-e1-nonbase-ak=a}).
    We apply 
    \Cref{cor:unsat-chains-alternate-solution}.\ref{itm:unsat-swap-t1} to $\cset$ with $i=k$, 
    to get the alternate solution $\Nu'$. 
    If non-solution edges of $\omega^1$ (not adjacent to $\psi^1$) are not solution edges of any triangle in $\Nu'$, then we can replace $\{\psi^1, \psi\}$ with $\{\omega^1, \omega, t\}$ in $\Nu'$ to get a larger packing solution, which is a contradiction. Note that by \Cref{lem:chains-half-edge-interactions}.\ref{itm:unsat-tail-no-share}, 
    The non-solution edges of $\omega$ with respect to $\Nu$ will still be non-solution edges in $\Nu'$.
    Hence, at least one non-solution edge of $\omega^1$ with respective $\Nu$ must become the solution edge of $\Nu'$.
    Using the same argument as above, we can conclude that applying
    \Cref{cor:unsat-chains-alternate-solution}.\ref{itm:unsat-swap-t1} to $\cset'$ with $i=k'$ 
    gives an alternate solution such that its triangles also contain a non-solution edge of $\omega^1$.
    Hence the swaps in \Cref{cor:unsat-chains-alternate-solution}.\ref{itm:unsat-swap-t1} to both $\cset, \cset'$ with $i=k$ and $i=k'$ respectively 
    must contain at least one non-solution edges (with respect to $\Nu$) of
    $\omega^1$. Moreover, as the non-solution edges (with respect to $\Nu$) of
    $\omega^1$ 
    must belong to the set of half-edges for
    $\cset, \cset'$ respectively, hence by \Cref{lem:chains-half-edge-interactions}.\ref{itm:two-chains-no-common-half-edge}, they must be different edges of $\omega^1$.

    Consider $\psi_{k-1}$. Notice that $\overbar{c^2 a} \in E(\psi_{k-1})$. Since $\{c^2, a\}$ dominates existing vertices, the third vertex $v_{k-1}$ of $\psi_{k-1}$ has to be a new different vertex.
    This means the edge $\overbar{v^1 a^1}$ cannot be in $\omega_{k-1}$.
    As $\overbar{c^2 a^1}$ is adjacent to $g_k$ it cannot be $e^1_{k-1}$ which together with \Cref{lem:chains-half-edge-interactions}.\ref{itm:unsat-singly-no-share-before}, implies that $\overbar{c^2 a^1} \notin \omega_{k-1}$.
    Hence, if $\omega_{k-1}$ is disjoint from $\omega$
    (which will be true for $k-1=0$ case by \Cref{lem:chains-half-edge-interactions}.\ref{itm:unsat-tail-no-share}), then we can swap
    $\{\psi_{k-1}, \psi_k, \psi, \psi^1\}$ with
    $\{\omega_{k-1}, t_k^1, \omega, \omega^1, t\}$ to get a larger packing solution, which contradicts the optimality of $\Nu$.
    Hence, $k-1>0$ and $\overbar{va} \in E(\omega_{k-1})$, because $\overbar{ca}$ cannot be in $\omega_{k-1}$ by \Cref{fact:chain-non-solution-edges}. This implies that the anchoring vertex $a_{k-1} = v$ and the base edge $b_{k-1} = \overbar{v_{k-1}}a$ (see \Cref{fig:uncovered-type-111-e1-nonbase-ak=a-detail-k-1}).
    \input{tikz_figures/uncovered-type-111-e1-nonbase-ak=a-detail-k-1}

    Consider $\psi_{k-2}$. Notice that $\overbar{c^2 v} \in E(\psi_{k-2})$.
    Hence by the same reasoning used for $\psi_{k-1}$, the third vertex $v_{k-2}$ of $\psi_{k-2}$ is a new different vertex.
    By the same reasoning used for $\omega_{k-1}$, $\overbar{v^1 a^1}$ and $\overbar{c^2 a^1}$ cannot be in $\omega_{k-2}$.
    Which together with \Cref{fact:chain-non-solution-edges} and the adjacency of triangle $\psi_{k-2}$ implies that $\omega_{k-2}$ is disjoint from both $\omega^1$ and $\omega$. Hence, we can replace
    $\{\psi_k, \psi_{k-1}, \psi_{k-2}, \psi, \psi^1\}$ with
    $\{t_k^1, t_{k-1}^1, \omega_{k-2}, \omega, \omega^1, t\}$ to get a larger packing solution, leading to a contradiction.
    Hence the case where $a_k = a$ is not possible either.
    All in all, the case where $e^1$ is a non-base edge is not possible.

    \input{tikz_figures/uncovered-type-111-e1-base}

    Suppose $e^1$ is a half-base edge
    (as in \Cref{fig:uncovered-type-111-e1-base}).

    This means $\psi_k, \psi$ are tails of $k,k'$ sized chains $\cset, \cset' \in \Chi_{unsat}$ respectively and $\psi^1$ is also part of some chain (possibly be $\cset$ or $\cset'$).
    Using \Cref{obs:tail-renaming}, we assume that $e_k=g_{k+1}$ and $e = g'_{k'+1}$.
    In this case, we apply
    \Cref{cor:unsat-chains-alternate-solution}.\ref{itm:unsat-swap-t1} with $i=k$ and $i=k'$ on both $\cset$ and $\cset'$ to get an alternate solution $\Nu'$.
    If $\psi^1$ is part of one of the chains $\cset, \cset'$, then $b^1$ will become a non-solution edge in $\Nu'$. This together with the fact that the swaps made $e,e_k$ non-solution edges, we can add $t$ to $\Nu'$ to get a larger packing solution, leading to a contradiction.
    Hence $\psi^1$ is part of a different chain $\cset''$. The chain $\cset''$ can either be satisfied or unsatisfied. If it is unsatisfied, then we 
    apply \Cref{cor:unsat-chains-alternate-solution}.\ref{itm:unsat-swap-t1} to $\cset''$ starting from $\psi^1$ to free the base edge $e^1$ 
    to get another alternate packing solution $\Nu''$.
    By \Cref{lem:chains-half-edge-interactions}.\ref{itm:two-chains-no-common-half-edge}, 
    $\Nu''$ is a valid packing solution, i.e., any pair of solution triangles in $\Nu''$ are edge-disjoint.
    Again, we can add $t$ to $\Nu''$ to get a larger packing solution, leading to a contradiction.
    Otherwise, in case when $\cset'' \in \Chi_{sat}$, if $\psi^1$ is not a tail triangle, then we can use \Cref{cor:sat-chains-alternate-solution}.\ref{itm:sat-swap-t1} to free up $e^1$, without
    using any edge of 
    any solution triangle in $\Nu'$ and finally add $t$ to reach at a contradiction.

    Finally, the last case is when $\psi^1$ is a tail triangle of a chain $\cset'' \in \Chi_{sat}$. We need to deal with this case in a more delicate way.
    Based on the {\em order} in which the chains $\cset, \cset', \cset''$ were constructed, we will arrive at a contradiction by showing different swaps in each case.
    
    First we assume that $\cset''$ is constructed before both $\cset, \cset'$. By \Cref{lem:chains-half-edge-interactions}.\ref{itm:sat-tail-share} and \Cref{lem:ordered-chains-half-edge-interactions}.\ref{itm:sat-half-edge-after}, $\omega^1$ contains exactly one half-edge of some chain constructed before $\cset''$ (including $\cset''$). Hence by \Cref{lem:chains-half-edge-interactions}.\ref{itm:two-chains-no-common-half-edge} the other edge (say $e'$) of $\omega^1$ could be a half-edge of at most one of the chains in $\cset, \cset'$. By renaming assume it does not contain any half-edge of $\cset$. In this case, we can simultaneously 
    apply 
    \Cref{cor:unsat-chains-alternate-solution}.\ref{itm:unsat-swap-t1} with $i=k$ to $\cset$ and apply
    \Cref{cor:sat-chains-alternate-solution}.\ref{itm:sat-swap-tail} to $\cset''$ to get an alternate solution $\Nu'$ while making $e_k, e^1$ non-solution edges. By \Cref{lem:chains-half-edge-interactions}.\ref{itm:unsat-tail-no-share},
    solution triangles in $\Nu'$ cannot contain non-solution edges (with respect to $\Nu$) of
    $\omega$. Hence we can swap $\psi$ with $\omega$ in $\Nu'$ to eventually free all edges of $t$ and hence finally add $t$ to get a larger packing solution reaching at a contradiction.
  
    Now we assume that at least one of the chains in $\cset, \cset'$ is constructed before $\cset''$. By renaming, assume $\cset$ is constructed before $\cset''$.
    In this case, we can simultaneously apply
    \Cref{cor:unsat-chains-alternate-solution}.\ref{itm:unsat-swap-t1} with $i=k$ for $\cset$ and apply
    \Cref{cor:sat-chains-alternate-solution}.\ref{itm:sat-swap-tail} to $\cset''$ to get an alternate solution $\Nu'$ and make $e_k, e^1$ non-solution edges. By \Cref{lem:chains-half-edge-interactions}.\ref{itm:unsat-tail-no-share},
    solution triangles in $\Nu'$ cannot contain non-solution edges (with respect to $\Nu$) of $\omega$. Hence we can swap $\psi$ with $\omega$ in $\Nu'$ to finally free all edges of $t$ and hence finally add $t$ to get a larger packing solution reaching at a contradiction.

    Since all cases lead to a contradiction, we conclude that $t$ must be initially be covered by $f_{fix}$.

\end{proof}

Now we prove that there cannot be any type-$[1, 1, 3]$ uncovered triangle. Note that this structure is very similar to type-$[1, 1]$ case as we will use the fact that at least one of the type-$1$'edge is null-edge and the other type-$1$'s edge is either non-base or half-base edge. Another similarity can be seen by exchanging the type-$3$ triangle with one of its singly-attached triangle without changing the type of the two type-$1$ triangles and making the third edge non-solution edge, hence making it a type-$[1, 1, 3]$ triangle. But we can still not use the proof for type-$[1, 1]$ case as a black-box because this kind of exchange can destroy the chain structure that we crucially use in the proof.

\begin{claim}
\label{clm:excluding-type-113}
All the type-$[1, 1, 3]$ triangle in $G$ are covered by charge function $f_{fix}$.
\end{claim}
\begin{proof}
    Let $t$ be a type-$[1, 1, 3]$ triangle in $G$ which is considered in the claim and is not covered by $f_{fix}$.
    It must be that at least two edges in $E(t)$ are null-edges and $t$ does not contain any full-base edge.
    Hence, it must be that
    $t$ is adjacent to at least one null-edge
    (with respect to $f_{fix}$) of a type-$1$ triangle which is tail triangle of an unsatisfied chain.
    Suppose $\psi_k$ be the type-$1$ tail triangle
    of chain $\cset \in \Chi_{unsat}$. By \Cref{obs:tail-renaming}, let $e_k = E(t)\cap E(\psi_k)$ be the edge $g_{k+1}$ of $\cset$.
    Let $\psi$ be another type-$1$ triangle and let $\psi^3$ be the type-$3$ triangle adjacent to $t$.

    We let $c = V(\psi) \cap V(\psi_k)$ be the common vertex between $\psi$ and $\psi_k$.
    Also, let $c^1 = V(\psi^3) \cap V(\psi)$ and $c^2 = V(\psi^3) \cap V(\psi_k)$.
    We let $v, v_k, v^3$ be $V(\psi) \setminus V(t), V(\psi_k) \setminus V(t), V(\psi^3) \setminus V(t)$, the vertex of each solution triangle not in $t$, respectively. Moreover, we let $a^3 = anchor(\psi^3)$
    and $a = anchor(\psi)$ (in case $|CL_{sin}(\psi)|>1$, pick any arbitrary anchoring vertex as $a$).
    Let $e = E(t) \cap E(\psi)$.

    By \Cref{prop:structures-hollow}
    there are two cases, first when $e$ is a non-base edge and second when $e$ is a half base-edge.

    First let $e$ be a non-base edge of $\psi$.
    By \Cref{prop:structures-hollow},
    it must be that the only anchoring vertex $a$ of $\psi$ is same as $a_k = anchor(\psi_k)$.
    Let $b$ be the base edge of $\psi$, there are two sub-cases here.

    \input{tikz_figures/uncovered-type-113-1anchoring.tex}
    First, when $c \notin V(b) \cup V(b_k)$.
    Since
    $\Delta_{c^1c^2a}$ forms a singly-attached triangle of $\psi^3$ based at the edge $e^3 = E(\psi^3) \cap E(t)$ and anchoring at $a^3$,
    it must be that $a^3 = a_k = a$ or we would be in the case where $\psi^3$ has two anchoring vertex, which is impossible by~\Cref{prop:structures-solution} (see \Cref{fig:uncovered-type-113-1anchoring}).
    Since $\psi_k$ is in a chain, there exists a triangle $\psi_{k-1}$ such that $(\psi_k, \psi_{k-1}) \in lend$, such that $\overbar{ac}$ is a non-base edge of $\psi_{k-1}$.
    Notice that since $(\psi, \psi_{k-1}) \in lend$, it must be that $\psi$ is not part of
    any chain. In this case, the edge $\overbar{c c^1}$ is a half-edge.
    This implies that $\psi^3$ is not in any chain and is not satisfied.
    Consider $\psi_{k-1}$, since $\{a,c\}$ dominates all existing vertices,
    the third vertex $v_{k-1}$ of $\psi_{k-1}$ must be a new different vertex.
    If the anchoring vertex $a_{k-1}$ is not an existing vertex
    in $V(\psi_k) \cup V(\psi) \cup V(\psi^3) \cup \{a\}$, then we can replace
    $\{\psi_k, \psi_{k-1}, \psi, \psi^3\}$ with
    $\{\omega_{k-1}, t_k^1, \Delta_{a c^2 v^3}, \Delta_{a c^1 v}, t\}$ to obtain
    a larger packing solution, which leads to a contradiction.
    In case when $k-1=0$, then $a_0$ cannot be an existing vertex except $v^3$ since there exists solution edges between any of them and $\{a,c\}$.
    In any case there is a singly-attached triangle $\omega_0$ of $\psi_0$ that is disjoint from the existing structure, hence the same argument leads to a contradiction.
    Hence, $k-1>0$ and $v^3$ must be the anchoring vertex of $\psi_{k-1}$. No other existing vertex can be the anchoring vertex, since
    having any of $v_k, c^2$ to be the anchoring vertex would violate \Cref{clm:chain-consecutive-disjoint-singly-twins} and
    having any of $v, c^1$ to be an anchoring vertex would imply that $(\psi_{k-1}, \psi) \in lend$,
    which contradicts the deduction that $\psi$ is not part of any chain. Moreover, $\overbar{a v_{k-1}}$ is the base edge $b_{k-1}$ of $\psi_{k-1}$ (see \Cref{fig:uncovered-type-113-1anchoring-detail-k-1}).

    \input{tikz_figures/uncovered-type-113-1anchoring-detail-k-1}

    Now consider $\psi_{k-2}$ containing edge $\overbar{cv^3}$. Since $\{c, v^3\}$ dominates all existing vertices in $V(\psi_{k-1}) \cup V(\psi_k) \cup V(\psi) \cup V(\psi^3) \cup \{a\}$,
    $v_{k-2}$ must be a new different vertex.
    Since any existing non-solution edges incident to $\psi_{k-2}$ is in $\omega_{k-1}$,
    it must be the case that $\omega_{k-2}$ is disjoint by \Cref{clm:chain-consecutive-disjoint-singly-twins},
    Now we can replace
    $\{\psi_k, \psi_{k-1}, \psi_{k-2}, \psi, \psi^3\}$ with
    $\{t_k^1, t_{k-1}^1,\omega_{k-2}, \Delta_{a c^2 v^3}, \Delta_{a c^1 v}, t\}$ to obtain
    a larger packing solution, contradicting the optimality of $\Nu$.
    So we conclude that the sub-case where $c \notin V(b) \cup V(b_k)$ is not possible.

    \input{tikz_figures/uncovered-type-113-pair-2anchoring.tex}

    The second case is when $b, b_k \notin E(t)$, $a=a_k$ and $c = V(b) \cap V(b_k)$.
    In this case, $a^3 \neq a_k$, otherwise $\psi$ and $\psi_k$ would become type-$3$
    (see \Cref{fig:uncovered-type-113-pair-2anchoring}).

    Let $\omega$ be the singly-attached of $\psi$.
    Also let $\omega^3$ be the singly-attached of $\psi^3$ not adjacent to $c_k = c^2 =  V(\psi^3) \cap V(\psi_k)$.
    Since the edge $E(\psi_k) \cap E(t)$ is a null-edge, there exists $\psi_{k-1}$ where $(\psi_k, \psi_{k-1}) \in lend$, such that $\overbar{c^2 a} \in E(\psi_{k-1})$.
    Since $\{c_k, a_k\}$ dominates $V(\psi) \cup V(\psi_k) \cup V(\psi^3) \cup \{a, a^3\}$, the third vertex $v_{k-1}$ of $\psi_{k-1}$, has to be a new different vertex.

    There are four non-solution edges adjacent to $\{c^2, a\}$, namely,
    $\{\overbar{a v_k}, \overbar{a c}, \overbar{a v}, \overbar{c^2 a^3}\}$.
    If $\omega_{k-1}$, the singly-attached triangle of $\psi_{k-1}$, does not contain any of these four edges, then we can replace
    $\{\psi_{k-1}, \psi_k, \psi, \psi^3\}$ with
    $\{\omega_{k-1}, \Delta_{a c^2 v_k}, \omega, \omega^3 \}$, hence this is not possible.
    The same swap works even if $\overbar{c^2 a^3} \in E(\omega_{k-1})$. A similar swap works, if $\psi_{k-1}$ is a type-$3$ triangle as its singly-attached triangles cannot contain any of $\{\overbar{a v_k}, \overbar{a c}, \overbar{a v}\}$ edges. Hence, $k-1>0$.

    Now, $a_{k-1}$ cannot be $v_k$ or $c$ by \Cref{clm:chain-consecutive-disjoint-singly-twins}, hence $\overbar{a v_k}, \overbar{a c} \notin E(\omega_{k-1})$.
    We deduced that, $\overbar{c^2 a^3} \notin E(\omega_{k-1})$, which implies that $a_{k-1}=v$ and $\overbar{v_{k-1}a_k} = b_{k-1}$.

    As $k-1>0$, hence $\psi_{k-2}$ exists, such that $\overbar{c^2 v} \in E(\psi_{k-2})$.
    Also, the third vertex $v_{k-2}$ of $\psi_{k-2}$ cannot be vertices adjacent to $\{c^2, v\}$, hence it is a new different vertex (see \Cref{fig:uncovered-type-113-pair-2anchoring-detail-k-2}).

    \input{tikz_figures/uncovered-type-113-pair-2anchoring-detail-k-2}

    There are three non-solution edges adjacent to $\{c^2, v\}$, namely,
    $\{\overbar{v a} , \overbar{v v_{k-1}}, \overbar{c^2 a^3}\}$.
    The anchoring vertex of $\psi_{k-2}$ cannot be $a=a_k$ or $v_{k-1}$ by
    \Cref{clm:chain-consecutive-disjoint-singly-twins},
    which implies $\overbar{av} \notin E(\omega_{k-2})$.
    Hence we can replace (even if $\overbar{c^2 a^3} \in E(\omega_{k-2})$)
    $\{\psi_{k-2}, \psi, \psi^3\}$ by
    $\{\omega_{k-2}, \Delta_{v c^1 c^2}, \omega, \omega^3\}$, where $\omega_{k-2}$ is a singly-attached triangle of $\psi_{k-2}$.
    The same swap works even if $\psi_{k-2}$ is a type-$3$ triangle.
    Hence, it cannot be the case that $c = V(b) \cap V(b_k)$.

    Note that it can't be the case that $c \in V(b) \cup V(b_k)$ but $c \notin V(b) \cap V(b_k)$, because the common anchoring vertex property would would imply that either $\psi$ or $\psi_k$ is a type-$3$ triangle, which is a contradiction.

    All in all, the case when $e$ is a non-base edge of $\psi$ is not possible.

    We will rule out the other case when $e$ is the base-edge of $\psi$. If $e$ is a full-edge, then $t$ is covered, so we can assume that $e$ is a half-edge.
    This implies that $\psi$ is some $j^{th}$ triangle of some $k'$-sized chain $\cset' \in \Chi$.
    Also, $\psi^3$ is an unsatisfied triangle in $\Nu_3 \setminus head(\Chi)$, otherwise $t$ will be covered.
    Recall that $e_k$ is a null-non-base edge of $\psi_k$. Hence, it is impossible that $a^3 = a$ since that would imply $b_k = e_k$, as it will support $\Delta_{c c^2 a^3}$. This implies that $\omega$ is disjoint from any singly-attached triangle of $\psi^3$.

    There are two possibilities for the base-edge $b_k$ of $\psi_k$.
    One is when $c \in V(b_k)$ and another is when $c \notin V(b_k)$.

    \input{tikz_figures/uncovered-type-113-halfbase-3anchoring-c-notin-bk.tex}

    First we consider the case where $c \notin V(b_k)$.
    In this case, note that $a$, the anchoring vertex of $\psi$ cannot be any vertex in $V(\psi_k) \cup V(\psi^3)$ since $V(\psi)$ dominates these vertices via solution edges. Moreover, $a \neq a_k$, otherwise $\psi_k$ is not type-$1$.

    Now we use one of the applicable  \Cref{cor:unsat-chains-alternate-solution} or \Cref{cor:sat-chains-alternate-solution} (depending on the structure of $\cset'$ and $\psi$) with $i=j$ on $\cset'$ to get an alternate solution $\Nu'$ and free up the base edge $e = \overbar{c c^1}$.
    No solution triangles in $\Nu'$ can contain any non-solution edges (with respect to $\Nu$) of
    $\omega_k$ and any singly-attached of $\psi^3$ by \Cref{lem:chains-half-edge-interactions}.\ref{itm:unsat-tail-no-share} and \Cref{lem:chains-half-edge-interactions}.\ref{itm:unsat-type-3}.
    Hence we can replace $\psi_k, \psi^3$ with $\omega_k, \Delta{c^1v^3a^3}$ in $\Nu'$ and add $t$, to obtain a larger valid packing solution, which leads to a contradiction.

    Hence, the case where $c \notin E(b_k)$ is not possible.

    Now we arrive at the final case when $c \in E(b_k)$.
    Recall that, $e_k$ is null-non-base edge of $\psi_k$, $e$ is a half-base edge of $\psi$ and $\psi^3$ is an unsatisfied type-$3$ triangle in $\Nu_3 \setminus head(\Chi)$.

    Note that $a_k \neq a^3$ and $a \neq a^3$, otherwise either $\psi_k$ or $\psi$ will not be type-$1$.

    There are two cases, one is when $a_k \neq a$ and another is when $a_k = a$.
    For illustrations,
    see \Cref{fig:uncovered-type-113-halfbase-3-anchoring-c-in-bk,fig:uncovered-type-113-halfbase-2-anchoring-pair-c-in-bk}. We argue both these cases in one go.

    \input{tikz_figures/uncovered-type-113-halfbase-3-anchoring-c-in-bk.tex}
    \input{tikz_figures/uncovered-type-113-halfbase-2-anchoring-pair-c-in-bk.tex}

    We use either \Cref{cor:unsat-chains-alternate-solution} or \Cref{cor:sat-chains-alternate-solution} on $\psi$ to free up the edge $e = \overbar{c c^1}$ 
    to get an alternate solution $\Nu'$.
    It may use the edge $\overbar{ca}$ when $a=a_k$.
    By \Cref{lem:chains-half-edge-interactions}.\ref{itm:unsat-tail-no-share}, solution triangles in $\Nu'$ cannot contain
    the edge $\overbar{av_k}$ of $\omega_k$.
    Hence, we can simultaneously 
    apply
    \Cref{cor:unsat-chains-alternate-solution}.\ref{itm:unsat-swap-t1}
    for $i=k$ to $\cset$
    to get an alternate solution $\Nu'$ which free up the edge $e_k = \overbar{c c^2}$ in addition to $e$.
    By \Cref{lem:chains-half-edge-interactions}.\ref{itm:unsat-type-3},
    no solution triangles in $\Nu'$ can contain any non-solution edges (with respect to $\Nu$) of any singly-attached triangle of $\psi^3$.
 
    Hence, we can replace $\psi^3$ with $\Delta_{a^3 c^1 v^3}$ in $\Nu'$ and add $t$ to obtain
    a larger valid packing solution, contradicting the optimality of $\Nu$.



    All in all, we conclude that such a $t$ does not exist in the first place, hence any type-$[1,1,3]$ triangle must be covered in $G$ by $f_{fix}$.
\end{proof}

\begin{claim}
\label{clm:excluding-type-133}
All the type-$[1, 3, 3]$ triangle in $G$  are covered by charge function $f_{fix}$.
\end{claim}
\begin{proof}
    Let $t$ be a type-$[1, 3, 3]$ triangle in $G$ which is considered in the claim and is not covered by $f_{fix}$.
    It must be that at least two edges in $E(t)$ are null-edges and $t$ does not contain any full-base edge.
    Let $\psi_i$ be the type-$1$ triangle.
    Let $\psi_1^3$ and $\psi_2^3$ be type-$3$ triangles.
    Let $b_i$ be the base-edge of $\psi_i$.
    By \Cref{prop:structures-hollow}, $b_i \in t$. Hence, $b_i$ has to be a half-base edge and $\psi_i$ must be part of some chain $\cset \in \Chi$.
    Moreover,
    $\psi_1^3$ and $\psi_2^3$ must not be in $head(\Chi)$ otherwise $t$ will be covered by $f_{fix}$. Hence,

  Let us name vertices.
  Let $c = V(\psi_i) \cap V(\psi_2^3)$, $c^1 = V(\psi_2^3) \cap V(\psi_1^3)$, $c^2 = V(\psi_1^3) \cap V(\psi_i)$.
  Let $a_i = anchor(\psi_i), a_1^3 = anchor(\psi_1^3), a_2^3 = anchor(\psi_2^3)$.

  There are two sub-cases, one when $a_i = a_1^3 = a_2^3$ and
  another when $a_i$ is different from at least one vertex out of $a_1^3$ or $a_2^3$.

  \input{tikz_figures/uncovered-type-133-1anchoring}
  First, let consider the case when $a_i = a_1^3 = a_2^3$ (see \Cref{fig:uncovered-type-133-1anchoring}).
  Notice that in this case, $\psi_i$ has to be a tail triangle of an unsatisfied chain,
  otherwise, one of $\psi_1^3$ or $\psi_2^3$
  will contain the half-edge contained in $\omega_i$ which is not possible by
  \Cref{lem:chains-half-edge-interactions}.\ref{itm:unsat-type-3}.
  Hence, we can apply
  \Cref{cor:unsat-chains-alternate-solution}.\ref{itm:unsat-swap-t1} starting with $\psi_i$ to $\cset$ to get an alternate solution $\Nu'$ while making $b_i$ a non-solution edge.
  By \Cref{lem:chains-half-edge-interactions}.\ref{itm:unsat-type-3}, solution triangles in $\Nu'$ will contain
  exactly one of the non-solution edges $\overbar{c^2a_i}$ or $\overbar{ca_i}$ of singly-attached triangles of $\psi_1^3$ and $\psi_2^3$.

  WLOG we can assume that $\overbar{c^2a_i}$ is 
  contained in $\Nu'$
  , hence we can swap $\psi_1^3$ and $\psi_2^3$ with
  $\Delta_{v_1c^1a_i}$ and $\Delta_{cv_2a_i}$ in $\Nu'$.
  This will free up all the edges of $t$ and we can add it to $\Nu'$ to get a larger packing solution, which contradicts the optimality of $\Nu$.

  \input{tikz_figures/uncovered-type-133-3anchoring}

  Secondly, we deal with the case when $a_i$ is different from at least one vertex out of $a_1^3$ or $a_2^3$. WLOG, we can assume that $a_i \neq a^3_1$ (see \Cref{fig:uncovered-type-133-3anchoring}).
  In this case, first we show that in fact $|\{a_i, a^3_1, a^3_2\}| = 3$. If not, then either $a_i = a^3_2$ or $a^3_1 = a^3_2$.
  In the former case, $\Delta_{c^1c^2a_i}$ will be a singly-attached triangle of $\psi_1^3$, which contradicts \Cref{prop:structures-solution} that type-$3$ triangles have exactly one anchoring vertex.
  In the latter case when $a^3_1 = a^3_2$, $\Delta_{cc^2a^3_1}$ will be a singly-attached triangle of $\psi_i$, it
  contradicts \Cref{prop:type-[1-1]}.\ref{itm:anchoring-vertex} since $\psi_i$ is part of a chain $\cset$ hence have exactly one anchoring vertex.
  Hence $|\{a_i, a^3_1, a^3_2\}| = 3$.
  In this case, we 
  apply one
  \Cref{cor:unsat-chains-alternate-solution,cor:sat-chains-alternate-solution} (based on the type of $\cset$ and $\psi_i$) to $\cset$ starting with $\psi_i$ to get an alternate solution $\Nu'$ which does not use the edge $b_i$. By \Cref{lem:chains-half-edge-interactions}.\ref{itm:unsat-type-3}, 
  $\Nu'$ will not contain
  any non-solution edge of the singly-attached triangles of the unsatisfied triangles $\psi_1^3, \psi_2^3$. Hence, we can swap $\psi_1^3$ and $\psi_2^3$ with any of their singly-attached triangles (for instance $\Delta_{ca^3_2v_2}$ and $\Delta_{c^2a^3_1v_1}$) respectively 
  in $\Nu'$
  to free all the edges of $t$. Clearly $\Delta_{ca^3_2v_2}$ and $\Delta_{c^2a^3_1v_1}$ are disjoint, hence we can swap $\psi_1^3$ and $\psi_2^3$  with $\Delta_{ca^3_2v_2}$ and $\Delta_{c^2a^3_1v_1}$ in $\Nu'$ and finally add $t$ to get a larger valid packing solution to reach at a contradiction.

  Overall, we reach at a contradiction in each case, hence such a triangle $t$ cannot exist.

\end{proof}

\subsection{Omitted proofs for \Cref{subsec:type-0}}
\label{subsec:proofs-type-0}

Similar to the way we proved \Cref{lem:main-type-1-3}, we will do an exhaustive case analysis here to prove  \Cref{lem:demanding-pairwise}.

\paragraph{Changes due to Simplification:}
All the triangles which were covered after the initial charge distribution and the {\em vanilla} Discharging-via-Chains construction are still covered if we replace the vanilla version with the {\em Satisfy-and-Truncate Discharging-via-Chains} construction. In addition to that if a triangle is adjacent to at least one of the type-$0$ triangles and at least one of the satisfied triangles in $\Nu$, then also it is covered.
Recall that, we maintain the {\em flexibility} (see~\Cref{obs:unsat-type-1-rotation,obs:unsat-type-3-rotation}) of credits for the tail triangles for chains in
$\Chi_{unsat}$ and the non-head triangles in $\Nu_3 \setminus head(\Chi)$. Any other triangle does not need this flexibility, since they are satisfied.

Hence we only need to look at all the {\em demanding triangles}
$\dset$, which may not be covered.

\begin{definition}[Demanding triangle]
\label{def:demanding-triangles}
Let $\dset \subseteq \tset$ be the set of all the non-solution triangles in $G$, such that any triangle in $\dset$:
\begin{enumerate}
    \item contains exactly one edge of type-$0$ triangle,
    \item all other solution edges are from one of the unsatisfied triangles in $tail(\Chi_{unsat}) \cup (\Nu_3 \setminus head(\Chi))$,
    \item does not contain any base-edge of any type-$1$ triangle in $\Nu_1$,
    \item does not contain any half non-solution edges of non-tail triangles, i.e., $\{e_0^1, e_0^2\} \cup \{e^1_i\}_{i \in [k-1]}$ of any $k$-sized chain $\cset \in \Chi$.

\end{enumerate}

\end{definition}
Note that any non-solution triangle adjacent to type-$0$ triangle not in $\dset$ is guaranteed to be covered.

Correspondingly, $\aset$ will be redefined as the set of \emph{free} solution triangles consisting of (1) all type-$0$ triangles in $\Nu_0$, (2) unsatisfied tail type-$1$ triangles in $tail(\Chi_{unsat})$, (3) type-$3$ triangles in $\Nu_3\setminus head(\Chi)$ which are not part of any chain. Let $\aset_i = \Nu_i \cap \aset$ be the set of type-$i$ triangles in $\aset$. Also, let $\aset_{13}:=\aset_1 \cup \aset_3$.

\subsubsection{Proof of \Cref{lem:demanding-pairwise}}
\label{subsubsec:proof-demanding-pairwise}

Let $\psi \in \Nu_0$ be a type-$0$ triangles such that $t, t' \in \dset(\psi)$ are any two arbitrary triangles. To prove this lemma, we need to show that $t, t'$ always share an edge.

Now we prove a structural lemma by using the chain structure which enables us to change the type of $t,t'$. Then we conclude that $t, t'$ should share an edge by proving three technical claims. Note that these changes are done only for the purpose of analysis and are not part of the algorithm.

Note that $type(t)$ and $type(t')$ can be any of five types namely type-$[0,1]$, type-$[0,3]$, type-$[0,1,1]$, type-$[0,1,3]$, and type-$[0,3,3]$.
Hence, if we directly try to prove this lemma, we will need to prove ${5 \choose 2} + 5 = 15$ claims
to prove that $t, t'$ share an edge, to cover all the combinations for types of $t$ and $t'$.

Fortunately, we can use the new structure of the chains in $\Chi_{unsat}$ (see~\Cref{obs:make-tail-type-3}) to alter the packing solution $\Nu$ into $\Nu'$ in such a way that any doubly-attached triangles of $\text{type-}[0,1]$ becomes a $\text{type}_{\Nu'}\text{-}[z, 3]$ triangle and; any hollow triangles of type-$[0,1,1]$ or type-$[0,1,3]$ becomes a $\text{type}_{\Nu'}\text{-}[z, 3, 3]$ triangle, where $z = type_{\Nu'}(\psi)$ after the swaps.
Moreover, the type of any type-$3$ triangles adjacent to $t, t'$ does not change. Also, the non-solution edges contained in $t, t'$ remains non-solution edges with respect to $\Nu'$. Note that the type of $\psi$ may no longer be type-$0$ after the swaps, hence $z$ appears in the type of $t, t'$. These changes also maintain the optimality of the alternate packing solution.
Hence in the end we just need to prove ~\Cref{clm:type-03-03,clm:type-03-033,clm:type-033-033} for any optimal packing solution to prove \Cref{lem:demanding-pairwise}.

\begin{lemma}
\label{lem:demanding-reduction}

Given an optimal packing $\Nu$, the set of chains $\Chi$ constructed by the Satisfy-and-Truncate Discharging-via-Chains construction and the set $\dset$ as defined in \Cref{def:demanding-triangles}. 
For any $\text{type}_{\Nu}\text{-}0$ triangle $\psi$, it is possible to perform a sequences of chain swaps to get an alternate packing solution $\Nu'$ such that for every demanding triangle $t \in \dset(\psi)$:
\begin{enumerate}
    \item If $t$ is a $\text{type}_{\Nu}
    [0, 1]$ triangle 
    then it becomes a $\text{type}_{\Nu'}\text{-}[z,3]$ triangle.
    \item If $t$ is a $\text{type}_{\Nu}
    [0,1,1]$ or $\text{type}_{\Nu}
    [0,1,3]$ triangle then it becomes a $\text{type}_{\Nu'}\text{-}[z,3,3]$ triangle.
\end{enumerate}
where $z = type_{\Nu'}(\psi)$ is the type of triangle $\psi$ after the swaps are performed.
\end{lemma}
\begin{proof}

  Let $t \in \dset(\psi)$ be any triangle which is of
  $\text{type}_{\Nu}\text{-}[0,1],
  \text{type}_{\Nu}\text{-}[0,1,1]$ or
  $\text{type}_{\Nu}\text{-}[0,1,3]$. 
  Let $\psi^1$ be any $\text{type}_{\Nu}\text{-}1$ triangle adjacent to any $t$.
  By definition of $\dset$,
  the edge $E(\psi^1) \cap E(t)$ must be a non-base edge.
  Moreover, $\psi^1$ is the tail triangle of an unsatisfied chain in $\Chi_{unsat}$. 
   This implies, every $\text{type}_{\Nu}\text{-}1$ triangle adjacent to any triangle $t \in \dset(\psi)$, is a tail triangle of a some unsatisfied chain in $\Chi_{unsat}$.
   Let $\Chi' \subseteq \Chi_{unsat}$ be the set of all these chains.
   We apply \Cref{obs:make-tail-type-3} to $\Chi'$ to get an alternate packing solution $\Nu'$. The observation implies that, all the $\text{type}_{\Nu}\text{-}1$ triangles adjacent to any $t \in \dset(\psi)$ become $\text{type}_{\Nu'}\text{-}3$ triangles.
   Also since any $\text{type}_{\Nu}\text{-}3$ triangle adjacent to any $t \in \dset(\psi)$ is in $\Nu_3 \setminus head(\Chi)$, hence it remains a $\text{type}_{\Nu'}\text{-}3$ triangle.
  
  The only thing which remains to be shown is that for any doubly-attached $\text{type}_{\Nu}\text{-}[0,1]$ triangle $t \in \dset(\psi)$, the non-solution edge in $t$ remains a non-solution edge in with respect to $\Nu'$. 
  This follows from the fact that every edge used 
  by the new solution $\Nu'$ given by
  \Cref{obs:make-tail-type-3} is a half-edge of some chain whereas by \Cref{def:demanding-triangles}, $t$ cannot contain any half non-solution edge.
  
\end{proof}



\paragraph{Proof of \Cref{lem:demanding-pairwise}}
Now assuming that the \Cref{clm:type-03-03,clm:type-033-033,clm:type-03-033}, we can show the proof of \Cref{lem:demanding-pairwise}.

Let $t, t' \in \dset(\psi)$ be two demanding triangles for some type-$0$ triangle $\psi$. Applying \Cref{lem:demanding-reduction}, we obtain an alternate optimal packing solution $\Nu'$ from $\Nu$ in such a way that, $type_{\Nu'}(t)$ and $type_{\Nu'}(t')$ are either $[z,3]$ or $[z,3,3]$ where $z = type_{\Nu'}(\psi)$.

By using one of the \Cref{clm:type-03-03,clm:type-033-033,clm:type-03-033} on $\Nu'$, $\psi$, $t$, and $t'$, we can conclude that $t$ and $t'$ must share an edge.

\begin{claim}
\label{clm:type-03-03}
Given an optimal packing solution $\Nu$. For any solution triangle $\psi$ of type-$z$, if there are two triangles $t,t'$ of type-$[z, 3]$ adjacent to $\psi$, then they share an edge.
\end{claim}
\begin{proof}
    Suppose not, then $E(t) \cap E(t') = \varnothing$. Let $\psi_\l$ and $\psi_r$ be the type-$3$ triangles adjacent to $t$ and $t'$ respectively.

    Let $S = \{\psi, \psi_\l, \psi_r\}$. In every case, we will construct $S'$ such that $\Nu \setminus S \cup S'$ is a valid triangle packing of size at least $|\Nu| + 1$, which contradicts the optimality of $\Nu$ and hence conclude that $t,t'$ must share an edge.

    \input{tikz_figures/demand-two-doubly-attached-one-type-3-common}

    Suppose $\psi^3 = \psi_\l = \psi_r$, then let $S = \{\psi, \psi_3\}$ and let $S' = \{t, t', \omega\}$ where $\omega$ is a singly-attached triangle of $\psi_3$ attached to the edge $E(\psi_3) \setminus (E(t) \cup E(t'))$. 
    Since $\psi$ and $\psi^3$ share a vertex, hence $anchor(\psi^3) \notin V(\psi)$. This implies that $\omega$ is disjoint from both $t$ and $t'$.
    
    Hence $\Nu \setminus S \cup S'$ is a triangle packing of size $|\Nu| + 1$, this is a contradiction (see \Cref{fig:demand-two-doubly-attached-one-type-3-common}).

    \input{tikz_figures/demand-two-doubly-attached-two-type-3}
    
    Hence $\psi_\l \neq \psi_r$.
    Let $v_\l = V(\psi) \cap V(\psi_\l)$ and $v_r = V(\psi) \cap V(\psi_r)$. It is possible that $v_\l = v_r$.
    Now, let $\omega_\l$ and $\omega_r$ be singly-attached triangles of $\psi_\l$ and $\psi_r$ such that $v_\l \notin V(\omega_\l)$ and $v_r \notin V(\omega_r)$ respectively.
    This implies that $V(\omega_\l) \cap V(\psi)=  V(\omega_r) \cap V(\psi) = \varnothing$.
    If $S' = \{t, t', \omega_\l, \omega_r\}$ is a set of four disjoint triangles, then by replacing $S$ with $S'$ in $\Nu$, we get a contradiction.
    Hence, at least two triangles in $S'$ must share an edge.
    $E(t) \cap E(t') = \varnothing$ since we assume that $t$ and $t'$ do not share any edge.
    The triangles $t$ and $\omega_\l$ cannot share an edge since any vertex of $\psi$ cannot be an anchoring vertex of $\psi_\l$.
    Similarly, $t'$ and $\omega_r$ cannot share an edge by the same reasoning.
    
    Let $e$ be the non-solution edges of $t$.
    Suppose $\omega_r$ and $t$ share an edge, then the edge that
    they share must be the non-solution edge $e$ since
    the solution edge of $\omega_r$ is in $E(\psi_r)$ and there is no solution edge of $\psi_r$ in $E(t)$.
    
    Hence, one of the vertices in $V(e)$ must be in $V(\psi_r)$ and the another vertex must be the anchoring vertex $a_r = anchor(\psi_r)$ of $\psi_r$.
    Since  $a_r \notin V(\psi)$, it must be that $v_r$ is $V(e) \cap V(\psi)$ and $a_r$ is $V(e) \cap V(\psi_\l)$.
    But this together with $e \in E(\omega_r)$ contradicts the fact that $V(\omega_r) \cap V(\psi) = \varnothing$.
    Hence $t$ and $\omega_r$ cannot share any edge.
    
    By symmetry, $\omega_\l$ and $t'$ cannot share any edge.
    
    The only remaining possibility is when $\omega_\l$ and $\omega_r$ share some non-solution edge.
    Let $a_r = anchor(\psi_r)$.
    There are two possibility,
    either $a_\l = a_r$ or $a_\l \in V(\psi_r)$ and $a_r \in V(\psi_\l)$.
    
    Suppose $a_\l = a_r$. In this case, for $\omega_\l$ and $\omega_r$ to share an edge, there should be common vertex $u =  V(\psi_\l) \cap V(\psi_r)$.
    Since $u \in V(\omega_\l) \cap V(\omega_r)$, hence $v_\l \neq u$ and $v_r \neq u$.
    Moreover, as any two solution triangles can share at most one vertex, which implies that $v_\l \neq v_r$.
    Notice that $\Delta_{v_\l v_ra_\l}$ is a singly-attached triangle of $\psi$.
    Hence, $z \in \{1, 3\}$.
    Since $t$ and $t'$ do not share any edge, $\overbar{v_\l v_r}$ belongs to at most one of $t$ and $t'$.
    WLOG, by renaming, let $t$ be the doubly-attached triangle such that $\overbar{v_\l v_r} \notin E(t)$.
    Notice that in this case, $t$ is either a type-$[3,3]$ triangle or a type-$[1,3]$ triangle where $E(t) \cap E(\Delta_{v_\l v_ra_\l}) = \varnothing$. Both cases cannot exist by \Cref{prop:structures-doubly-attached}.
    Hence, the case where $a_\l = a_r$ is not possible.
   
    Suppose $a_\l \in V(\psi_r)$ and $a_r \in V(\psi_\l)$.
    Notice that in this case, $V(\psi_\l) \cap V(\psi_r) = \varnothing$, since the edges from $a_\l$ to $V(\psi_r) \setminus a_\l$ are solution edges and edges from $a_\l$ to $V(\psi_\l)$ are non-solution edges.
    This implies that $v_\l \neq v_r$.
    There are two additional doubly-attached triangles, namely,
    $\Delta_{v_\l v_ra_\l}$ and $\Delta_{v_\l v_ra_r}$.
    Since $t$ and $t'$ do not share an edge, $\overbar{v_\l v_r}$ belongs to at most one of $t$ and $t'$.
    WLOG, by renaming, let $t$ be the doubly-attached triangle such that $\overbar{v_\l v_r} \notin E(t)$.
    This implies, that $V(e) \cap V(\psi) = v_x \notin \{v_\l,v_r\}$.
    If $V(e) \cap V(\psi_\l) \neq a_r$, then there are two disjoint doubly-attached triangles 
    $t$ and $\Delta_{v_\l v_ra_r}$
    adjacent to $\psi$ and $\psi_\l$, which is not possible by the argument
    used for $\psi_\l = \psi_r$ case.
    If $V(e) \cap V(\psi_\l) = a_r$, then
    $\Delta_{v_x v_ra_r}$ is a singly-attached triangle of $\psi$, hence $z \in \{1, 3\}$. This implies that,
    $\Delta_{v_\l v_ra_\l}$ is either type-$[3,3]$ triangle or type-$[1,3]$ triangle where $E(\Delta_{v_xv_ra_r}) \cap E(\Delta_{v_\l v_ra_\l}) = \varnothing$. Both cases cannot exist by \Cref{prop:structures-doubly-attached}.
    Hence, the case where $a_\l \neq a_r$ is not possible.

    Since every case leads to a contradiction, then the claim must be true.

\end{proof}

\begin{claim}
\label{clm:type-033-033}
Given an optimal packing solution $\Nu$. For any solution triangle $\psi$ of type $z$, if there are two triangles $t,t'$ of type-$[z, 3, 3]$ adjacent to $\psi$, then they share an edge.
\end{claim}
\begin{proof}
    Suppose not, then $E(t) \cap E(t') = \varnothing$. Let $e = E(t) \cap E(\psi)$ and $e' = E(t') \cap E(\psi)$ be two edges of $\psi$ shared with $t$ and $t'$, respectively. Let $v = V(e) \cap V(e')$ be the shared vertex between two edges. Let $\psi_\l$ and $\psi_{\l t}$ be the two type-$3$ triangles adjacent to $t$. We name them in such a way that $v \in V(\psi_{\l t})$. Similarly, we name $\psi_r$ and $\psi_{rt}$ to be the two type-$3$ triangles adjacent to $t'$
    such that $v \in V(\psi_{rt})$.
   
    The structure of the proof is similar to the proof of the claim above. Let $S = \{\psi, \psi_\l, \psi_{\l t}, \psi_r, \psi_{rt}\}$. In each case, we will try to construct $S'$ in such a way that $\Nu' = \Nu \setminus S \cup S'$ is a valid triangle packing that is larger than $\Nu$, which contradicts the optimality of $\Nu$.

    Now we name more vertices (see \Cref{fig:demand-two-hollow-two-type-0-3-3} for illustration).
    Let $v_\l = V(e) \setminus v$ and let $v_r = V(e') \setminus v$.
    Similarly, let $v_{\l t} = V(\psi_{\l t}) \cap V(t) \setminus V(\psi)$ and let $v_{rt} = V(\psi_{rt}) \cap V(t') \setminus V(\psi)$.
    We let $u_{\l\l} = V(\psi_\l) \setminus \{v_\l, v_{\l t}\}$, $u_{rr} = V(\psi_r) \setminus \{ v_r, v_{rt}\}$, $u_{\l t} = V(\psi_{\l t}) \setminus \{v_{\l t}, v\}$, and $u_{rt} = V(\psi_{rt}) \setminus \{v_{r t}, v\}$ be the third vertex of each triangle respectively.
    Also, let $a_\l = anchor(\psi_\l)$, $a_r = anchor(\psi_r)$, $a_{\l t} = anchor(\psi_{\l t})$, and $a_{rt} = anchor(\psi_{r t})$.

    Notice that $\psi_\l \neq \psi_r$ as they share different vertices with $\psi$. By similar reasoning, $\psi_\l \neq \psi_{rt}$ and $\psi_r \neq \psi_{lt}$.

    \input{tikz_figures/demand-two-hollow-two-type-0-3-3-psi_lt=psi_rt}
    
    It might still be the case that $\psi_{\l t} = \psi_{rt}$ (in this case, $v_{\l t} = u_{rt}$ and $v_{rt} = u_{\l t}$. See~\Cref{fig:demand-two-hollow-two-type-0-3-3-psi_lt=psi_rt}).
    
    If $\psi_{\l t} = \psi_{r t} = \psi_t$, then $S = \{ \psi, \psi_\l, \psi_r, \psi_{t}\}$.
    Let $\omega_\l, \omega_{t}, \omega_r$ be singly-attached triangles having base edge $\overbar{u_{\l\l} v_\l}, \overbar{v_{\l t} v_{r t}}, \overbar{v_r u_{rr}}$, respectively. If $S' = \{ t, t', \omega_\l, \omega_t, \omega_r \}$ is a set of five edge-disjoint triangles, then we are done. Otherwise, some triangles must share an edge. $t$ and $t'$ are disjoint by our assumption.
    $E(\{t, t'\}) \cap E(\{\omega_\l, \omega_t, \omega_r\}) = \varnothing$ as $E(\{t, t'\})$ consists of all solution edges and these solution edges do not overlap with base-edges of $\{\omega_\l, \omega_t, \omega_r\}$ by the way they are chosen.
    Moreover, $\omega_t$ cannot overlap with either of $\omega_\l$ or $\omega_r$ as
    their base-edges do not share any vertex
    and the anchoring vertex $a_t = anchor(\psi_t)$ cannot be in $V(\psi_\l) \cup V(\psi_r)$ (as $\psi_t$ shares one vertex with each of two triangles).
    The only remaining case is when $E(\omega_\l) \cap E(\omega_r) \neq \varnothing$.
    Also, $a_\l = v_{r t}$ (or the symmetric case $a_r = v_{\l t}$) is not possible since 
    $\overbar{v_{\l t} v_{rt}} \in E(\psi_t)$ and having $a_\l = v_{r t}$ (or $a_r = v_{\l t}$) would enforce this edge to be a non-solution edge, which is a contradiction.
    
    Hence, it must be that
    either $a_\l = u_{rr}$ and $a_r = u_{\l\l}$ (see~\Cref{subfig:a_l=u_rr-a_r=u_ll})
    or $u = u_{\l\l} = u_{rr}$ and $a = a_\l = a_r$ (see~\Cref{subfig:u_rr=u_ll-a_l=a_r}).
    
    In case when $a_\l = u_{rr}$ and $a_r = u_{\l\l}$,
    notice that the triangle $\Delta_{v_{\l t} v_{r t} u_{rr}}$
    is a doubly-attached triangle of type-$[3,3]$ which cannot exist in any optimal packing solution by \Cref{prop:structures-doubly-attached}.
    Hence, this is not possible.
    
    In the second case, when $u = u_{\l\l} = u_{rr}$ and $a = a_r = a_\l$, note that the triangle $\Delta_{v_\l v_r a}$ is a singly-attached triangle of $\psi$, hence $z \in \{1, 3\}$.
    In this case $t$ be the doubly-attached triangle such that $\overbar{v_\l v_r} \notin E(t)$.
    Notice that in this case, $t$ is either a type-$[1,3,3]$ triangle or a type-$[3,3,3]$ triangle where $E(t) \cap E(\Delta_{v_\l v_ra}) = \varnothing$. Both cases cannot exist by \Cref{prop:structures-hollow}.
    Hence, this case is also not possible.

    \input{tikz_figures/demand-two-hollow-two-type-0-3-3}
    
    Now we consider that case when $\psi_{\l t} \neq \psi_{r t}$ (see \Cref{fig:demand-two-hollow-two-type-0-3-3}). As a reminder, $S= \{\psi, \psi_\l, \psi_r, \psi_{\l t}, \psi_{r t} \}$ in this case.
    Let us first rule out the cases where some type-$3$ triangles share vertices not in $V(t) \cup V(t')$.
    For any two type-$3$ triangles that already share vertex, namely,
    $\{\psi_\l, \psi_{\l t}\}, \{\psi_{\l t}, \psi_{r t}\}, \{\psi_{r t}, \psi_r\}$, they cannot share any other vertex otherwise their solution edges will overlap.
    The only remaining cases are $u_{\l\l} = u_{r t}$, $u_{\l\l} = v_{r t}$, or $u_{\l\l} = u_{rr}$. 
    All other cases are symmetric to these three cases, so omitting them.
    Note that at a time only one such overlap can happen.
    If $u_{\l\l} = u_{r t}$, then the triangle $\Delta_{v_{\l t} v u_{r t}}$ is a type-$[3,3,3]$ triangle which cannot exist by \Cref{prop:structures-hollow}.
    Likewise, if $u_{\l\l} = v_{r t}$, then the type-$[3,3,3]$ triangle $\Delta_{v_{\l t} v v_{r t}}$ cannot exist by \Cref{prop:structures-hollow}.
    \input{tikz_figures/demand-two-hollow-two-type-0-3-3-psi_lt-neq-psi_rt-u_ll=u_rr.tex}
    Suppose $u = u_{\l\l} = u_{rr}$ 
    (see \Cref{fig:demand-two-hollow-two-type-0-3-3-psi_lt-neq-psi_rt-u_ll=u_rr}),
    then let
    $\omega_{\l t}, \omega_{r t}, \omega_{rr}$ be the singly-attached triangle based
    at $\overbar{v_{\l t} u_{\l t}}, \overbar{v u_{r t}}, \overbar{v_{r t} u}$ accordingly. Moreover, let $t'' = \Delta_{v_\l v_r u}$ be an additional type-$[*,3,3]$ triangle.
    If $S' = \{t, t', t'', \omega_{\l t}, \omega_{r t}, \omega_{rr}\}$ is a set
    of six disjoint triangles, then we are done as we can construct a triangle packing larger than $\Nu$ by replacing $S$ with $S'$ in $\Nu$.
    Otherwise, some of the triangles in $S'$  must overlap.
    By assumption $t, t'$ do not overlap.
    $t''$ cannot be overlapped with $t, t'$ as they have different solution edges.
    Similarly, $E(\{t, t', t''\}) \cap E(\{\omega_{\l t}, \omega_{r t}, \omega_{rr}\}) = \varnothing$ as they use different solution edges and $E(\{t, t', t''\})$ contains only solution edges.
    $\omega_{\l t}$ and $\omega_{r t}$ are disjoint as their base-edges are disjoint by choice and the anchoring vertex of $\psi_{\l t}$ cannot be in $V(\psi_{r t})$ and the anchoring vertex of $\psi_{r t}$ cannot be in $V(\psi_{\l t})$.
    By similar reasoning, $\omega_{r t}$ and $\omega_{rr}$ are disjoint.
    Suppose $\omega_{\l t}$ and $\omega_{rr}$ are not disjoint, then the anchoring vertex $a_r = anchor(\psi_r)$ is either $v_{\l t}$ or $u_{\l t}$. In any case, either $\Delta_{v_{r t} v v_{\l t}}$ or $\Delta_{v_{r t} v u_{\l t}}$ is a type-$[3,3]$ doubly-attached triangle, which cannot exist by \Cref{prop:structures-doubly-attached}.

    All in all, we conclude that, that $u_{\l\l}$ must not be any of $u_{r t}, v_{r t}, u_{rr}$. Symmetrically, $u_{rr}$ must not be any of $u_{\l t}, v_{\l t}, u_{\l\l}$.
    Hence the solution triangles look like in \Cref{fig:demand-two-hollow-two-type-0-3-3} such that $|V(\psi) \cup V(\psi_\l) \cup V(\psi_r) \cup V(\psi_{\l t}) \cup V(\psi_{r t})| = 9$.
    Let $\omega_{\l\l}, \omega_{\l t}, \omega_{r t}, \omega_{rr}$ by the singly-attached triangles based at
    $\overbar{u_{\l\l } v_\l}, \overbar{u_{\l t} v}, \overbar{u_{r t} v_{r t}}, \overbar{v_r u_{rr}}$, in respective order.
    If $S' = \{t, t', \omega_{\l\l}, \omega_{\l t}, \omega_{r t}, \omega_{rr}\}$ is
    a set of six disjoint triangles, then we are done by contradiction.
    Otherwise, some triangles in $S'$ must be overlapped.
    By similar argument as before, $t$ and $t'$ are disjoint and
    $E(\{t, t'\}) \cap E(\{\omega_{\l\l}, \omega_{\l t}, \omega_{r t}, \omega_{rr}\}) = \varnothing$.
    Also, any pair of singly-attached triangles 
    in $\{\omega_{\l\l}, \omega_{\l t}, \omega_{r t}, \omega_{rr}\}$ corresponding to any two type-$3$ triangles 
    $\psi',\psi'' \in \{\psi_\l, \psi_r, \psi_{\l t}, \psi_{rt}\}$, such that $V(\psi')\cap V(\psi'') \neq \varnothing$ 
    are disjoint by choice of their base-edges. In particular, $\{\omega_{\l\l}, \omega_{\l t}\}, \{\omega_{\l t}, \omega_{r t}\}, \{\omega_{r t}, \omega_{rr}\}$.
    If $\omega_{\l t}$ and $\omega_{rr}$ overlap, then
    the only way is that anchoring vertex $a_{\l t} = anchor(\psi_{\l t})$ must be $u_{rr}$ 
    and $a_r = u_{\l t}$.
    In this case, the triangle $\Delta_{v v_{rt} u_{rr}}$ is a type-$[3,3]$ triangle, which cannot exist by \Cref{prop:structures-doubly-attached}.
    The same argument works for $\omega_{r t}$ and $\omega_{\l\l}$
    , as symmetrically in this case $a_{r t} = anchor(\psi_{r t})$ must be $u_{\l\l}$, which is not possible.

    The only remaining case is when $\omega_{\l\l}$ and $\omega_{rr}$ overlap.
    This implies, $a_\l = anchor(\psi_\l) = u_{rr}$ and $a_r = anchor(\psi_r) = u_{\l\l}$.
    Let $\omega_{\l\l}'$ be the singly-attached triangle based at $\overbar{v_{\l t} u_{\l\l}}$ and let $t_d$ be the doubly-attached triangle $\Delta_{v_\l v_r u_{rr}}$.
    $\omega_{\l\l}'$ and $t_d$ are disjoint. Moreover, they will not be overlapped with $\omega_{\l t}, \omega_{r t}, t, t'$ by the choice of edges
    and by the fact we proved in previous case that $a_{\l t} \neq u_{rr}$.
    Let $S' = \{t, t', \omega_{\l t}, \omega_{r t}, \omega_{\l\l}', t_d\}$.
    By replacing $S$ with $S'$ in $\Nu$, we create a larger triangle packing, which is a contradiction.

    Hence we conclude that $t$ and $t'$ must share an edge.
\end{proof}

\begin{claim}
\label{clm:type-03-033}
Given an optimal packing solution $\Nu$. For any solution triangle $\psi$ of type $z$, if there are two triangles, $t$ of type-$[z, 3]$ and $t'$ of type-$[z, 3, 3]$ adjacent to $\psi$, then they share an edge.
\end{claim}
\begin{proof}
  
  Suppose not, then $E(t) \cap E(t') = \varnothing$. Let $e = E(t) \cap E(\psi)$ and $e' = E(t') \cap E(\psi)$. Let $v = V(e) \cap V(e')$ be the shared vertex between two edges. Let $\psi_\l$ be the type-$3$ triangle adjacent to $t$ and let $\psi_r, \psi_t$ be two type-$3$ triangles adjacent to $t'$.
  We name these triangles in such a way that $V(\psi) \cap V(\psi_t) = v$.

  The structure of the proof is similar to the two claims above. Let $S = \{ \psi, \psi_\l, \psi_t, \psi_r \}$ be solution triangles in our consideration. We are trying to construct $S'$ in such a way that $\Nu' = \Nu \setminus S \cup S'$ is a triangle packing that is larger than $\Nu$, which is a contradiction since $\Nu$ is an optimal triangle packing.

  Let's name more vertices. Let $v_\l = V(e) \setminus v$, let $v_r = V(e') \setminus v$, and let $v_t = V(t') \setminus V(\psi)$.
  Let $u_\l = V(\psi_\l) \setminus V(t), u_t = V(\psi_t) \setminus V(t'), u_r = V(\psi_r) \setminus V(t')$.
  Also, let $u = V(t) \setminus V(\psi)$.
  Now there are two cases in which $V(\psi_\l) \cap V(\psi)$ is either $v$ or $v_\l$.
 
  \input{tikz_figures/demand-one-doubly-attached-one-hollow-type-03-type-033-common-v_l-psi_l-psi.tex}

    First, let us consider the case when $V(\psi_\l) \cap V(\psi) = v_\l$ 
    (see \Cref{fig:demand-one-doubly-attached-one-hollow-type-03-type-033-common-v_l-psi_l-psi}).
   
    In this case, $\psi_\l \neq \psi_t$ and $\psi_\l \neq \psi_r$ since they share different vertices of $\psi$.
    Now we argue that they cannot share vertices (excepts $v_t$ that is already shared by $\psi_r$ and $\psi_t$).
    Since $V(\psi_r) \cap V(\psi_t) = v_t$, and since they are
    different solution triangles, this is the only vertex they share.
    Since $v_\l \in V(\psi_\l) \cap V(\psi)$,
    $\{u, u_\l\} \notin V(\psi)$.
    Since $v_\l \notin \{v_r, v_t\}$, $v_\l \notin V(\psi_r) \cup V(\psi_t)$.
    The only remaining cases are when $\{u, u_\l\} \cap \{u_t, v_t, u_r\} \neq \varnothing$.
    $u$ cannot be $u_t$ or $v_t$ since $\overbar{vu}$ is a non-solution edge and $\overbar{vu_t}$ (or $\overbar{vv_t}$) is a solution edge.
    Suppose $u = u_r$, then the $\Delta_{vv_tu_r}$ is a type-$[3,3]$ triangle, which cannot exist by \Cref{prop:structures-doubly-attached}.
    Hence $u \notin \{u_t, v_t, u_r\}$.
    Finally, we argue about $u_\l$.
    Suppose $u_\l = u_t$ or $u_\l = v_t$, then the triangle $\Delta_{u_\l vu}$ is a type-$[3,3]$ triangle, which again
    is not possible by \Cref{prop:structures-doubly-attached}.

    Suppose $u_\l = u_r$, then let $S' = \{t, t', \Delta_{v_\l v_r u_r}, \omega_t,\omega_\l \}$ where $\omega_t$ and $\omega_\l$ are singly-attached triangles of $\psi_t$ and $\psi_\l$ based at $\overbar{u_t v_t}$ and $\overbar{u u_\l}$ accordingly.
    Clearly, $\{t, t', \Delta_{v_\l v_r u_r}\}$ are disjoint, since they all use different solution edges. 
    Also, $\omega_t, \omega_\l$ are disjoint from $\{t', \Delta_{v_\l v_r u_r}\}$ because their solution edges are different.
    What remains to show is that $\omega_t,\omega_\l$ do not overlap. 
    If they do, then the only way it is possible is by having $a_t = u$ and $a_\l = u_t$, because other vertices belongs to some solution triangles having common vertex with $\psi_\l$ (or $\psi_t$) so cannot be $a_\l$ (or $a_t$) respectively. 
    But in this case, $\Delta_{u u_\l v_t}$ is a type-$[3,3]$ triangle, which cannot exist by \Cref{prop:structures-doubly-attached}.
    Hence, $\omega_t,\omega_\l$ are disjoint. 
    Moreover, $t$ is disjoint from both $\omega_t,\omega_\l$.
    To see this, we first notice that their solution edges are different by choice.
    Also, since $v$ cannot be $a_\l$, $\overbar{uv} \notin E(\omega_\l)$.
    For $\omega_t$, because the base-edge $\overbar{u_t v_t}$ do not share any vertex with $\overbar{uv}$, hence $\overbar{uv} \notin E(\omega_t)$.
    Hence $S'$ is a set of edge-disjoint triangles.
    Since $S'$ is a set of five disjoint triangles, by replacing
    $\Nu$ with $\Nu' = \Nu \setminus S \cup S'$ is a larger
    triangle packing, which is not possible by the optimality of
    $\Nu$.
    Now we conclude that $|V(\psi_\l)\cup V(\psi) \cup V(\psi_r) \cup V(\psi_t)| = 8$.

    Consider $S' = \{t, t', \omega_t, \omega_{\l}', \omega_r\}$ where $\omega_{\l}'$ is the singly-attached triangle of $\psi_\l$ based at $\overbar{u_\l v_\l}$, $\omega_r$ is the singly-attached triangle of $\psi_r$ based at $\overbar{v_r u_r}$ and $\omega_t$ is the singly-attached triangle based at $\overbar{u_t v_t}$.
    If $S'$ is a set of five disjoint triangles, then we are done by replacing $S$ with $S'$ in $\Nu$.
    If not, then two triangles share an edge.
    By our assumption, $t$ and $t'$ cannot share an edge.
    Also, $t$ and $\omega_t$ cannot share an edge since their solution edges are from different triangles and $
    V(\overbar{uv}) 
    \cap V(\overbar{u_t v_t}) = \varnothing$ which implies that their non-solution edges cannot be shared. 
    By similar reasoning, $t$ and $\omega_r$ cannot share an edge.
    Since $v$ cannot be an anchoring vertex of $\psi_\l$, it implies that $t$ and $\omega_\l'$ cannot share an edge .
    Clearly, $t'$ cannot share an edge with $\omega_t$ or $\omega_r$ or $\omega_\l'$ since they have different solution edges and $t'$ does not contain any non-solution edge.
    Also, $\omega_r$ and $\omega_t$ cannot share an edge since their solution edges are vertex-disjoint and their anchoring vertices cannot be in $V(\psi_t) \cup V(\psi_r)$.
    If $\omega_t$ and $\omega_\l'$ share an edge, then it has to be that $a_t = anchor(\psi_t) = u_\l$, but this is not possible since it would imply that
    $\Delta_{u u_\l v}$ is a singly-attached triangle of $\psi_\l$, which in turns imply that $\psi_\l$ has more than one anchoring vertices, which is impossible by \Cref{prop:structures-solution}. Note that $a_t$ cannot be $v_\l$ since $\overbar{v v_\l}$ is a solution edge.
    The only remaining case is when $\omega_\l'$ share an edge with $\omega_r$. This case is possible only when $a_r = anchor(\psi_r) = u_\l$ and $a_\l = anchor(\psi_\l) = u_r$.
    We let $S' = \{t, t', \omega_t, \omega_\l, \Delta_{v_\l v_r u_r}\}$ 
    (recall that $\omega_\l = \Delta_{u u_\l a_\l}$)
    be the set of five disjoint triangles. By replacing
    $S$ with $S'$ in $\Nu$, we get a larger triangle packing which
    is a contradiction.
    We conclude that the case where $V(\psi_\l) \cap V(\psi) = v_\l$ is not possible.

    \input{tikz_figures/demand-one-doubly-attached-one-hollow-type-03-type-033-common-v-psi_l-psi.tex}
    Now, we consider the other case where $V(\psi_\l) \cap V(\psi) = v$ 
    (see \Cref{fig:demand-one-doubly-attached-one-hollow-type-03-type-033-common-v-psi_l-psi}).
    Similar to the previous case,
    we first rule out the possibilities that $\psi_\l, \psi_r, \psi_t$ share vertices (except $v_t$).
    By similar argument, the only vertex shared between $\psi_r$ and $\psi_t$ is $v_t$. Also, the only vertex shared between $\psi_\l$ and $\psi_t$ is $v$.
    Since $\psi_r$ shares vertices with $\psi$ and $\psi_t$,
    the only remaining cases are when $u_r = u$ or $u_r = u_\l$.
    In both cases, the triangle $\Delta_{v v_t u_r}$ will become a
    type-$[3,3,3]$ triangle which cannot exist by \Cref{prop:structures-hollow}.
    Hence, we conclude that $|V(\psi) \cup V(\psi_\l) \cup V(\psi_r) \cup V(\psi_t)| = 8$, as shown in \Cref{fig:demand-one-doubly-attached-one-hollow-type-03-type-033-common-v-psi_l-psi}.
    Let $\omega_\l$, $\omega_t$, $\omega_r$ be the singly-attached triangles
    based at $\overbar{uu_\l}$, $\overbar{vu_t}$, $\overbar{v_r u_r}$ accordingly.
    We claim that $S' = \{t, t', \omega_\l, \omega_t, \omega_r\}$ is a set of five disjoint triangles.
    If this is the case, then by replacing $S$ with $S'$ in $\Nu$,
    we create a larger triangle packing solution which is a contradiction.
    By assumption, $E(t) \cap E(t') = \varnothing$ .
    By similar argument regarding solution edges and anchoring vertices,
    $E(t) \cup E(t')$ are disjoint from $E(\omega_\l) \cup E(\omega_t) \cup E(\omega_r)$.
    Since the anchoring vertex of $\psi_t$ cannot be in $V(\psi_r)$ and vice versa, and since solution edges of $\omega_t$ and $\omega_r$ are different, $\omega_t$ is disjoint from $\omega_r$.
    By similar argument, $\omega_\l$ is disjoint from $\omega_t$.
    To show that $\omega_\l$ is disjoint from $\omega_r$, we look at $a_\l = anchor(\psi_\l)$
    , as it has to be one of $v_t$, $v_r$, or $u_r$ for $\omega_\l$ and $\omega_r$ to share an edge. By a similar argument as before, $a_\l \neq v_t, v_r$.
    By \Cref{prop:structures-doubly-attached}, $a_\l$ cannot be $u_r$, otherwise $\Delta_{v v_t u_r}$ will be a type-$[3,3]$ triangle, which is impossible.
    Hence $a_\l \notin V(\psi_r)$ which together with the fact that $\omega_\l$ and $\omega_r$ have different solution edges, imply that they are disjoint.
    Overall, we conclude that $S'$ is a set of five disjoint triangles.

    Since all the cases lead to a contradiction, this implies that
    $t$ and $t'$ must share an edge.

\end{proof}

\subsection{Table of notations}
\label{subsec:notations}
\begin{table}[H]
\centering
\begin{tabular}{||p{3.5cm}||p{12cm}|} 
\hline
Notation & Description \\
\hline \hline 
$G$ & initial graph with a fixed drawing \\ \hline
$\tset$ & set of all triangles in $G$ \\ \hline
$\nu(G)$ & optimal triangle packing number for $G$ \\ \hline
$\tau(G)$ & optimal triangle covering number for $G$ \\ \hline
$\tau^*_k(G)$ & optimal $1/k$-fractional triangle covering number for $G$ \\ \hline
$\Nu \subseteq \tset$ & set of optimal triangle packing solution  \\ \hline
$\psi$ & any solution triangle $\psi \in \Nu$   \\ \hline
$\omega$ & any singly-attached triangle $\omega \in \tset \setminus \Nu$   \\ \hline
$CL(\psi)$ & set of triangles in $\tset \setminus \Nu$ sharing an edge with $\psi$ \\ \hline
$\CL_{sin}(\psi) \subseteq \CL(\psi)$ & singly-attached triangles sharing the unique solution edge with $\psi$ \\ \hline
$\CL_{dou}(\psi) \subseteq \CL(\psi)$ & doubly-attached triangles sharing one of the two solution edges with $\psi$  \\ \hline
$\CL_{hol}(\psi) \subseteq \CL(\psi)$ & hollow triangles sharing one of the three solution edges with $\psi$ \\ \hline
$anchor(t)$ & the vertex in $V(t) \setminus V(\psi)$ of a singly-attached triangle $t \in CL_{sin}(\psi)$  \\ \hline
$base(\psi) \subseteq E(\psi)$ & edges adjacent to any singly-attached triangle in $CL_{sin}(\psi)$  \\ \hline
$\text{type-}i$ & $type(\psi) = i$, if $|base(\psi)| = i$  \\ \hline
$\Nu_i \subseteq \Nu$ & set of type-$i$ triangle   \\ \hline
$\psi^i$ & any type-$i$ solution triangle in $\Nu_i$   \\ \hline
$\Nu_1^1 \subseteq \Nu_1$ & set of every type-$1$ triangle $\psi^1 \in \Nu_1$ such that $|CL_{sin}(\psi^1)| = 1 $   \\ \hline
$anchor(\psi)$ & $:= anchor(t)$ for $t \in CL_{sin}(t)$, if $CL_{sin}(\psi) = \{t\}$ or $type(\psi) = 3$ \\ \hline
$\text{type-}[i,j]$ & $type(t) = [i,j]$, if $t$ is a doubly-attached triangle adjacent to $\psi_1, \psi_2 \in \Nu$ such that $type(\psi_1) = i, type(\psi_2) = j$ for $i\leq j$ \\ \hline
$\text{type-}[i,j,k]$ & $type(t) = [i,j,k]$, if $t$ is hollow triangle adjacent to $\psi_1, \psi_2, \psi_3 \in \Nu$ such that $type(\psi_1) = i, type(\psi_2) = j, type(\psi_3) = k$ for $i \leq j \leq k$ \\ \hline
null-edge & any edge $e \in G$, such that $f(e)=0$ \\ \hline
half-edge & any edge $e \in G$, such that $f(e)=\frac{1}{2}$ \\ \hline
full-edge & any edge $e \in G$, such that $f(e)=1$ \\ \hline
$lend$ & lend relation in $ \subseteq \Nu_1 \times (\Nu_1 \cup \Nu_3) $ \\ \hline
$(\psi^1, \psi) \in lend$ & if $|CL_{sin}(\psi^1)| = 1$, $|V(\psi^1) \cap V(\psi)| = 1$, and $anchor(\psi^1) \in V(\psi)$  \\ \hline
twin-doubly attached triangles& $t, t' \in CL_{dou}(\psi^1) \cap CL_{dou}(\psi)$ such that $(\psi^1, \psi) \in lend$ and $anchor(\psi^1) \in V(t) \cap V(t')$ \\ \hline
$gain(\psi^1, \psi)$ & $:= E(t) \cap E(\psi) = E(t) \cap E(t')$, where $t,t'$ are twin-doubly attached triangles corresponding to $(\psi^1, \psi) \in lend$ \\ \hline
$k$-sized chain $\cset \in \Chi$ & A chain $\cset = \{\psi_k, \psi_{k-1}, \dots, \psi_0\}$ \\ \hline
$\Chi_{sat}$,$\Chi_{unsat}$ & Set of satisfied/unsatisfied chains\\ \hline
$\Chi^3_{sat} = \Nu_3 \cap \Chi_{sat}$ & Set of $0$-sized satisfied chains\\  \hline
$\Chi = \Chi_{unsat} \cup \Chi_{sat} $ & Set of all the chains \\ \hline
$\dset$ & Set of {\em demanding} triangles covered in our Discharge-and-Pin algorithm\\  \hline
$\aset$ & Set of solution triangles whose credits we {\em discharge} or {\em pin} to cover $\dset$ \\  \hline

\hline
\end{tabular}
\caption{Table of notations}
\label{tab:notations}
\end{table}

\subsubsection{Table of chains notations}
\label{subsubsec:chain-notations}
\begin{table}[H]
\centering
\begin{tabular}{||p{2.5cm}||p{13cm}|} 
\hline
Notation & Description \\
\hline \hline 
$tail(\cset)$ & The last type-$1$ triangle $\psi_k$ in $\cset$\\ \hline
$head(\cset)$ & The type-$3$ triangle $\psi_0$ in $\cset$.\\ \hline
$tail(\Chi')$ & For any $\Chi' \subseteq \Chi$, set of all the tail triangles.\\ \hline
$head(\Chi')$ & For any $\Chi' \subseteq \Chi$, set of all the head triangles.\\ \hline
$\omega_i$ & The singly-attached triangle of $\psi_i$ for $i>0$ \\ \hline
$\omega_0^1, \omega_0^2,\omega_0^3$ & The three singly-attached triangles of $\psi_0$ \\ \hline
$\omega_0 = \omega^3_0$ & The singly-attached triangles of $\psi_0$ such that $V(\psi_0 \cap \psi_1) \notin V(\omega_0)$ \\ \hline
$b_i$ & The base edge $base(\psi_i)$ for $i>0$ \\ \hline
$a_i$ & $:= anchor(\psi_i) \in V(\psi_{i-1})$ for $i\geq 0$ \\ \hline
$g_i$ & The gain edge $gain(\psi_i, \psi_{i-1}) \in E(\psi_{i-1})$ for $i>0$ \\ \hline
$g_{k+1}$ & The null-non-base edge of $\psi_k$ \\ \hline
$h_i$ & $:= E(\psi_i) \setminus \{b_i, g_{i+1}\}$ for $i>0$ \\ \hline
$c_i$ & The common vertex $\psi_i \cap \psi_{i-1}$ for $i>0$ \\ \hline
$t^1_i, t^2_i$ & twin-doubly attached triangles corresponding to $(\psi_i, \psi_{i-1}) \in lend$ for $i>0$, such that $g_{i+1} \in E(t^2_i)$\\ \hline
$e^1_i, e^2_i$ & The two non-solution edges of $\omega_i$ for $i>0$, such that $e^j_i \in t^j_i$ for $j \in \{1, 2\}$  \\ \hline
\hline
\end{tabular}
\caption{Table of notations for a $k$-sized chain $\cset = \{\psi_k, \psi_{k-1}, \dots, \psi_0\}$}
\label{tab:chain-notations}
\end{table}

\fi

\end{document}


\maketitle

\section{Multi-transversals of order two}
\label{sec:order-2}
The demanding cases.
   \input{tikz_figures/blue-0-3-demanding}
    \input{tikz_figures/blue-0-1-demanding}
    \input{tikz_figures/hollow-0-3-3-demanding}
    \input{tikz_figures/hollow-0-1-3-demanding}
    \input{tikz_figures/hollow-0-1-1-demanding}
    
\begin{lemma}[Demand Lemma]
\label{lem:demanding-lemma}
For any triangle $\psi \in \Nu_0$ of type-$0$, there exists at most one demanding edge in $E(\psi)$.
\end{lemma}

\clearpage

\section{Omitted Claims from Section~\ref{sec:order-2}}

Let us first define some notations that will be used throughout the proof.
First, for any triangle $t$, as in other sections, we let $V(t)$ be set of vertices of this triangle.
Also, for set of triangles $T$, we let $V(T)$ be set of vertices of these triangles.
We might abuse notation and use $t$ as $\{t\}$. For example, for two triangles
 $t_1$ and $t_2$, $V(t_1 \cup t_2) = V(t_1) \cup V(t_2)$ is the set of vertices of these two triangles.

\subsection{Proof of Lemma~\ref{lem:demanding-lemma}}

\begin{proof}
   First, as mentioned in the distribution scheme, there are five local structures that we can see when a triangle has a demanding edge. Let $\psi$ be the type-$0$ triangle. Let $t$ be the triangle next to $\psi$ where the demand is needed. Then $t$ is either
   \begin{enumerate}
      \item \cat{type-$[0,3]$ doubly-attached triangle} See fig~\ref{fig:blue-0,3-demanding}.
      \item \cat{type-$[0,1]$ doubly-attached triangle} In this case, the edge shared with the type-$1$ triangle must be drained edge.
      See fig~\ref{fig:blue-0,1-demanding}.
      \item \cat{type-$[0,3,3]$ hollow triangle} See fig~\ref{fig:hollow-0,3,3-demanding}.
      \item \cat{type-$[0,1,3]$ hollow triangle} As in the previous case, the edge shared with the type-$1$ triangle must be poor edge. See fig~\ref{fig:hollow-0,1,3-demanding}.
      \item \cat{type-$[0,1,1]$ hollow triangle} the two other solution edges of $t$ are poor edges. See fig~\ref{fig:hollow-0,1,1-demanding}. Note that the type-$3$ triangles which we see in the figure cannot be merged into one triangle as $\psi$ will no longer be type-$0$.
    \end{enumerate}

    \begin{claim}
    \label{clm:disjoint-type3}
      For any of the above cases, all the type-$3$ triangles involved in either the doubly attached or hollow triangle directly or making base edge of some type-$1$ poor are disjoint.
   \end{claim}
   \begin{proof}
      Either it is not possible to draw the case with common type-$3$ triangle or it violates the fact that 
     $|\CL_{sin}(\psi)| = 0$.
      \input{tikz_figures/hollow-0-1-1-with-itself}
   \end{proof}

    We will show that it is impossible to have a type-$0$ triangle with two demanding edges. For contradiction suppose the lemma is not true, then we have a type-$0$ triangle $\psi$ with two demanding edges $e_{\l}$ and $e_r$. Think of $e_{\l}$ as the {\em left} demanding edge and $e_r$ as the {\em right} demanding edge. We will follow this naming convention in the proof and the figures to help visualize various cases. 
    
    On each side, we will see one of five cases above.

    One might categorize these structures by the types of triangles adjacent to $\psi$, which are combinations of doubly-attached and hollow triangles. We name these two triangles $t_\l$ and $t_r$. Remember that both $t_\l$ and $t_r$ demand credit from $\psi$ (so that $\psi$ has two demanding edges).

    \begin{lemma}
      \label{lem:both-doubly-attached-demanding}
      If $t_\l$ is a doubly-attached triangle, then $t_r$ cannot be a doubly-attached triangle.
    \end{lemma}

    \begin{proof}
      Suppose not, then $t_\l$ and $t_r$ are doubly-attached triangles sharing demanding edges $e_\l \neq e_r$ with $\psi$. Let $\psi_\l$ be the other solution triangle that share an edge with $t_\l$ and $\psi_r$ be the other solution triangle that share an edge with $t_r$. Let $v_\l$ be the common vertex of $\psi$ and $\psi_\l$. Let $v_r$ be the common vertex of $\psi$ and $\psi_r$ (possibly $v_r = v_\l$). Similarly, let $u_\l$ to be the other vertex of $\psi$ in $t_\l$ and $u_r$ to be the other vertex of $\psi$ in $t_r$ (possibly $u_r = u_\l$). Note that either $v_r = v_\l$ or $u_r = u_\l$ but not both. 
      If $u_r = u_\l$ and $v_r \neq v_\l$, we call $u_r = u_\l$ as $u$ else we call $v_r = v_\l$ as $v$. Note that $e_\l = \overbar{u_\l v_\l}$ and $e_r = \overbar{u_r v_r}$ in both the cases. Let $w_\l$ be the second common vertex of $\psi_\l$ and $t_\l$ and; $w_r$ be the the second common vertex of $\psi_r$ and $t_r$. Finally, let $v_{\l\l}$ be the third vertex of $\psi_\l$ and $v_{rr}$ be the third vertex of $\psi_r$ (see fig~\ref{fig:demand-two-doubly-attached-two-type-3}).
      Up to renaming $\psi_\l$ and $\psi_r$, there are three cases.
      \begin{enumerate}
         \item \cat{$\psi_\l$ and $\psi_r$ are type-$3$}
         There are several sub-cases here.
         
         In case when $u_r \neq u_\l$ and $v$ is a common vertex of $\psi, \psi_\l, \psi_r$, then it could be that $\psi_\l = \psi_r$ (say $\psi^3$). Then $v_{rr} = w_\l$ and $v_{\l\l} = w_r$ (see fig~\ref{fig:demand-two-doubly-attached-one-type-3-common}). But this is impossible because there is a solution improving $2$-swap by replacing the set of triangles $\sset=\{\psi, \psi^3\}$ by the bigger set $\sset'= \{t_\l, t_r, t'\}$,
         where $t'$ is the singly-attached triangles of $\psi^3$  attached to edge $E(\psi) \setminus E(t_l \cup t_r)$.
         
         \input{tikz_figures/demand-two-doubly-attached-one-type-3-common}

         In the case when $u$ is a common vertex and $v_r \neq v_\l$, it is impossible that $\psi_\l = \psi_r$ because $\overbar{v_\l v_r}$ is in $E(\psi)$. 
        
         Now both the cases when $\psi_\l \neq \psi_r$ are impossible because there is a solution improving $3$-swap by replacing the set of triangles $\sset=\{\psi, \psi_\l, \psi_r\}$ by the bigger set $\sset'= \{t'_\l, t_\l, t_r, t'_r\}$,
         where $t'_\l$ and $t'_r$ are the singly-attached triangles of $\psi_\l$ and $\psi_r$ attached to edge $\overbar{w_\l v_{\l\l}}$ and $\overbar{w_r v_{rr}}$ respectively. It is easy to that the triangles in $\sset'$ are edge disjoint even in the worst case when $anchor(\psi_\l) = anchor(\psi_r)$. For illustration see fig~\ref{fig:demand-two-doubly-attached-two-type-3}.

         \input{tikz_figures/demand-two-doubly-attached-two-type-3}

         \item \cat{$\psi_\l$ is type-$3$ and $\psi_r$ is type-$1$}
         Since $\psi_r$ is type-$1$, the edge $e_r = \overbar{v_r w_r}$ shared between $t_r$ and $\psi_r$ must be a poor base edge. This also indicates that there exists a doubly-attached triangle $t_{rr}$ which share an edge with $\psi_r$, a type-$3$ triangle $\psi_{rr}$ and non-solution edge of the only singly attached-attached triangle of $\psi_r$. Note that, $v_{rr}$ will be the common vertex between $\psi_r$ and $\psi_{rr}$. Let $w_{rr}$ be the other vertex of $\psi_{rr}$ contained in $t_{rr}$ and $x_{rr}$ be the third vertex of $\psi_{rr}$ (see fig~\ref{fig:demand-two-doubly-attached-one-type-3-one-type-1}). If $\psi_\l \neq \psi_{rr}$, then in both cases when $v_\l=v_r=v$ or when $u_\l=u_r=u$ are impossible since there exists a solution improving $4$-swap by replacing the set of triangles $\sset=\{\psi, \psi_\l, \psi_r, \psi_{rr}\}$ by set $\sset'= \{t'_\l, t_\l, t_r, t_{rr}, t'_r\}$, where $t'_\l$ and $t'_r$ are the singly-attached triangles of $\psi_\l$ and $\psi_{rr}$ attached to edge $\overbar{w_\l v_{\l\l}}$ and  $\overbar{w_{rr} x_{rr}}$ respectively. It is easy to that the triangles in $\sset'$ are edge disjoint even in the worst case when $anchor(\psi_\l) = anchor(\psi_{rr})$. For illustration see fig~\ref{fig:demand-two-doubly-attached-one-type-3-one-type-1}).

         \input{tikz_figures/demand-two-doubly-attached-one-type-3-one-type-1}

         Otherwise, $\psi_\l = \psi_{rr}$ (say $\psi^3$). 
         In this case, $v_\l$ and $v_r$ cannot be the same since then for $\psi_r$ the $anchor(\psi_r)$ needs to be either $w_\l$ or $v_{\l\l}$ and; the base edge needs to be $\overbar{v w_r}$, but the edges $\overbar{v w_\l}$ and $\overbar{v v_{\l\l}}$ both are solution edges of $\psi_\l$ hence cannot be non-solution edges for the singly-attached triangles of $\psi_r$.
         
         Now we need to deal with the case when $\psi_\l = \psi_{rr} = \psi^3$ and $u_\l=u_r=u$.
         In this case, $\psi^3$ share the vertex $v_\l$ with $\psi$ to create $t_\l$ and $v_{rr}$ vertex with $\psi_r$ since $e_r=\overbar{v_rw_r}$ is the poor base edge of $\psi_r$. Note that the third vertex of $\psi^3$ will be $w_\l = w_{rr}$ (say $w$). Now there will be non-solution edge $uw$ to complete the doubly-attached $t_\l$ and; edges $v_rw$ and $w_rw$ to complete the singly attached triangle of $\psi_r$. But it is impossible since $\psi$ cannot support any triangle like $\Delta_{uwv_r}$. For illustration see fig~\ref{fig:demand-two-doubly-attached-one-type-3-one-type-1}.
         
         \begin{figure}[ht]
    \centering
      \subfloat{
\begin{tikzpicture}[scale=0.6, every node/.style={scale=0.8}]
\draw [fill=lightgray] (0,0) -- (1.5,2) -- (3,0) -- (0,0);
\draw [fill=lightgray]  (3,0) -- (6,0) -- (4.5,2) -- (3,0);
\draw [line width=2pt, fill=lightgray] (0,0) to [out=-20,in=200] (6,0) -- (3, -2) -- (0,0) ;
\GraphInit[vstyle=Normal]
\tikzset{VertexStyle/.style = {shape=circle, fill=white, draw=black, inner sep=1pt, minimum size=20pt}}

\Vertex[Math,L=v_\ell]{a}
\Vertex[Math,L=u, y=2,x=1.5]{b}
\Vertex[Math,L=v_r,y=0,x=3]{c}
\Vertex[Math,L=w_r,y=2,x=4.5]{d}
\Vertex[Math,L=v_{rr},y=0,x=6]{e}
\Vertex[Math,L=w, y=-2,x=3]{f}

\SetUpEdge[
    lw = 2pt,
    color = black,
]
\Edges(a,b,c,a)
\Edges(c,d,e,c)
\SetUpEdge[
    lw = 1pt,
    color = black,
    style = dashed
]
\Edge(b)(d)
\Edge(d)(f)
\Edge(a)(f)
\tikzstyle{EdgeStyle}=[bend right=20]
\Edge(b)(f)
\Edge(c)(f)
\end{tikzpicture}
}

    \caption{$\psi=\Delta^0_{v_\ell v_r u}$ is type-$0$ and $\psi_r= \Delta^1_{v_r w_r v_{rr}}$ is type-$1$ and $\psi^3=\psi_\ell = \psi_{rr} = \Delta^3_{v_\ell v_{rr} w}$ is type-$3$.  
    The three doubly-attached triangles are 
    $t_\ell = \Delta_{v_\ell u w}$, $t_r = \Delta_{v_r w_r v_{rr}}$ and $t_{rr} = \Delta_{v_{rr} w_r w}$.
    \textbf{Impossible because $\psi$ cannot support any triangle like $\Delta_{u v_r w}$.}}
    \label{fig:demand-two-doubly-attached-one-type-3-one-type-1-common-type-3}
 \end{figure}







         
         \item \cat{$\psi_\l$ and $\psi_r$ are type-$1$} In this case, we name the triangles $t_{rr}$ and $\psi_{rr}$ and their vertices as in the previous case. We also name $\psi_{\l \l}$ and $t_{\l \l}$ similarly (see fig~\ref{fig:demand-two-doubly-attached-two-type-1}). Note that $\psi_\l$ and $\psi_r$ cannot be the same as $t\l$ and $t_r$ need to be adjacent to the poor base edges of $\psi_\l$ and $\psi_r$ respectively.
         If $\psi_{\l \l} \neq \psi_{rr}$, then in both cases when $v_\l=v_r=v$ or when $u_\l=u_r=u$ are impossible since there exists a solution improving $5$-swap by replacing the set of triangles $\sset=\{\psi, \psi_\l, \psi_{\l\l}, \psi_r, \psi_{rr}\}$ by set $\sset'= \{t'_\l, t_{\l\l}, t_\l, t_r, t_{rr}, t'_r\}$, where $t'_\l$ and $t'_r$ are the singly-attached triangles of $\psi_{\l \l}$ and $\psi_{rr}$ attached to edge $\overbar{w_{\l \l} x_{\l\l}}$ and  $\overbar{w_{rr} x_{rr}}$ respectively. It is easy to that the triangles in $\sset'$ are edge disjoint even in the worst case when $anchor(\psi_{\l \l}) = anchor(\psi_{rr})$. For illustration see fig~\ref{fig:demand-two-doubly-attached-two-type-1}).
         
        \input{tikz_figures/demand-two-doubly-attached-two-type-1}
            
         Otherwise $\psi_{\l \l} = \psi_{rr} =\psi^3$. Note that $\psi^3$ must contain the vertices $v_{\l\l}$ and $v_{rr}$ to make the edges $\overbar{v_\l w_\l}$ and $\overbar{v_r w_r}$ poor base edges for triangles $\psi_\l$ and $\psi_r$ respectively. It should also contain $anchor(\psi_\l)$ and $anchor(\psi_r)$. Let the third vertex of $\psi^3$ be $w$. 
         
         In case when $v_\l = v_r = v$, the anchoring vertex for the singly attached triangles of $\psi_\l$ and $\psi_r$ both has to be $w$, because $\overbar{v w_\l}$ and $\overbar{v w_r}$ are suppose to be the poor base edges and $vv_{rr}$ and $vv_{\l\l}$ are already solution edges. This also implies that $w \notin \{v, u_\l, u_r, w_\l, w_r\}$. But this case is impossible since there exists a solution improving $4$-swap by replacing the set of triangles $\sset=\{\psi, \psi_\l, \psi_r, \psi^3\}$ by set $\sset'= \{t_{\l\l}, t_\l, t_r, t_{rr}, t'\}$, where $t'$ is the singly-attached triangles of $\psi^3$ attached to edge $\overbar{v_{\l \l} v_{rr}}$. It is easy to that the triangles in $\sset'$ are edge disjoint even in the worst case when $anchor(\psi^3)$ is common with one of the only possible vertices out of $\{u_\l, u_r\}$. For illustration see fig~\ref{fig:demand-two-doubly-attached-two-type-1-common-type-3-v}.
         
        \input{tikz_figures/demand-two-doubly-attached-two-type-1-common-type-3-v}
            
         The only case left is when $u_\l=u_r=u$ given that $\psi_{\l \l} = \psi_{rr} =\psi^3$. In this case the third vertex $w$ cannot be any of the already defined vertices in $\{v_\l, v_r, w_\l, w_r\}$ as $\psi^3$ should be edge disjoint from other solution triangles in $\{\psi, \psi_\l, \psi_r\}$. In this case, both the anchoring vertex for the singly attached triangles of $\psi_\l$ and $\psi_r$ cannot be $w$, otherwise the edge $\overbar{v_\l w}$ and $\overbar{v_r w}$ will lead to an attachment for $\psi$ which is not possible as it is type-$0$. For illustration see fig~\ref{fig:demand-two-doubly-attached-two-type-1-common-type-3-attachment}.

        \begin{figure}[ht]
   \centering
   \subfloat{
\begin{tikzpicture}[scale=0.6, every node/.style={scale=0.8}]
\draw [fill=lightgray] (0,0) -- (1.5,2) -- (3,0) -- (0,0);
\draw [fill=lightgray]  (3,0) -- (6,0) -- (4.5,2) -- (3,0);
\draw [fill=lightgray] (6,0) -- (7.5,2) -- (9,0) -- (6,0);
\draw [line width=2pt, fill=lightgray] (0,0) to [out=-20,in=200] (9,0) -- (4.5, -2) -- (0,0) ;
\GraphInit[vstyle=Normal]
\tikzset{VertexStyle/.style = {shape=circle, fill=white, draw=black, inner sep=1pt, minimum size=20pt}}

\Vertex[Math, L=v_{\ell\ell}]{a}
\Vertex[Math, L=w_\ell, y=2,x=1.5]{b}
\Vertex[Math, L=v_\ell, y=0,x=3]{c}
\Vertex[Math, L=u ,y=2,x=4.5]{d}
\Vertex[Math, L=v_r, y=0,x=6]{e}
\Vertex[Math, L=w_r, y=2,x=7.5]{f}
\Vertex[Math, L=v_{rr}, y=0,x=9]{g}
\Vertex[Math, L=w, y=-2,x=4.5]{h}

\SetUpEdge[
   lw = 2pt,
   color = black,
]
\Edges(a,b,c,a)
\Edges(c,d,e,c)
\Edges(e,f,g,e)
\SetUpEdge[
   lw = 1pt,
   color = black,
   style = dashed
]
\Edge(b)(d)
\Edge(d)(f)
\Edge(c)(h)
\Edge(e)(h)
\tikzstyle{EdgeStyle}=[bend left=70]
\Edge(h)(b)
\tikzstyle{EdgeStyle}=[bend right=70]
\Edge(h)(f)
\end{tikzpicture}
   }
    \caption{$\psi^3=\psi_{\ell\ell}=\psi_{rr}=\Delta^3_{v_{\ell\ell} v_{rr} w}$ is type-$3$; $\psi=\Delta^0_{v_\ell v_r u}$ is type-$0$ and; $\psi_\ell = \Delta^1_{v_\ell w_\ell v_{\ell\ell}}$ and $\psi_r = \Delta^1_{v_r w_r v_{rr}}$ are type-$1$. The poor base edges of $\Delta^1_{v_\ell w_\ell v_{\ell\ell}}$ and $\Delta^1_{v_r w_r v_{rr}}$ are adjacent to doubly-attached triangles $t_\ell = \Delta_{v_\ell u w_\ell}$ and $t_r = \Delta_{v_r u w_r}$ respectively. Since $\psi^3$ is responsible to make the two type-$1$ triangles' base edge poor using $w$ as anchor, hence there exists the non-solution edges $\overbar{v_\ell w}$ and $\overbar{v_r w}$. \textbf{Impossible because $\psi$ cannot support any triangle like $\Delta_{v_\ell v_r w}$.}}
   
   \label{fig:demand-two-doubly-attached-two-type-1-common-type-3-attachment}
\end{figure}

         So either $anchor(\psi_\l) = v_{rr}$ or $anchor(\psi_r) = v_{\l\l}$ (or both). 
         
         First let us argue for the case when one of the anchoring vertex is $w$. By renaming, we can assume $anchor(\psi_\l) = w$, which implies $anchor(\psi_r) = v_{\l\l}$ and $w$ cannot be same as $u$. But this case is impossible since there exists a solution improving $4$-swap by replacing the set of triangles $\sset=\{\psi, \psi_\l, \psi_r, \psi^3\}$ by set $\sset'= \{t_{\l\l}, t_\l, t_r, t_{rr}, t'\}$, where $t'$ is the singly-attached triangles of $\psi^3$ attached to edge $\overbar{w v_{rr}}$. It is easy to that the triangles in $\sset'$ are edge disjoint. For illustration see fig~\ref{fig:demand-two-doubly-attached-two-type-1-common-type-3-anchor-w}.

        \begin{figure}[ht]
   \centering
   \subfloat{
\begin{tikzpicture}[scale=0.6, every node/.style={scale=0.8}]
\draw [fill=lightgray] (0,0) -- (1.5,2) -- (3,0) -- (0,0);
\draw [fill=lightgray]  (3,0) -- (6,0) -- (4.5,2) -- (3,0);
\draw [fill=lightgray] (6,0) -- (7.5,2) -- (9,0) -- (6,0);
\draw [line width=2pt, fill=lightgray] (0,0) to [out=-20,in=200] (9,0) -- (4.5, -2) -- (0,0);
\GraphInit[vstyle=Normal]
\tikzset{VertexStyle/.style = {shape=circle, fill=white, draw=black, inner sep=1pt, minimum size=20pt}}

\Vertex[Math, L=v_{\ell\ell}]{a}
\Vertex[Math, L=w_\ell, y=2,x=1.5]{b}
\Vertex[Math, L=v_\ell, y=0,x=3]{c}
\Vertex[Math, L=u ,y=2,x=4.5]{d}
\Vertex[Math, L=v_r, y=0,x=6]{e}
\Vertex[Math, L=w_r, y=2,x=7.5]{f}
\Vertex[Math, L=v_{rr}, y=0,x=9]{g}
\Vertex[Math, L=w, y=-2,x=4.5]{h}
\Vertex[Math, L=x, y=-4,x=4.5]{x}

\SetUpEdge[
   lw = 2pt,
   color = black,
]
\Edges(a,b,c,a)
\Edges(c,d,e,c)
\Edges(e,f,g,e)
\SetUpEdge[
   lw = 1pt,
   color = black,
   style = dashed
]
\Edge(b)(d)
\Edge(d)(f)
\Edge(a)(f)
\Edge(c)(h)
\Edge(h)(x)
\Edge(a)(x)
\Edge(g)(x)
\tikzstyle{EdgeStyle}=[bend right=70]
\Edge(b)(h)
\tikzstyle{EdgeStyle}=[bend right=15]
\Edge(a)(e)
\end{tikzpicture}
}
    \caption{$\psi^3=\psi_{\ell\ell}=\psi_{rr}=\Delta^3_{v_{\ell\ell} v_{rr} w}$ is type-$3$; $\psi=\Delta^0_{v_\ell v_r u}$ is type-$0$ and; $\psi_\ell = \Delta^1_{v_\ell w_\ell v_{\ell\ell}}$ and $\psi_r = \Delta^1_{v_r w_r v_{rr}}$ are type-$1$. The poor base edges of $\Delta^1_{v_\ell w_\ell v_{\ell\ell}}$ and $\Delta^1_{v_r w_r v_{rr}}$ are adjacent to doubly-attached triangles $t_\ell = \Delta_{v_\ell u w_\ell}$ and $t_r = \Delta_{v_r u w_r}$ respectively. The common type-$3$ triangle $\Delta^3_{v_{\ell\ell} v_{rr} w}$ is the one responsible for making the two type-$1$ triangles' $\psi_\ell$ and $\psi_r$ base edge poor using vertices $w$ and $v_{\ell\ell}$ as anchors respectively. The other two doubly-attached triangles are  $t_{\ell\ell} = \Delta_{v_{\ell\ell} w_\ell w}$ and $t_{rr} = \Delta_{v_{rr} w_r v_{\ell \ell}}$. The singly attached triangle of $\psi^3$ used for the swap is $t' = \Delta_{w v_{rr} x}$.
   \textbf{Impossible because there is an improving $4$-swap.}}
   \label{fig:demand-two-doubly-attached-two-type-1-common-type-3-anchor-w}
\end{figure}

        Finally the case left to argue is when $u_\l=u_r=u$ given that $\psi_{\l \l} = \psi_{rr} =\psi^3$. Moreover, $anchor(\psi_\l) = v_{rr}$ and $anchor(\psi_r) = v_{\l\l}$. This is the only case where $w$ could possibly be $u$. Note that this is a special case since $t_{\l\l}$ and $t_{rr}$ share a common edge $v_{\l\l} v_{rr}$, hence we cannot use them both for an improving swap. But still there exists another solution improving $4$-swap by replacing the set of triangles $\sset=\{\psi, \psi_\l, \psi_r, \psi^3\}$ by set $\sset'= \{t_{\l\l}, t_\l, t_r, t'_{rr}, t'\}$, where $t'$ is the singly-attached triangles of $\psi^3$ attached to edge $\overbar{w v_{rr}}$ and $t'_{rr} = \Delta_{v_{\l\l} v_\l v_r}$. It is easy to that the triangles in $\sset'$ are edge disjoint even in the worst case when $w=u$. For illustration see fig~\ref{fig:demand-two-doubly-attached-two-type-1-common-type-3-anchors-vll-vrr}.

        \begin{figure}[ht]
   \centering

\subfloat{
\begin{tikzpicture}[scale=0.6, every node/.style={scale=0.8}]
\draw [fill=lightgray] (0,0) -- (1.5,2) -- (3,0) -- (0,0);
\draw [fill=lightgray]  (3,0) -- (6,0) -- (4.5,2) -- (3,0);
\draw [fill=lightgray] (6,0) -- (7.5,2) -- (9,0) -- (6,0);
\draw [line width=2pt, fill=lightgray] (0,0) to [out=-20,in=200] (9,0) -- (4.5, -2) -- (0,0) ;
\GraphInit[vstyle=Normal]
\tikzset{VertexStyle/.style = {shape=circle, fill=white, draw=black, inner sep=1pt, minimum size=20pt}}

\Vertex[Math, L=v_{\ell\ell}]{a}
\Vertex[Math, L=w_\ell, y=2,x=1.5]{b}
\Vertex[Math, L=v_\ell, y=0,x=3]{c}
\Vertex[Math, L=u ,y=2,x=4.5]{d}
\Vertex[Math, L=v_r, y=0,x=6]{e}
\Vertex[Math, L=w_r, y=2,x=7.5]{f}
\Vertex[Math, L=v_{rr}, y=0,x=9]{g}
\Vertex[Math, L=w, y=-2,x=4.5]{h}
\Vertex[Math, L=x, y=-4,x=4.5]{x}

\SetUpEdge[
   lw = 2pt,
   color = black,
]
\Edges(a,b,c,a)
\Edges(c,d,e,c)
\Edges(e,f,g,e)
\SetUpEdge[
   lw = 1pt,
   color = black,
   style = dashed
]
\Edge(b)(d)
\Edge(d)(f)
\Edge(a)(f)
\Edge(b)(g)
\Edge(h)(x)
\Edge(a)(x)
\Edge(g)(x)

\tikzstyle{EdgeStyle}=[bend right=15]
\Edge(a)(e)
\Edge(c)(g)
\end{tikzpicture}
}
    \caption{$\psi^3=\psi_{\ell\ell}=\psi_{rr}=\Delta^3_{v_{\ell\ell} v_{rr} w}$ is type-$3$; $\psi=\Delta^0_{v_\ell v_r u}$ is type-$0$ and; $\psi_\ell = \Delta^1_{v_\ell w_\ell v_{\ell\ell}}$ and $\psi_r = \Delta^1_{v_r w_r v_{rr}}$ are type-$1$. The poor base edges of $\Delta^1_{v_\ell w_\ell v_{\ell\ell}}$ and $\Delta^1_{v_r w_r v_{rr}}$ are adjacent to doubly-attached triangles $t_\ell = \Delta_{v_\ell u w_\ell}$ and $t_r = \Delta_{v_r u w_r}$ respectively. The common type-$3$ triangle $\Delta^3_{v_{\ell\ell} v_{rr} w}$ is the one responsible for making the two type-$1$ triangles' $\psi_\ell$ and $\psi_r$ base edge poor using vertices $v_{rr}$ and $v_{\ell\ell}$ as anchors respectively. The other two doubly-attached triangles are  $t_{\ell\ell} = \Delta_{v_{rr} w_\ell v_{\ell \ell}}$ and $t_{rr} = \Delta_{v_{rr} w_r v_{\ell \ell}}$ sharing the edge $\overbar{v_{rr} v_{\ell \ell}}$. The singly attached triangle of $\psi^3$ used for the swap is $t' = \Delta_{w v_{rr} x}$ and the new doubly attached triangle is $t'_{rr} = \Delta_{v_{\ell\ell} v_\ell v_r}$. \textbf{Impossible because there is an improving $4$-swap.}}
    
   \label{fig:demand-two-doubly-attached-two-type-1-common-type-3-anchors-vll-vrr}
\end{figure}




      \end{enumerate}
   \end{proof}

    \begin{lemma}
      \label{lem:one-doubly-attached-demanding}
      If $t_\l$ is doubly-attached, then $t_r$ cannot be hollow.
    \end{lemma}
    \begin{proof}
      Suppose not, then by renaming, let $t_\l$ and $t_r$ be the doubly-attached triangle and hollow triangle respectively sharing demanding edges $e_\l \neq e_r$ with $\psi$. Let $\psi_\l$ be the other solution triangle that share an edge with $t_\l$. 
      \sumii{Let $\psi_\l$ be the other solution triangle that share an edge with $t_\l$ and; $\psi_r, \psi_h$ be the other two-solution triangles that share an edge with $t_r$. Let $v_\l$ be the common vertex of $\psi$ and $\psi_\l$. Let $v_r, v_h$ be the common vertex of $\psi$ and $\psi_r$ and; $\psi$ and $\psi_h$ respectively. Let $u_\l$ to be the other vertex of $\psi$ in $t_\l$ which is incident to the non-solution edge of $t_\l$. By renaming, we can assume that either $v_r = v_\l$(say $v$) or $u_\l = v_h$ but not both.}\sumi{This naming will be in sync with the two-doubly proof.}
      Let $\psi_h$ be a solution triangle that share an edge with $t_r$ and a vertex with $t_\l$ and let $\psi_r$ be the other solution triangle that share an edge with $t_r$.
      \input{tikz_figures/demand-one-doubly-attached-one-type-3-3-hollow-examples.tex}
      Let $v_\l$ be the common vertex of $\psi$ and $\psi_\l$. Let $v_r$ be the common vertex of $\psi$ and $\psi_r$. Let $v_h$ be the common vertex of $\psi$ and $\psi_h$.
      Let $u$ be the vertex of $\psi$ adjacent to non-solution edge of $t_\l$. Note that $u = v_h$ or $v_\l = v_h$. 
      Let $w_\l$ be the remaining vertex of $t_\l$ and $w_r$ be the remaining vertex of $t_r$. The remaining vertices of $\psi_\l$, $\psi_r$, $\psi_h$ are named $v_{\l\l}$, $v_{rr}$, $v_{rh}$, respectively.
      
      Note that $\psi_\l$, $\psi_r$, $\psi_h$ can be type-$1$ or type-$3$. If it is type-$1$, then its base edge must be drained. In that case, we will name additional vertices using the similar manner. For example, if $\psi_\l$ ($\psi_r, \psi_h$) is type-$1$, the $\overbar{v_\l w_\l}$ ($\overbar{v_r w_r}, \overbar{v_h w_r}$) must be drained base edge. The anchoring vertex of $\psi_\l$ ($\psi_r, \psi_h$) will be named $w_{\l\l}$ ($w_{rr}, w_{rh}$). The type-$3$ triangle which contains $\overbar{v_{\l\l} w_{\l\l}}$ ($\overbar{v_{rr} w_{rr}}, \overbar{v_{rh} w_{rh}}$ ) is $\psi_{\l\l}$ ($\psi_{rr}, \psi_{rhr}$). The third vertex of $\psi_{\l\l}$ ($\psi_{rr}, \psi_{rhr}$) is $x_{\l\l}$ ($x_{rr}, x_{rh}$).
      See the Figure~\ref{fig:demand-one-doubly-attached-one-type-3-3-hollow-examples} for illustration. Note that some of these vertices can actually be the same vertex.
      
      Now we will prove the lemma for each sub-case separately. 
      
     \begin{enumerate}
         \item \label{itm:demand-doubly-hollow-3-3-3} $\psi_\l, \psi_r, \psi_h$ are all type-$3$. First, notice that $\psi_{\l} \neq \psi_r$ and $\psi_h \neq \psi_{r}$ (or at least one edge will intersect $E(\psi)$).
         In one case where $v_\l = v_h$, $\psi_\l = \psi_h$ is possible.
         Moreover, $a_h = anchor(\psi_h) \neq a_r = anchor(\psi_r)$ or $\psi$ will not be type-$0$. Let $\sset =\{\psi_\l, \psi, \psi_r, \psi_h \}$. For most sub-cases, we will construct the set $\sset^\prime$ of $5$ disjoint triangles. Once we have $\sset^\prime$, we reach a contradiction as substituting $\sset$ with $\sset^\prime$ in $\Nu$ yields a larger packing solution which contradict the fact that $\Nu$ is the optimal packing solution. 
         
        \begin{enumerate}
            \item $u = v_h$
                \begin{enumerate}
                    \item $a_\l \notin \{v_{rh}, w_r, v_{rr}\}$: Note that $a_\l \notin V(\psi)$ either. Hence, we can select 
                    $$\sset^\prime = \{
                        t_\l,
                        t_r,
                        \Delta_{v_{rh}w_ra_h},
                        \Delta_{w_r v_{rr} a_r},
                        \Delta_{w_\l v_{\l\l}, a_\l}
                    \}.$$
                    \item $a_\l = v_{rh}$: In this case, $a_h = w_\l$ because $\Delta_{v_h v_{rh} w_\l}$ is a singly-attached triangle of $\psi_h$ and any type-$3$ triangle must have only one anchoring vertex.
                    Hence, we can select
                    $$\sset^\prime = \{
                        t_\l,
                        t_r,
                        \Delta_{a_\l, v_\l, v_{\l\l}},
                        \Delta_{v_{rh}w_ra_h},
                        \Delta_{w_r v_{rr} a_r}
                    \}.$$
                    \item $a_\l = w_{r}$: In this case, $a_h = w_\l$ because $\Delta_{v_h w_r w_\l}$ is a singly-attached triangle of $\psi_h$ and any type-$3$ triangle must have only one anchoring vertex.
                    Hence, we can select
                    $$\sset^\prime = \{
                        t_\l,
                        t_r,
                        \Delta_{a_\l, v_\l, v_{\l\l}},
                        \Delta_{v_{rh}w_ra_h},
                        \Delta_{v_r v_{rr} a_r}
                    \}.$$
                    \item $a_\l = v_{rr}$: Notice that  $\Delta_{v_\l v_r v_{rr}}$ becomes a doubly-attached triangle, which mean $\overbar{v_\l v_r}$ is also a demanding edge. By renaming, this is impossible by Lemma~\ref{lem:both-doubly-attached-demanding}.
                        
                \end{enumerate}
            \item $v_\l = v_h$
                \begin{enumerate}
                    \item $\psi_\l = \psi_h$: In this case, $|\sset| =3$, so we only need $\sset^\prime$ of size 4. Note that $a_\l = a_h \neq a_r$ or $\psi$ would not be type-$0$. Hence, we can select
                    $$\sset^\prime = \{
                        t_\l,
                        t_r,
                        \Delta_{w_\l w_r a_l},
                        \Delta_{v_r v_{rr} a_r}
                    \}.$$
                    \item $a_\l \notin \{v_{rh}, w_r, v_{rr}\}$: Note that $a_\l \notin V(\psi)$ either. Hence, we can select 
                    $$\sset^\prime = \{
                        t_\l,
                        t_r,
                        \Delta_{v_{rh}w_ra_h},
                        \Delta_{w_r v_{rr} a_r},
                        \Delta_{w_\l v_{\l\l}, a_\l}
                    \}.$$
                    \item $a_\l = v_{rh}$: Impossible as $\overbar{v_\l v_{rh}} \in E(\psi_h)$.
                    \item $a_\l = w_{r}$: Impossible as $\overbar{v_\l w_r} \in E(\psi_h)$.
                    \item $a_\l = v_{rr}$: In this case, $\Delta_{v_\l v_r v_{rr}}$ becomes a doubly-attached triangle which also demand credit from $\overbar{v_\l v_r}$, hence this is impossible by Lemma~\ref{lem:both-doubly-attached-demanding}. 
                \end{enumerate}
                
        \end{enumerate}
         
         \item $\psi_\l, \psi_h$ are type-$3$ and $\psi_r$ is type-$1$.
         
         \begin{enumerate}
             \item $u = v_h$ In this case, note that $\psi_\l, \psi_r, \psi_h$ are three different triangles. Consider $\psi_{rr}$, notice that $\psi_{rr} \neq \psi_h$ as $V(\psi_r) \cap V(\psi_{rr}) \neq V(\psi_r) \cap V(\psi_h)$. Moreover, $a_{rr} \neq a_h$ or $\psi_r$ is a type-$2$ triangle. $\psi_{rr} \neq \psi_\l$ or $w_\l = w_{rr}$ and $\psi$ would not be type-$0$. Let $t_1$ 
             \begin{enumerate}
                 \item $a_{rr} \notin \{u, w_\l, a_\l\}$: In this case, by substituting $\psi_{rr}$ with either $\Delta_{v_{rr}x_{rr}a_{rr}}$ or $\Delta_{w_{rr}x_{rr}a_{rr}}$, 
                 we create a situation in \ref{itm:demand-doubly-hollow-3-3-3}. Hence, by reusing the statement above, we know this will lead to a contradiction.
                 \item $a_{rr} = a_\l$:
             \end{enumerate}
                
             \item $v_\l = v_h$
         \end{enumerate}
         

         
         
     \end{enumerate}

    \end{proof}

    \begin{lemma}
    \label{lem:both-hollow-demanding}
      If $t_\l$ is hollow, then $t_r$ cannot be hollow.
    \end{lemma}
    \begin{proof}
      Suppose not, then $t_\l$ and $t_r$ are hollow triangles sharing demanding edges $e_\l \neq e_r$ with $\psi$. Let $\psi_\l, \psi_{\l h}$ be the other two-solution triangles that share an edge with $t_\l$ and; $\psi_r, \psi_{rh}$ be the other two-solution triangles that share an edge with $t_r$. By renaming, we can assume that $v_\l$ is the common vertex between $\psi$ and $\psi_\l$, $v_r$ is the common vertex between $\psi$ and $\psi_r$ and $v_h$ is the common vertex of the three triangles $\psi, \psi_{\l h}, \psi_{rh}$. Let $w_\l, w_r$ be the third vertex of $t_\l, t_r$ respectively. Similarly let $v_{\l \l}, v_{rr}, v_{\l h}, v_{rh}$ be the third vertex of $\psi_\l, \psi_r, \psi_{\l h}, \psi_{r h}$ respectively.
    \input{tikz_figures/demand-two-hollow-two-type-0-3-3.tex}
    \end{proof}

    The above three lemmas together imply that $t$ cannot have two demanding edges $e_\l$ and $e_r$.
\end{proof}